\documentclass[preprint,aps,showpacs,prb,amsmath,amssymb,floatfix]{revtex4-1}
\usepackage{graphicx}
\usepackage{dcolumn}
\usepackage{bm}
\usepackage{rotating}
\usepackage{float}
\usepackage{color}
\usepackage{subfigure}
\pdfoutput=1
\begin{document}

\preprint{Version 4 \date{\today} }
\title{
      First-principles study of magnetism, structure and chemical order in small FeRh alloy clusters
      }
\author{Junais Habeeb Mokkath}\author{G. M. Pastor} 
 \affiliation{%
Institut f{\"u}r Theoretische Physik, 
Universit{\"a}t Kassel, 
Heinrich Plett Stra{\ss}e 40, 34132 Kassel, Germany
      }
\date{\today}

\begin{abstract}
The structural, electronic and magnetic properties of small 
${\rm Fe}_m {\rm Rh}_n$ clusters having $N = m+n \leq 8$ atoms 
are studied in the framework of a generalized-gradient approximation 
to density-functional theory.
The correlation between structure, chemical order, and 
magnetic behavior is analyzed as a function of size and composition. 
For $N = m+n \leq 6$ a thorough sampling
of all cluster topologies has been performed, while for $N = 7$ and $8$  
only a few representative topologies are considered. In all cases the entire 
concentration range is systematically investigated.
All the clusters show ferromagnetic-like order in the optimized structures. As a result, the average magnetic moment per atom $\overline\mu_N$ increases 
monotonously, which is almost linear over a wide range of concentration with Fe content.
A remarkable enhancement of the local Fe moments beyond 3~$\mu_B$ is
observed as result of Rh doping. This is a 
consequence of the increase in the number of Fe $d$ holes, 
due to charge transfer from Fe to Rh, combined with 
the extremely reduced local coordination.
The Rh local moments, which are important already in the
pure clusters ($N\le 8$) are not significantly enhanced
by Fe doping. However, the overall stability of magnetism,
as measured by the energy gained upon spin polarization, 
increases when Rh is replaced by Fe. The composition 
dependence of the electronic structure and the influence of spin-orbit 
interactions on the cluster stability are discussed.

\end{abstract}
%

%

\maketitle

\section{Introduction}
\label{sec:introd}

Alloying elements with complementary qualities in order to tailor 
their physical behavior for specific technological purposes
has been a major route in material development since 
the antiquity. Cluster research is no exception to this trend. 
After decades of systematic studies of the size and structural
dependence of the most wide variety of properties of monoelement 
particles, the interest has actually been moving progressively 
over the past years towards investigations on finite-size binary 
alloys.\cite{faraday_vol}  The magnetism of transition-metal (TM) 
clusters opens numerous possibilities and challenges in this 
context.\cite{zitoun02,dennlerCoRh,IrenaPdCu,mukulCoMn,peter09,Antonis04,georg08,bansmann,antoniak06,yin,knick07,%
farle08,gruner08,four08,four09,dupuis08} 
For example, one would like to understand
how to modify the magnetic characteristics of clusters, in particular 
the saturation magnetization and the magnetic anisotropy energy (MAE),
as it has been done in solids. This would indeed
allow one to design new nanostructured materials from a
microscopic perspective. Nevertheless, it also true that 
controlling composition, system size, and magnetic  
behavior sets serious difficulties for both experiment and theory. 

Pure TM clusters such as Fe$_N$, Co$_N$ and Ni$_N$ show spin moments, 
orbital moments, and MAEs that are enhanced with respect to the 
corresponding periodic 
solids.\cite{pastor89,bucher91,billas93,prl_mae,apsel96,knick01,prb_gin} 
Still, the possibilities of optimizing the cluster magnetic behavior 
by simply tuning the system size have been rather disappointing, 
particularly concerning the MAE, which remains relatively small 
---despite being orders of magnitude larger than in solids\cite{prl_mae}--- 
due to the rather weak spin-orbit (SO) coupling in the $3d$ atoms. 
This is one of the motivations for alloying $3d$ TMs with $4d$ and 
$5d$ elements which, being heavier, are subject to stronger SO 
interactions. In this context it is useful to recall that large 
nanoparticles and three dimensional 
solids of these elements are non-magnetic. However, at very 
small sizes the $4d$ and $5d$ clusters often develop a finite spontaneous
low-temperature magnetization, due to the reduction of local 
coordination and the resulting $d$-band 
narrowing.\cite{galicia,onset,coxRh,reddy,pedroRh}
The first experimental observation of this important finite-size 
effect has been made by Cox {\em et al.}
by performing Stern-Gerlach-deflection measurements 
on Rh$_N$ clusters. In this work the average magnetic moments per 
atom $\overline\mu_N$ = $0.15$--$0.80 \mu_B$ have been experimentally determined
for $N \le 30$--$50$ atoms.\cite{coxRh} In view of these contrasting 
features one expects that $3d$-$4d$ and $3d$-$5d$
alloy clusters should show very interesting 
structural, electronic and magnetic behaviors. 

The purpose of this paper is to investigate the ground-state properties 
of the small FeRh clusters in the framework of Hohenberg-Kohn-Sham's 
density functional 
theory.\cite{hks} Besides the general interest of the problem 
from the perspective of $3d$-$4d$ nanomagnetism, these clusters
are particularly appealing because of the remarkable phase diagram 
of FeRh bulk alloys.\cite{phdiag} In the case of ${\rm Fe}_{50} {\rm Rh}_{50}$
the magnetic order at normal pressure and low temperatures is 
antiferromagnetic (AF). As the temperature increases this $\alpha''$ 
phase undergoes a first order transition to a ferromagnetic (FM) 
state, the $\alpha'$ phase, which is accompanied by a change in 
lattice parameter. The corresponding transition temperature 
$T_c^{\alpha'\alpha''}$ increases rapidly with increasing external 
pressure P, eventually displacing the FM $\alpha'$ phase completely 
for $P\ge 7$~GPa 
($T_c^{\alpha'\alpha''}\simeq290 K$ for ${\rm Fe}_{50} {\rm Rh}_{50}$ 
at normal pressure). Moreover, $T_c^{\alpha'\alpha''}$ decreases very 
rapidly with decreasing Rh content. 
At low pressures the FM $\alpha'$ phase undergoes a FM 
to paramagnetic (PM) transition at ($T_C \simeq 670 K$).\cite{phdiag} 
In addition, the properties of $\alpha$-FeRh bulk alloys have been the subject 
of first principles and model theoretical investigations.\cite{Entel_FeRh}
In particular these show that the relative stability of the
FM and AF solutions depends strongly on the interatomic distances.
Such remarkable condensed-matter effects enhance the
appeal of small FeRh particles as specific example of
$3d$-$4d$ nanoscale alloy. Investigations of their magnetic properties as a function of 
size, composition, and structure are therefore of fundamental
importance.

The remainder of the paper is organized as follows. 
In Sec.~\ref{sec:teo} the main details of the theoretical background
and computational procedure are presented. This includes in particular 
a description of the strategy used for exploring the cluster energy 
landscape as a function of geometrical conformation and chemical order. 
The results of our calculations for FeRh clusters having $N\le 8$ 
atoms are reported in Secs.~\ref{sec:stmagn} and \ref{sec:comp}.
First, we focus on the interplay between structure, chemical order and magnetism 
in the most stable geometries for different cluster sizes.  
Second, we analyze the concentration dependence of the cohesive energy, 
the local and average magnetic moments, and the spin-polarized electronic
structure. Finally, we conclude in Sec.~\ref{sec:concl} with a summary of 
the main trends and an outlook to future extensions.

\section{Computational aspects}
\label{sec:teo}

The calculations reported in this work have been performed in the
framework of Hohenberg-Kohn-Sham's density functional theory,\cite{hks}
as implemented in the Vienna ab initio simulation package (VASP).\cite{vasp} 
The exchange and correlation energy is described by using both the spin-polarized 
local density approximation (LDA) and 
Perdew and Wang's generalized-gradient approximation (GGA).\cite{pw91} 
The VASP solves the spin-polarized Kohn-Sham equations in an augmented 
plane-wave basis set, taking into account the core electrons within the 
projector augmented wave (PAW) method.\cite{paw} 
This is an efficient frozen-core all-electron approach which allows to 
incorporate the proper nodes of the Kohn-Sham orbitals in the core region
and the resulting effects on the electronic structure, total energy and 
interatomic forces. The $4s$ and $3d$ orbitals of Fe, and the $5s$ and 
$4d$ orbitals of Rh are treated as valence states. The wave functions 
are expanded 
in a plane wave basis set with the kinetic energy cut-off $E_{max}=268$~eV. 
In order to improve the convergence of the solution of the selfconsistent 
KS equations the discrete energy levels are broadened by using a Gaussian smearing
$\sigma = 0.02$~eV. The validity of the present choice of computational parameters 
has been verified.\cite{foot-acc} The PAW sphere radii for Fe and Rh are $1.302$~{\AA} and $1.402$~{\AA}, 
respectively. A simple cubic supercell is considered with 
the usual periodic boundary conditions. The linear size of the cell is $a$ = $10$--$22$~{\AA}, so that 
any pair of images of the clusters are well separated and 
the interaction between them is negligible.
Since we are interested in finite systems, the reciprocal space summations are restricted 
to the $\Gamma$ point.

Although the potential advantages of alloying magnetic $3d$ elements 
with highly-polarizable $4d$ or $5d$ elements 
can be grasped straightforwardly, the problem 
involves a number of serious practical challenges.
Different growth or synthesis conditions can lead to different 
chemical orders, which can be governed not just by energetic reasons
but by kinetic processes as well. For instance, one may have to deal with
segregated clusters having either a $4d$ core and a $3d$ outer shell or 
vice versa. Post-synthesis manipulations can induce different 
degrees of intermixing, including for example surface diffusion 
or disordered alloys. 
Moreover, the inter atomic distances are also expected to 
depend strongly on size and composition. Typical TM-cluster bond-lengths 
are in fact $10$--$20$\% smaller than in the corresponding bulk crystals. 
Taking into account that itinerant $\rm 3d$-electron magnetism
is most sensitive to the local and chemical environments of the 
atoms,\cite{onset,faradayCoRh,phm,garyclus} 
it is clear that controlling the distribution of the elements 
within the cluster   
is crucial for understanding magnetic nanoalloys.

Systematic theoretical studies of binary-metal clusters are hindered by the
diversity of geometrical conformations, ordered and disorder arrangements, 
as well as segregation tendencies that have to be taken into account. This 
poses a serious challenge to both first-principles and model 
approaches. In order to determine the interplay between cluster structure, 
chemical order and magnetism in FeRh clusters we have performed a 
comprehensive set of electronic calculations for clusters having 
$N \le 8$ atoms. In the present paper we focus on the most stable 
cluster structure and magnetic configuration,
which are determined by exploring the ground-state energy landscape.\cite{tobe}
This is a formidable task, since one needs to consider a large, most possibly complete and unbiased set of 
initial structures. Such a thorough geometry optimization must include not only the representative cluster
geometries or topologies, but also all relevant chemical orders. This requires taking into account
all distributions of the Fe and Rh atoms for any given size and composition.
These two aspects of the problem of determining the structure of nanoalloys are discussed in more detail in the following. 

The different cluster topologies are sampled
by generating all possible graphs for $\rm N\le 6$ 
atoms as described in Ref.~\onlinecite{phm} (see also Ref.~\onlinecite{Wang}).
For each graph or adjacency matrix it is important to verify 
that it can be represented by a true structure in $\rm D\le 3$ dimensions. 
A graph is acceptable as a cluster structure, only if a set of atomic 
coordinates $\vec R_i$ with $i = 1, \dots, N$ exists, such that the 
interatomic distances $R_{ij}$ satisfy the conditions
$R_{ij}=R_0$ if the sites $i$ and $j$ are connected in the graph 
(i.e., if the adjacency matrix element $A_{ij} =1$) 
and $R_{ij}> R_0$ otherwise (i.e., if $A_{ij} =0$). 
Here $R_0$ refers to the nearest neighbor (NN) distance, which 
at this stage can be regarded as the unit of length, assuming for 
simplicity that it is the same for all clusters.
Notice that for $N\le 4$ all graphs are possible cluster 
structures. For example, for $N = 4$, the different structures are the tetrahedron, rhombus,
square, star, triangular racket and linear chain.\cite{phm} However, for $N\ge 5$ there are graphs, i.e., topologies, which cannot
be realized in practice. For instance, it is not possible to have five 
atoms being NNs from each other in a three dimensional space. Consequently, for $N\ge 5$ there are less real structures than mathematical graphs. 
The total number of graphs (structures) is 21 (20), 112 (104), and 853 (647)
for $N = 5, 6$, and $7$, respectively.\cite{phm} 

For clusters having $N\le 6$ atoms all these topologies have
indeed been taken as starting points of our structural relaxations.
Out of this large number of different initial configurations 
the unconstrained relaxations using VASP lead to only a few geometries, 
which can be regarded as stable or metastable isomers. For larger clusters
($N = 7$ and $8$) we do not aim at performing a full global optimization. Our purpose here is to explore
the interplay between magnetism and chemical order as a function of composition
for a few topologies that are representative of open and close-packed structures. Taking
into account our results for smaller sizes, and the available information
on the structure of pure Fe$_N$ and Rh$_N$ clusters, we have restricted the 
set of starting topologies for the unconstrained relaxation of FeRh
heptamers and octamers to the following: bicapped trigonal bipyramid, capped octahedra, and pentagonal bipyramid for $N = 7$, and
tricapped trigonal bipyramid, bicapped octahedra, capped pentagonal bipyramid 
and cube for $N = 8$. Although, the choice of topologies for $N = 7$ and $8$ is quite restricted, it includes
compact as well as more open structures. Therefore, it is expected to shed light on the dependence 
of the magnetic properties on the chemical order and composition.

The dependence on concentration is investigated systematically for each topology  
of Fe$_m$Rh$_n$ by varying $m$ and for each size $N = m+n \leq 8$, including the pure Fe$_N$ and Rh$_N$ limits. Moreover, we 
take into account all possible non-equivalent distributions of the $m$ Fe 
and $n$ Rh atoms within the cluster. In this way, any {\em a priori}
assumption on the chemical order is avoided. Obviously, such an
exhaustive combinatorial search increasingly complicates the computational 
task as we increase the cluster size, and as we move away from pure clusters 
towards alloys with equal concentrations. Finally, in order to
perform the actual density-functional calculations we set for 
simplicity all NN distances in the starting cluster geometry equal to the Fe bulk  
value\cite{foot_R0} $R_0 = 2.48$~{\AA}. Subsequently, a fully unconstrained geometry optimization is performed from 
first principles by using the VASP.\cite{vasp} 
The atomic positions are fully relaxed by means of conjugate gradient 
or quasi-Newtonian methods, without imposing any symmetry constraints, 
until all the force components are smaller than the threshold
$5$~meV/{\AA}. The convergence criteria are set 
to $10^{-5}$~eV/{\AA} for the energy gradient, and 
$5\times10^{-4}$~{\AA} for the atomic displacements.\cite{vasp_manual} 
The same procedure applies to all considered clusters 
regardless of composition, chemical order, or total magnetic moment.
Notice that the diversity of geometrical structures and atomic arrangements 
often yields many local minima on the ground-state energy surface, which complicates 
significantly the location of the lowest-energy configuration.

Lattice structure and magnetic behavior are intimately related in TMs,
particularly in weak ferromagnets such as Fe and its alloys.\cite{foot_struct}
On the one side, the optimum structure and chemical order depend on the
actual magnetic state of the cluster as given by the average magnetic 
moment per atom $\overline\mu_N$ and the magnetic order. On the other side, 
the magnetic behavior is known to be different for different structures
and concentrations. Therefore, in order to rigorously determine the 
ground-state magnetic properties of FeRh clusters, we have varied systematically the value of the total 
spin polarization of the cluster $S_z$ by performing fixed spin-moment (FSM) calculations in the whole physically 
relevant range. Let us recall that $S_z = (\nu_\uparrow - \nu_\downarrow)/2$
where $\nu_\uparrow$($\nu_\downarrow$) represents the number of electrons in the majority (minority)
states. In practice we start from the non-magnetic 
state ($S_z^{min} = 0$) and increase $S_z$ until the local spin moments
are fully saturated, i.e., until the Fe moments in the PAW sphere
reach $\mu_{\rm Fe}\simeq 4 \mu_B$ and the Rh moments 
$\mu_{\rm Rh}\simeq 2.5 \mu_B$ (typically, $S^{max}_{z} \gtrsim 3N/2$).
The above described global geometry optimizations  are performed 
independently for each value of $S_z$. These FSM study
provides a wealth of information on the isomerization energies, the spin-excitation energies, and their 
interplay. These are particularly interesting for a 
subtle magnetic alloy such as FeRh, and would therefore deserve to be analyzed in some more detail. 
In the present paper we shall focus
on the ground-state properties by determining for each considered 
Fe$_m$Rh$_n$ the most stable structural and magnetic
configuration corresponding to energy minimum as a function of $S_z$
and of the atomic positions.\cite{tobe}

Once the optimization with respect to structural and magnetic degrees of 
freedom is achieved, we derive the binding energy per atom 
$E_B = [m E({\rm Fe}) + n E({\rm Rh}) - E({\rm Fe}_m {\rm Rh}_n) ]/ N$
in the usual way by referring the total energy $E$ to the corresponding 
energy of $m$ Fe and $n$ Rh isolated atoms. Moreover, for each stationary 
point of the total energy surface (i.e., for each relaxed structure having 
a nearly vanishing $\|{\vec{\nabla}E}\|$)
we determine the vibrational frequencies from the diagonalization of 
the dynamical matrix. The latter is calculated from finite differences 
of the analytic gradients of the total energy. In this way we can 
rule out saddle points to which the local optimization procedure
happens to converge on some occasions. Only configurations which 
correspond to true minima are discussed in the following. 
Finally, a number of electronic and magnetic properties 
---for example, the magnetic energy 
$\Delta E_m = E(S_z \! = \! 0) - E(S_z)$, the local 
magnetic moments $\mu_{i}$ integrated within the Wigner-Seitz (WS) or Bader atomic 
cells of atom $i$,\cite{bader,thesis_jl} and the spin polarized 
density of electronic states (DOS) 
$\rho_{\sigma}(\varepsilon)$--- are derived from the self-consistent
spin-polarized density and Kohn-Sham spectrum.

\section{Structure and magnetism}
\label{sec:stmagn} 
\label{sec:2} 
In this section we discuss the ground-state structure, 
chemical order, binding energy, and magnetic moments of Fe$_m$Rh$_n$ 
clusters having $N = m + n\le 8$ atoms. The main emphasis is here on understanding how the various
electronic, structural and magnetic properties depend on the chemical composition of the alloy.
First, each cluster size $N$ is analyzed separately, since a strong 
dependence on $N$ is expected in the small size, non-scalable regime.
Comparisons between the various $N$ are stressed by means 
of cross-references between different subsections.
In addition the main trends as a function of size and concentration
are summarized in Sec.~\ref{sec:comp}. 
\subsection{FeRh dimers}
Despite being the simplest possible systems, dimers allow to infer very useful 
trends on the relative strength, charge transfers and magnetic order
in the various types of bonds which are 
found in FeRh alloy clusters. The results summarized 
in Table~\ref{tab:2} show that the FeRh bond
yields the highest cohesive energy, followed by the 
Rh$_2$ bond, the Fe$_2$ bond being the weakest. The particular
strength of the heterogeneous bond is confirmed by the fact
that the corresponding vibrational frequency is the highest. 
The bond length, however, follows the trend of the atomic 
radius which, being larger for Rh, gives 
$d_{\rm RhRh}>d_{\rm FeRh}>d_{\rm FeFe}$. 
Quantitatively, the binding energy per atom $E_B^\mathrm{GGA} = 1.35$~eV obtained
for Fe$_2$ within the GGA is smaller than
the LDA result $E_B^\mathrm{LDA} = 2.25$~eV \cite{compa_stru} 
although it still remains larger than the experimental value 
$E_B^\mathrm{expt} = 0.65$~eV reported in Ref.~\onlinecite{mosk}.
The calculated vibrational frequency 
$\nu_{0}({\rm Fe}_2) = 288~{\rm cm^{-1}}$ is consistent with 
previous experimental results [$\nu_{0}({\rm Fe}_2) = 299.6~{\rm cm^{-1}}$ 
from Ref.~\onlinecite{mosk} and $\nu_{0}({\rm Fe}_2) = 300\pm 15~{\rm cm^{-1}}$
from Ref.~\onlinecite{Leo}].
Our result for $E_B$ and $\overline\mu_N$ of Rh$_2$ coincide
with previous GGA calculations by B. V. Reddy {\em et al.}\cite{Rh_nayak}
These are however larger than the experimental values 
$E_B^\mathrm{expt}({\rm Rh}_2) = 1.46$~eV derived from Knudsen effusion,\cite{Rh_Cocke}
$E_B^\mathrm{expt}({\rm Rh}_2) = 0.70 \pm 0.15$~eV derived from resonance Raman in 
Ar matrices\cite{Rh_wang} and $E_B^\mathrm{expt}({\rm Rh}_2) = 1.203$~eV 
derived from the resonant two-photon ionization.\cite{Rh_lang} 
The calculated vibrational frequency of 
$\nu_{0}({\rm Rh}_2)^\mathrm{GGA} = 224~{\rm cm^{-1}}$ should be compared
with the experimental value $\nu_{0}({\rm Rh}_2)^\mathrm{expt} = 283.9~{\rm cm^{-1}}$ 
reported in Ref.~\onlinecite {Rh_wang}.
\begin{table}
\caption{(Color online)\label{tab:2}
Structural, electronic and magnetic properties of FeRh dimers. 
Results are given for the binding energy $E_B$ (in eV), 
the magnetic stabilization energy $\Delta E_{m}$ = $E(S_z \!= \!0) - E(S_z)$ 
(in eV), the average interatomic distance $d_{\alpha\beta}$ (in {\AA})
between atoms $\alpha$ and $\beta$ ($\alpha, \beta = {\rm Fe}$ or Rh), 
the average spin moment per atom $\overline\mu_{N} = 2 S_z /N$ (in {$\mu_B$}), 
the local spin moment $\mu_{\alpha}$ (in {$\mu_B$}) at the Fe or Rh atoms, 
and the vibrational frequency $\nu_{0}$ (in $\rm cm^{-1}$).
        }
\renewcommand{\arraystretch}{0.5}
\begin{tabular}{|c|c|c|c|c|c|c|c|c|} 
\hline 
\hline
            &                                          &     &     &       &        &          &               &         \\ 
Cluster     &Struct. &$E_B$&$\Delta{E_m}$&$d_{\alpha\beta}$  &$\overline\mu_{N}$&$\mu_{\rm Fe}$&$\mu_{\rm Rh}$ &$\nu_{0}$\\ \hline
            &                                          &     &     &       &        &          &               &         \\
Fe$_2$      &\includegraphics[scale=0.30]{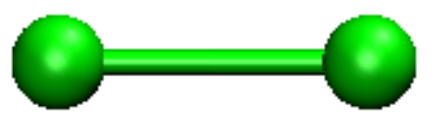} &1.35 &0.77&1.98   &3.00      &2.82      &               &288      \\
            &                                          &     &     &       &        &          &               &         \\
FeRh        &\includegraphics[scale=0.25]{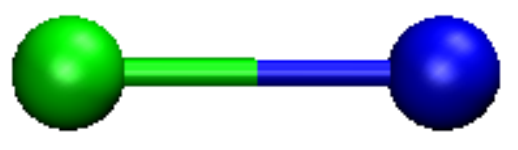}&1.95 &0.24&2.07   &2.50    &3.34      &1.33           &359      \\
            &                                          &     &     &       &        &          &               &         \\
Rh$_2$      &\includegraphics[scale=0.22]{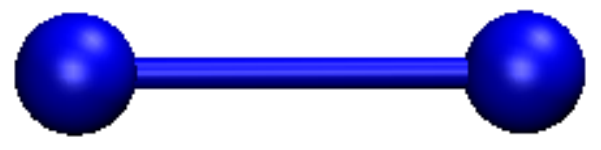} &1.65 &0.00 &2.21   &2.00       &          &1.83           &224      \\
            &                                          &     &     &       &        &          &               &         \\
\hline \hline
\end{tabular}
\end{table}

The stability of magnetism, as measured by the difference in the 
total energy $\Delta E_m$ of the non-magnetic ($S_z =0$)
and optimal magnetic solutions, is largest for Fe$_2$ and smallest for 
Rh$_2$. The same trend holds for the average magnetic moment 
per atom which decreases linearly from $\overline\mu_2 = 3 \mu_B$ to $2 \mu_B$
as one goes from Fe$_2$, to FeRh, to Rh$_2$. These average magnetic moments
per atom correspond to a full polarization of all $d$ electrons in the WS spheres: $\nu_d \simeq 7$ 
for Fe and $\nu_d \simeq 8$ for Rh, where $\nu_d$ stands for the number of valence $d$ electrons of the corresponding atom. 
The local magnetic moments $\mu_\alpha$ ($\alpha \equiv $ Fe and Rh), 
are obtained by integrating the spin density within the PAW 
spheres which have the radius $r_{\rm PAW}({\rm Fe}) =  1.3$~{\AA} 
for Fe and $r_{\rm PAW}({\rm Rh}) =  1.4$~{\AA} for Rh. 
In the pure dimers, the local moments 
$\mu_{\rm Fe} = 2.82 \mu_B$ and $\mu_{\rm Rh} = 1.83 \mu_B$
are close to the respective total moment per 
atom $\overline\mu_2 = 3 \mu_B$ and $2 \mu_B$, which indicates
that the spin-density $m(\vec r)$ = $n_{\uparrow}(\vec r) - n_{\downarrow}(\vec r)$ is quite localized
around the atoms. Actually, the differences between 
$\mu_\alpha$ and $\overline\mu_N$ give a measure of 
the small spill-off effect in $m(\vec r)$. 
Taking this into account, the results for $\mu_\alpha$ 
in the FeRh dimer seem quite remarkable.
Here the Fe local moment is significantly enhanced
with respect to the Fe$_2$ or Fe-atom value, while
the Rh moment is reduced by a similar amount ($\Delta\mu_{Fe}$ = 0.52 $\mu_B$ and $\Delta\mu_{Rh}$ = -0.50 $\mu_B$, see Table~\ref{tab:2}).
This is mainly the consequence of a transfer of $d$ electrons
from Fe to Rh, which allows the Fe atom to develop a
larger spin moment, due to the larger number of $d$ holes. 
This occurs at the expense of the moment at the Rh atom, which has less $d$ holes to polarize.
An integration of the electronic density in the 
Bader cells\cite{bader} shows that $0.33$ electrons are transferred
from the Fe to the Rh atom in FeRh. This behavior is qualitatively 
in agreement with the higher Pauling 
electronegativity $\chi$ of the Rh atom ($\chi_{\rm Fe}$ = 1.83 
and $\chi_{\rm Rh}$ = 2.28).\cite{pauling} 
\subsection{FeRh trimers}
\label{sec:3}

The results for trimers are summarized in Table \ref{tab:3}.
As expected, the lowest energy isomers are found to be 
triangles for all compositions. According to our calculations the ground state of
${\rm Rh}_3$ is an equilateral triangle ($\rm D_{3h}$) with $E_B = 2.31$~eV, 
bond length $d = 2.37$~{\AA} and average magnetic moment
$\overline\mu_3 = 1 \mu_B$. The local magnetic moments 
$\mu_\alpha = 0.93 \mu_B$ in the WS cells align parallel to each other and are almost as large as 
$\overline\mu_3$. These results are consistent with those reported
in previous GGA studies of Rh$_3$
($E_B = 2.35$ ~eV, $d = 2.45$~{\AA} and $\overline\mu_3 = 1 \mu_B$).\cite{Rh_nayak} 
A single Fe substitution yields an isosceles 
FeRh$_2$ with an elongated base composed of the two Rh atoms.
Notice that the bond-length $d_{\rm RhRh} = 2.57$~{\AA} is 
larger than in Rh$_3$. The linear isomer of the 
form Rh-Fe-Rh, i.e., with only FeRh bonds, 
lies 0.4~eV above the optimal structure. 
It is the only true local minimum among the linear FeRh trimers.
The other linear structures (Rh-Rh-Fe, Fe-Rh-Fe, and Fe-Fe-Rh) are 
all found to be saddle points connecting triangular minima of the potential energy surface (PES).
Further Fe substitution yields a isosceles Fe$_2$Rh in
which the FeFe bond is the shortest. One observes, as in the dimers, 
that the interatomic distances follow the trends in the atomic radii.
Finally, for ${\rm Fe}_3$, the calculated lowest-energy structure is a Jahn-Teller
distorted isosceles triangle with two longer bonds ($d_{12} = d_{13} = 2.30$~{\AA}) and 
a shorter one ($d_{23}= 2.07$~{\AA}). The calculated average magnetic moment of ${\rm Fe}_3$
is $\overline\mu_3 = 3.33 \mu_B$. These results coincide with
previous GGA studies\cite{Fe3_GGA} predicting $d_{12} = d_{13} = 2.33$~{\AA} and $d_{23} = 2.09$~{\AA}.
In contrast, LDA calculations \cite{compa_stru} yield an equilateral Fe$_3$ 
with average magnetic moment $\overline\mu_3 = 2.66 \mu_B$
and $d = 2.10$~{\AA}. By using the spin-polarized LDA, we also obtain an equilateral triangle similar to the one reported 
in Ref.~\onlinecite{compa_stru}. In contrast, in the GGA one finds that the  
equilateral triangle (D$_{3h}$) is unstable with respect to a Jahn-Teller distortion.
The isosceles shape of Fe$_3$ can therefore be interpreted as a consequence of exchange and correlation effects.
Moreover, we have analyzed the GGA Kohn-Sham spectrum in the equilateral structure and 
found a high degeneracy at the Fermi energy, which is consistent with the interpretation
that the distortion is triggered by a Jahn-Teller effect. 

Concerning the composition dependence of $E_B$ one observes a
non-monotonous behavior as for $N=2$, which indicates 
that the FeRh bonds are the strongest. 
The lowest vibrational frequency follows a similar trend, 
despite the larger mass of Rh. Notice that
FeRh$_2$ is somewhat more stable than Fe$_2$Rh, since the bonds
between Rh atoms are in general stronger than between Fe atoms.
Finally, one may also notice that the energy gain $\Delta E_m$ 
associated to magnetism only plays a quantitative role 
in the relative stability of triangular and linear FeRh$_2$.
$\Delta E_m$ is actually larger for the linear chain than for 
the triangle. Therefore, the later remains the most stable
structure even in the non-magnetic case, although with 
somewhat different bond lengths.

\begin{table}
\caption{(Color online)\label{tab:3} 
Structural, electronic and magnetic properties of FeRh trimers. 
Results are given for the binding energy per atom $E_B$ (in eV), 
the magnetic stabilization energy per atom $\Delta E_{m}$ = [$E(S_z \!= \!0) - E(S_z)$]/N 
(in eV), the average interatomic distance $d_{\alpha\beta}$ (in {\AA})
ordered from top to bottom as $d_{\rm FeFe}$, $d_{\rm FeRh}$ and $d_{\rm RhRh}$, 
the average spin moment per atom $\overline\mu_{N} = 2 S_z /N$ (in {$\mu_B$}) , 
the local spin moment $\mu_{\alpha}$ (in {$\mu_B$}) at the Fe or Rh atoms, 
and the lowest vibrational frequency $\nu_{0}$ (in $\rm cm^{-1}$).
           }
\renewcommand{\arraystretch}{0.1}
\begin{tabular}{|c|c|c|c|c|c|c|c|c|} 
\hline 
\hline
         &                                                      &       &      &      &      &      &    &          \\ 
Cluster  &Struct. &$E_B$&$\Delta{E_m}$&$d_{\alpha\beta}$&$\overline\mu_{N}$&$\mu_{\rm Fe}$&$\mu_{\rm Rh}$&$\nu_{0}$ \\ \hline
         &                                                      &       &      &             &      &    &     &    \\
Fe$_3$   &\raisebox{-0.4cm}{\includegraphics[scale=0.3]{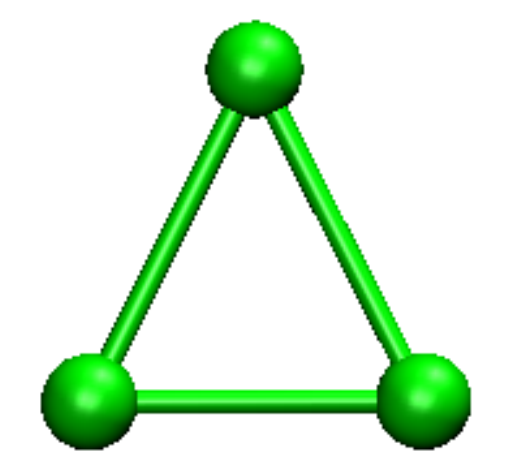}}  &1.80  &0.69  &2.22   &3.33  &2.99&     &267 \\
         &                                                      &       &      &      &      &      &    &          \\
Fe$_2$Rh &\raisebox{-0.4cm}{\includegraphics[scale=0.24]{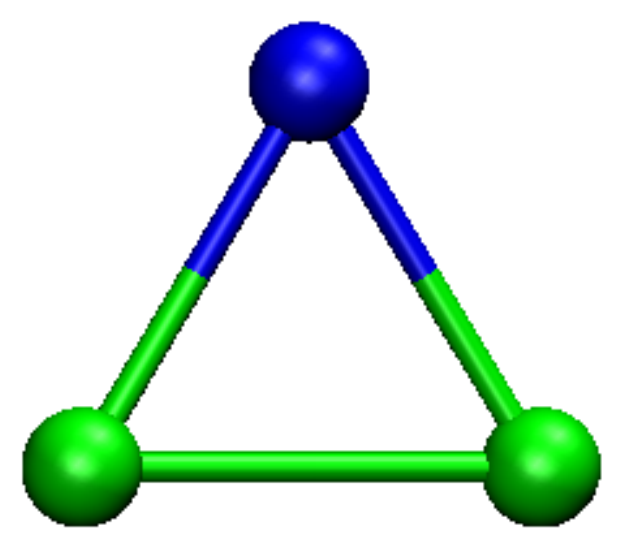}}&2.24 &0.32  &2.25  &3.00  &3.35&1.21 &235 \\[-0.33cm] 
         &                                                      &       &      &2.35  &      &      &    &          \\[0.4cm]
         &                                                      &       &      &      &      &      &    &          \\
FeRh$_2$ &\raisebox{-0.3cm}{\includegraphics[scale=0.28]{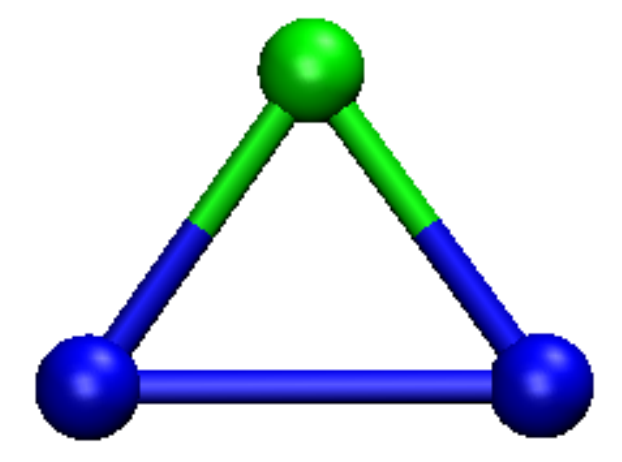}}&2.45 &0.05  &2.21  &2.00  &3.27&1.18 &287 \\[-0.25cm] 
         &                                                      &       &      &2.57  &      &      &    &          \\[0.4cm] 
         &                                                      &       &      &      &      &      &    &          \\
Rh$_3$   &\raisebox{-0.4cm}{\includegraphics[scale=0.33]{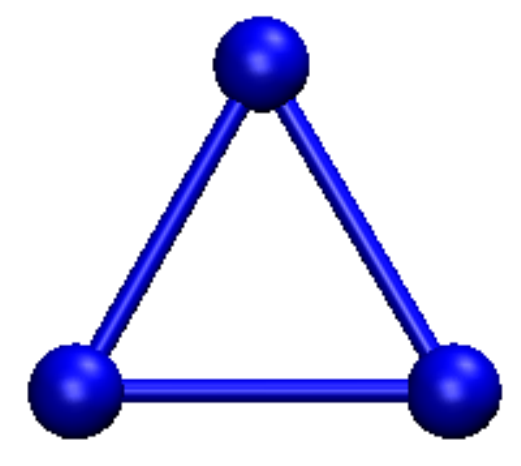}}&2.31  &0.02  &2.37  &1.00  &    &0.93 &210 \\ \hline 
\hline 
\end{tabular}
\end{table} 

The average magnetic moment per atom $\overline\mu_3$ amounts to $1 \mu_B$ for 
Rh$_3$. In the alloys it increases monotonously with Fe doping, reaching 
$\overline\mu_3 = 10/3 \mu_B$ for Fe$_3$. The local magnetic 
moments $\mu_\alpha$ always show a FM-like coupling.
They are all identical in Rh$_3$, which is
consistent with the $C_3$ point-group symmetry. 
In the pure clusters $\mu_\alpha$ is always close to 
$\overline\mu_3$. This indicates that the 
spin polarization is dominated by electrons occupying localized states
and that spill-off contributions are not important.
For example, in the case of Fe$_3$, one finds $\mu_1 = 3.23\mu_B$
and $\mu_2 = \mu_3 = 2.87\mu_B$, the latter corresponding to the 
pair of atoms forming the shorter bond.
On the other side, the average local moments $\mu_{\rm Rh}=0.93 \mu_B$ 
in Rh$_3$ should be compared with $\overline\mu({\rm Rh}_3) = 1 \mu_B$.
As soon as FeRh bonds are present, for mixed compositions,
the local Fe moments are enhanced beyond $3\mu_B$. This is mainly
due to a charge transfer from Fe to Rh, leading to an increase in the number 
of Fe $d$ holes as already observed in the dimer. Quantitatively, 
the local $\mu_{\rm Fe}$ and $\mu_{\rm Rh}$ in mixed trimers
are similar, though somewhat smaller, to the corresponding 
values in the FeRh dimer. Notice, moreover, the enhancement of the Rh local
moments in Fe$_2$Rh and FeRh$_2$ as compared to pure
Rh$_3$. This reflects the importance of the proximity of Fe on the magnetic behavior  
of the Rh atoms.

\subsection{FeRh tetramers}
\label{sec:4} 

The most stable FeRh tetramers are all tetrahedra and the first
low-lying isomers are rhombi (see Table \ref{tab:4}).
The distribution of the atoms within the optimal
topology does not play a role since all sites are equivalent in a   
tetrahedron. In the case of Rh$_4$ we obtain a nonmagnetic undistorted tetrahedron
having $E_B = 2.75$ eV and bond length $d = 2.45$~{\AA}. The closest isomer is found to be a bent rhombus with an average 
bond length $d = 2.35$~{\AA}. Similar results have been obtained
in previous studies on Rh clusters.\cite{Rh4_ref} Notice, however, that Bae {\em et al.}\cite{ycbau} have obtained a bend rhombus as the ground-state structure for Rh$_4$ also by using VASP.  
This discrepancy is likely to be a consequence of the different choice of the pseudopotential and cutoff energy E$_{max}$.
In our calculations we considered the PAW method and E$_{max}$ = 268 eV, while in Ref.~\onlinecite{ycbau} one used ultrasoft pseudopotentials and E$_{max}$ = 205.5 eV.   

\begin{table}
\centering
\caption{(Color online)\label{tab:4} 
Structural, electronic and magnetic properties of FeRh tetramers 
as in Table \protect\ref{tab:3}.
        }
\renewcommand{\arraystretch}{0.3}
\begin{tabular}{|c|c|c|c|c|c|c|c|c|}
\hline 
\hline
            &                                   &     &             &       &         &        &       &                               \\ 
Cluster     &Struct.     &$E_B$&$\Delta{E_m}$   &$d_{\alpha\beta}$  &$\overline\mu_{N}$&$\mu_{\rm Fe}$ &$\mu_{\rm Rh}$ &$\nu_{0}$      \\ 
\hline      &                                   &     &             &       &         &        &       &                               \\
Fe$_4$      &\raisebox{-0.4cm}{\includegraphics[scale=0.22]{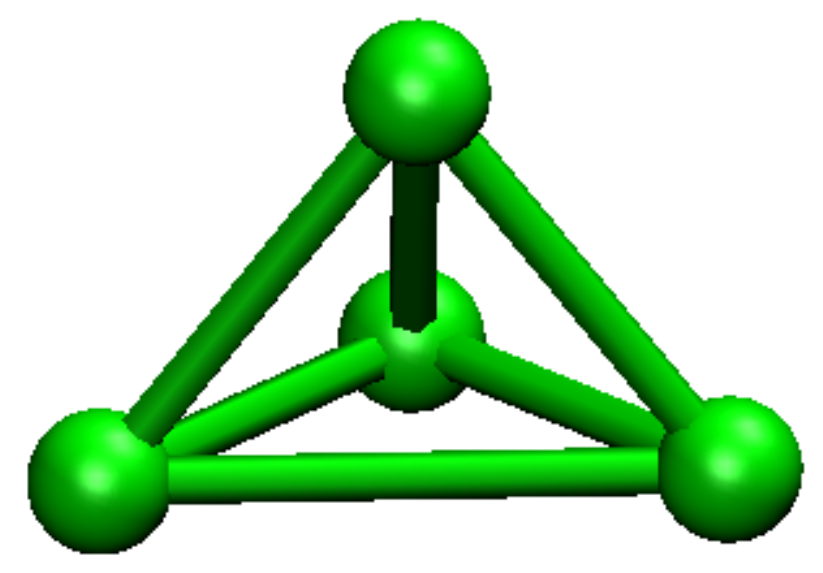}}   &2.21    &0.35    &2.28   &3.50   &3.08 &      &279       \\
            &                                   &     &             &       &         &        &       &                               \\
Fe$_3$Rh    &\raisebox{-0.4cm}{\includegraphics[scale=0.2]{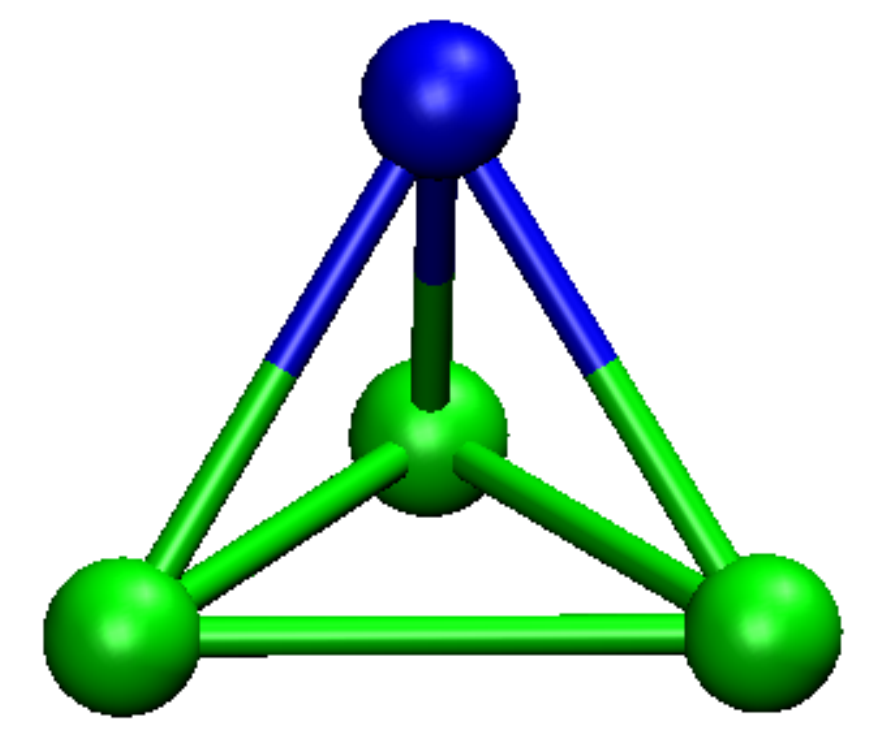}}&2.49    &0.58    &2.34   &3.00   &3.18 & 1.03 & 232      \\[-0.3cm]
            &                                   &     &             &2.40   &         &        &       &                               \\[0.2cm]   
Fe$_2$Rh$_2$&\raisebox{-0.4cm}{\includegraphics[scale=0.2]{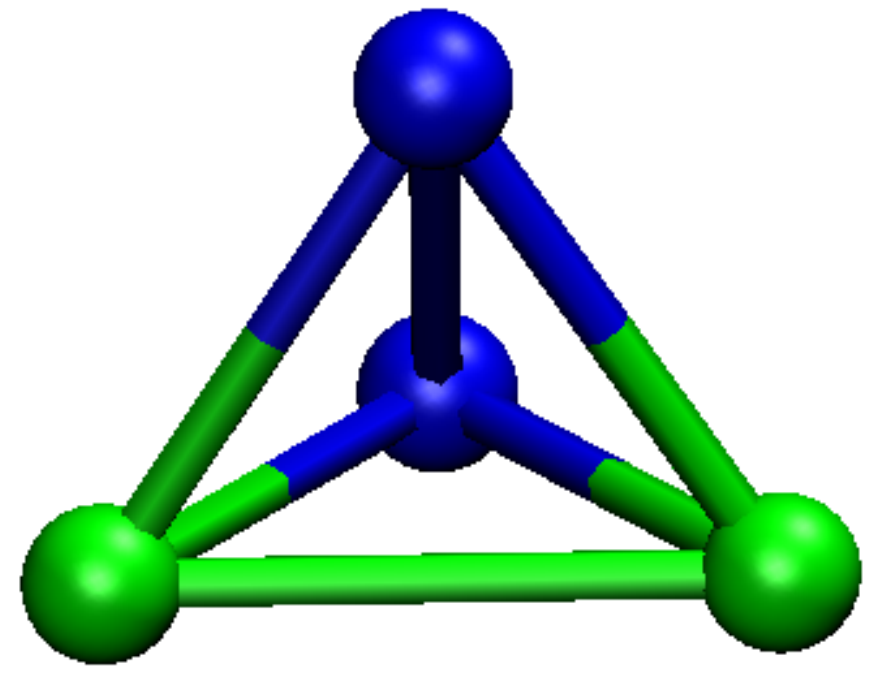}}&2.74    &0.37    &2.52   &2.50   &3.39 & 1.03 & 243      \\[-0.3cm]
            &                                   &     &             &2.31   &         &        &       &                               \\
            &                                   &     &             &2.72   &         &        &       &                               \\[-0.1cm]
FeRh$_3$    &\raisebox{-0.4cm}{\includegraphics[scale=0.2]{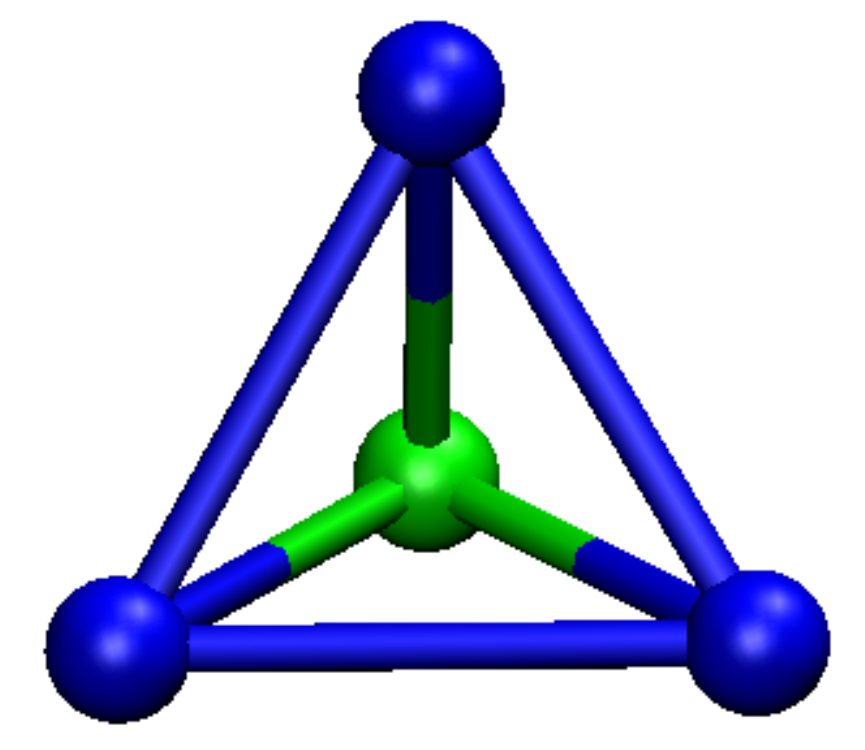}} &2.76    &0.21    &2.30   &1.75   &3.25 & 1.12 & 289      \\[-0.3cm]
            &                                   &     &             &2.60   &         &        &       &                               \\[0.2cm]
Rh$_4$      &\raisebox{-0.4cm}{\includegraphics[scale=0.23]{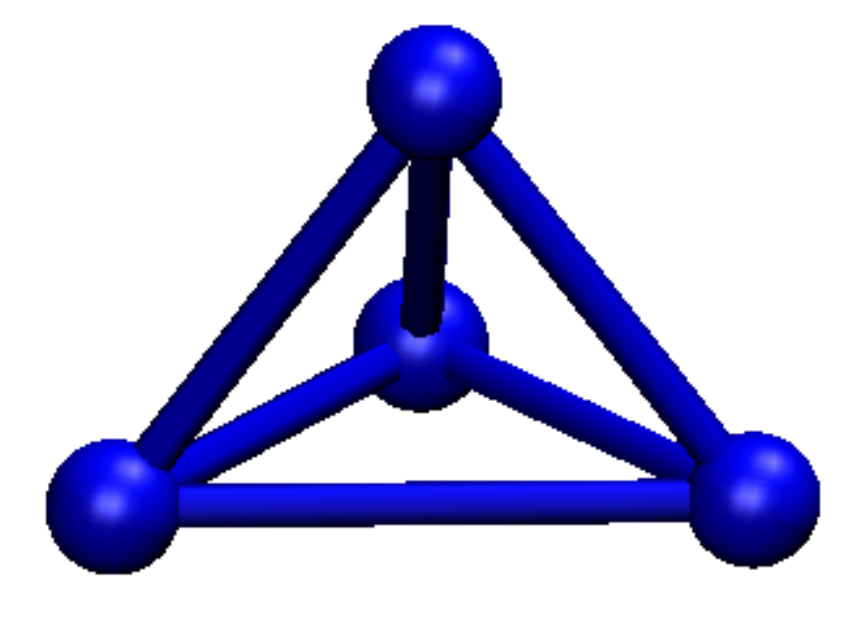}}   &2.75    &0.00    &2.45   &0.00   &     & 0.00 & 201      \\
            &                                   &     &             &       &         &        &       &                               \\
\hline \hline
\end{tabular}
\end{table}

The binding energy of the alloys shows a characteristic non-monotonous dependence
on concentration, which was also found in smaller clusters.
In fact Fe$_2$Rh$_2$ and FeRh$_3$ are the most stable tetramers 
with $E_B=2.74$~eV and $E_B=2.76$~eV, respectively. This confirms 
that the FeRh bonds are the strongest. It is worth noting that these
trends are not altered qualitatively if magnetism is neglected, i.e.,
if one considers $E_B$ for $S_z =0$.
In addition, it is interesting to follow how $E_B$ changes from Rh$_4$ to Fe$_4$. 
The stability of the clusters can be qualitatively related to the 
number of homogeneous and heterogeneous bonds by counting them for each of the  
clusters shown in Table \ref{tab:4}. For instance, FeRh$_3$, which 
is the most stable composition, has 
3 FeRh and 3 RhRh bonds. Replacing a Rh by an Fe to obtain Fe$_2$Rh$_2$
implies replacing 2 RhRh bonds by a stronger FeRh and a weaker FeFe bond.
Therefore, $E_B$ does not change significantly. The fact that $E_B$ depends weekly on composition for Rh rich tetramers shows that FeRh and RhRh bonds are comparably strong in 
these clusters. 

Concerning the magnetic moments one observes 
a approximately linear dependence of $\overline\mu_N$ as a function 
of Fe content. In general, the substitution of a Rh by and Fe atom
results in an increase of the total moment
$2S_z$ by 3 or 4$\mu_B$, or equivalently, $\Delta\overline\mu_N = (0.75$--$1)\mu_B$ 
(see Table \ref{tab:4}). The magnetic order is always FM-like.
In the alloys the local moments $\mu_{\rm Fe}$
show the above mentioned enhancement, which is due to a Fe-to-Rh $d$-electron
charge transfer that increases the number of $d$ holes and allows for
the development of $\mu_{\rm Fe}\simeq 3.2$--$3.4 \mu_B$.
In addition, the presence of Fe in Fe$_m$Rh$_n$ enhances the Rh local 
moments as compared to pure Rh$_4$. 

\subsection{FeRh pentamers}
\label{sec:5} 

In Table \ref{tab:5} the results for FeRh pentamers are summarized.
Although all possible cluster topologies (20 structures) were considered
as starting geometries for each composition, only the most highly 
coordinated trigonal bipyramid (TBP) and the square pyramid (SP) 
are found to be most stable geometries. The low coordinated structures transform into compact structures 
after the relaxation. Except for $\rm Rh_5$, which optimal 
structure is a SP, all the other FeRh pentamers have the TBP 
as ground-state geometry. The trend in the composition dependence
of the binding energy $E_B$ of pentamers confirms the behavior we started to observe for $N = 4$.
Indeed, in the Fe-rich limit $E_B$ increases rapidly with increasing Rh content, as the weakest FeFe bonds are replaced by FeRh bonds. Later on,
near 50$\%$ concentration and in the Rh-rich limit, the composition dependence is weak since FeRh and RhRh bonds are comparably strong (see Table \ref{tab:5}). 
In particular for Rh-rich compositions, replacing Fe by Rh atoms
no longer results in weaker binding.  
In other words, FeRh bonds are no longer primarily preferred.
This is possibly a consequence of the increasing coordination 
number, which enhances the role of electron delocalization and 
band formation, thus favoring the larger Rh hybridizations. 

\begin{table}
\centering
\caption{(Color online)\label{tab:5} 
Structural, electronic and magnetic properties of FeRh pentamers as in 
Table \protect\ref{tab:3}.
        }
\renewcommand{\arraystretch}{0.3}
\begin{tabular}{|c|c|c|c|c|c|c|c|c|} 
\hline 
\hline
            &                                     &     &      &    &         &         &        &                                              \\ 
Cluster     &Struct. &$E_B$&$\Delta{E_m}$&$d_{\alpha\beta}$&$\overline\mu_{N}$&$\mu_{\rm Fe}$&$\mu_{\rm Rh}$&$\nu_{0}$                          \\ 
\hline      &                                     &     &      &    &         &         &         &                                             \\
Fe$_5$      &\raisebox{-0.8cm}{\includegraphics[scale=0.23]{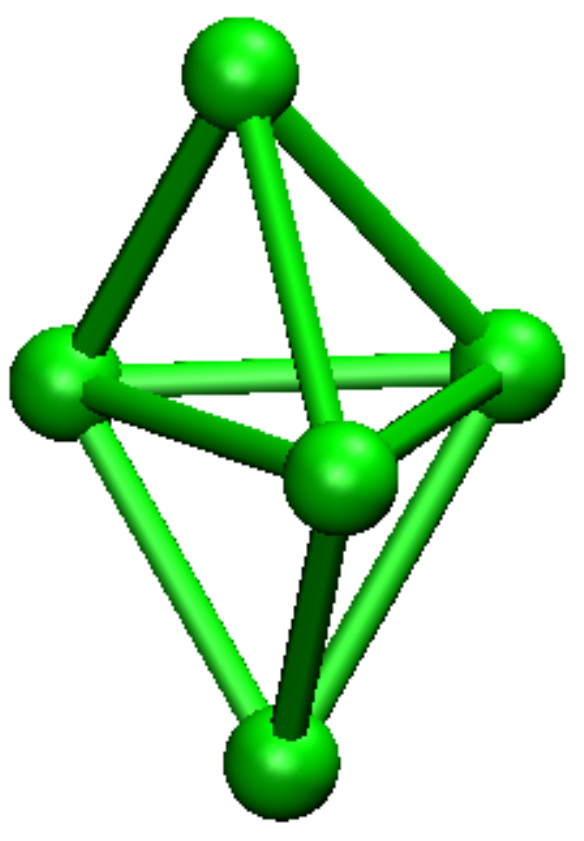}}     &2.51    &1.00     &2.41    &3.20      &2.93    &     &227       \\
            &                                     &     &      &    &         &         &         &                                             \\ 
Fe$_4$Rh    &\raisebox{-1.0cm}{\includegraphics[scale=0.2]{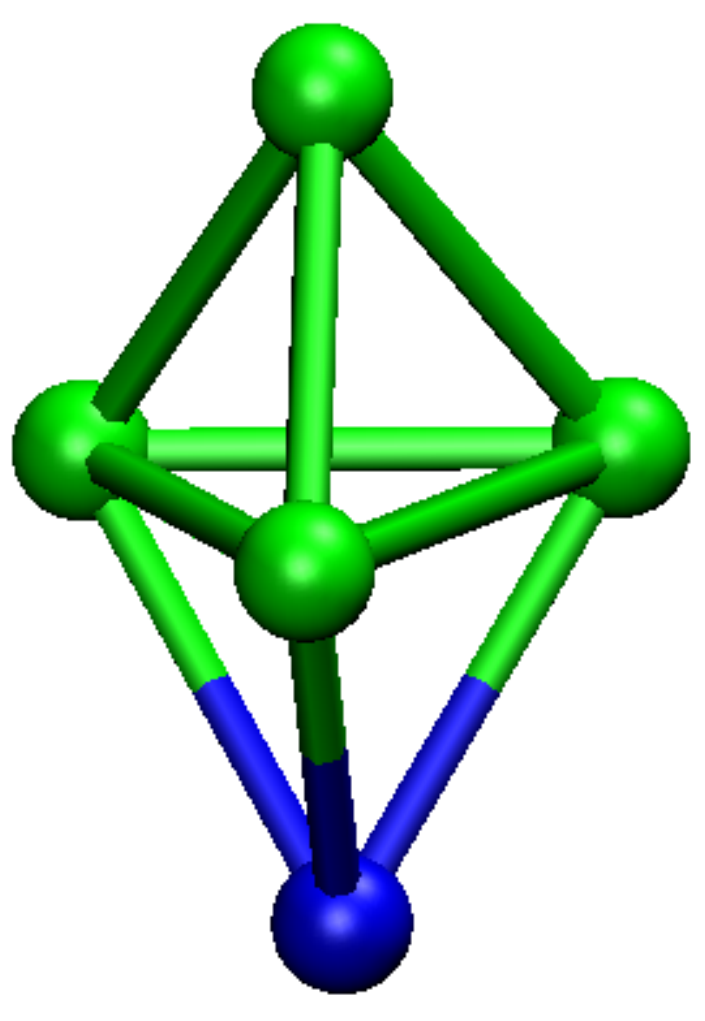}}   &2.76    &1.01     &2.30    &3.00       &3.09    &1.06 &243       \\[-0.8cm] 
            &                                     &     &      &2.47&         &         &         &                                             \\[0.7cm]
Fe$_3$Rh$_2$&\raisebox{-0.9cm}{\includegraphics[scale=0.2]{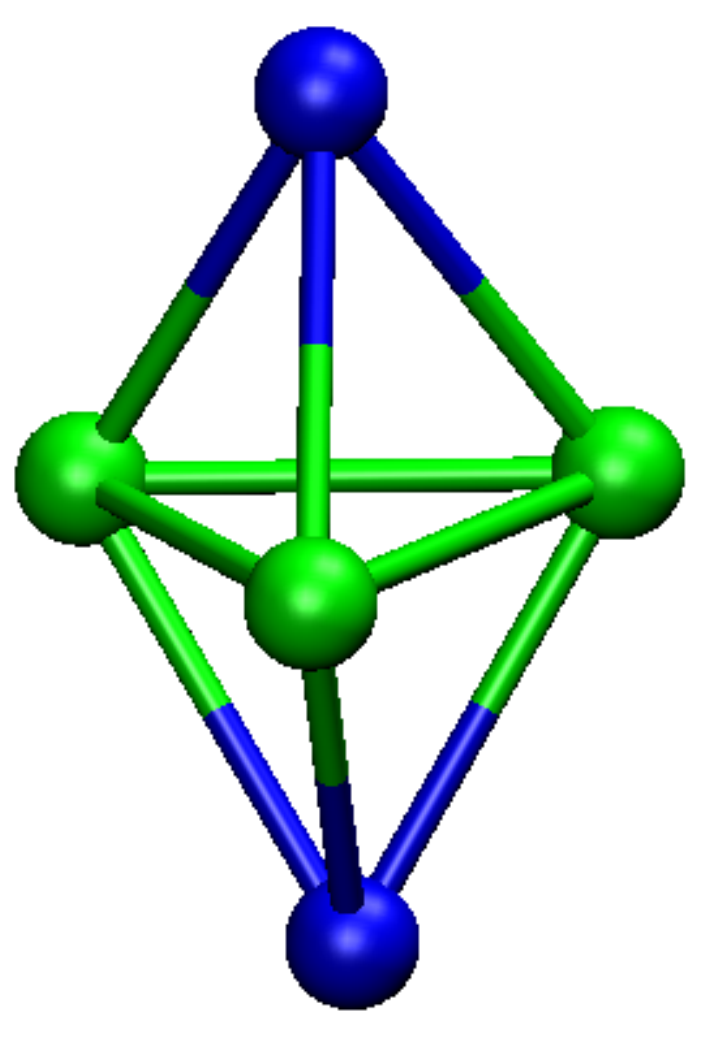}}  &2.96    &0.92     &2.37    &2.40      &3.13    &0.75 &244       \\[-0.7cm]
            &                                     &     &      &2.39&         &         &         &                                             \\[0.6cm]
Fe$_2$Rh$_3$&\raisebox{-0.9cm}{\includegraphics[scale=0.2]{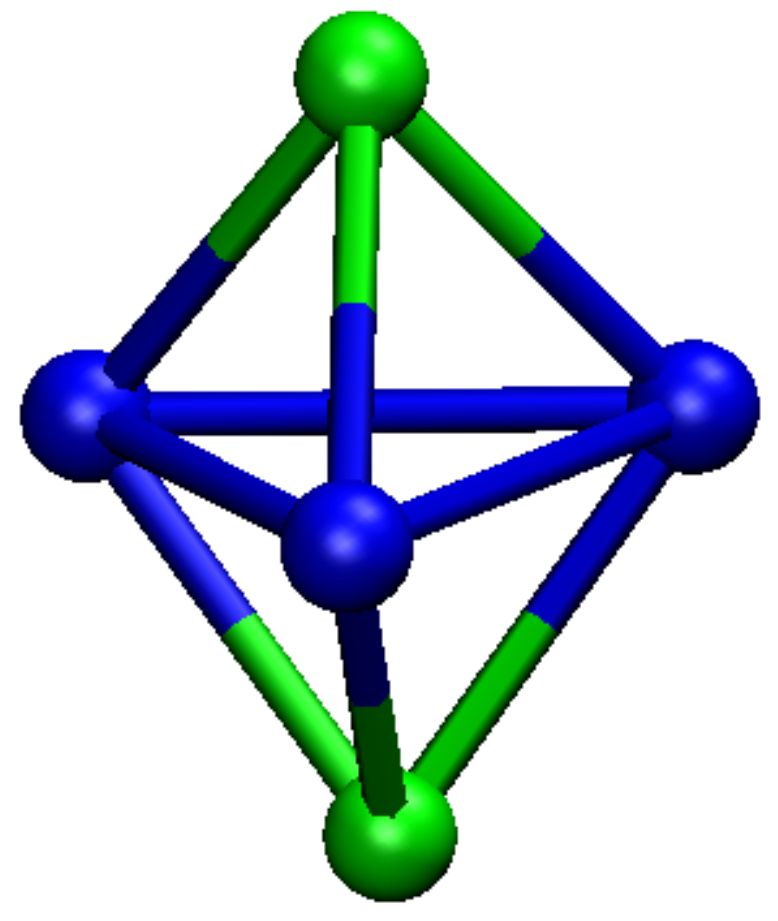}}  &3.06    &0.55     &2.35    &2.20      &3.36    &1.08 &260       \\[-0.7cm]
            &                                     &     &      &2.71&         &         &         &                                             \\[0.6cm]
Fe$_1$Rh$_4$&\raisebox{-0.9cm}{\includegraphics[scale=0.2]{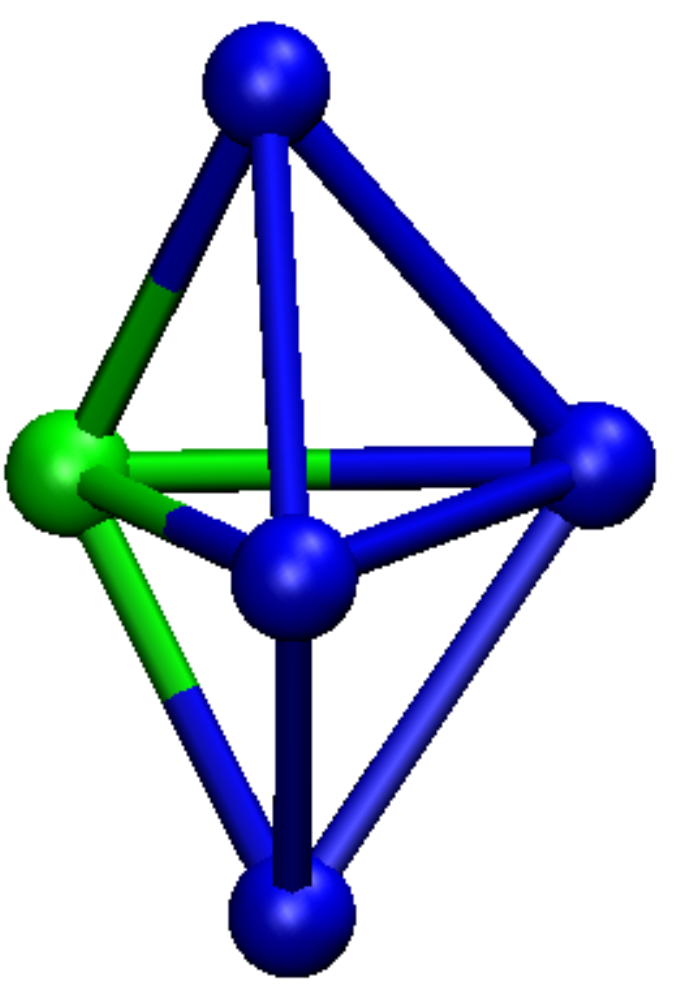}}   &3.01    &0.33     &2.39    &1.20      &3.31    &0.57 &251       \\[-0.7cm]
            &                                     &     &      &2.51&         &         &         &                                             \\[0.6cm]
Rh$_5$      &\raisebox{-0.5cm}{\includegraphics[scale=0.17]{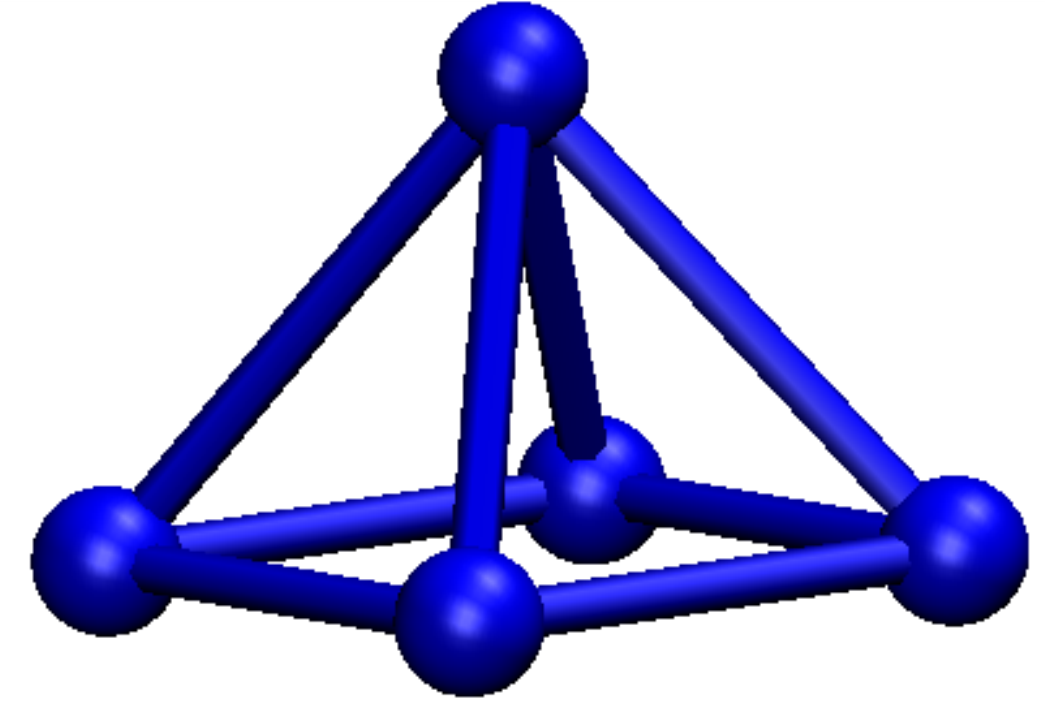}}    &3.03    &0.70     &2.48    &1.00         &        &0.95 &113       \\
            &                                     &     &      &    &         &         &         &                                             \\
\hline \hline
\end{tabular}
\end{table}

The calculated optimal structure of Rh$_5$, a square pyramid,
coincides qualitatively with previous DFT calculations.\cite{Rh_nayak}
Nevertheless, we obtain a binding energy that is 0.07~eV per atom lower than in Ref.~\onlinecite{Rh_nayak}. 
Substituting one Rh atom by Fe yields FeRh$_4$ and changes the optimal cluster topology to the 
more compact TBP. The SP remains a local minimum of the ground-state
energy surface, which is only $3$~meV per atom less stable than the optimal TBP geometry.
The average magnetic moment $\overline\mu({\rm FeRh}_4) = 1.2\mu_B$ is enhanced
with respect to Rh$_5$ due to the contribution of a large Fe local 
moment $\mu_{\rm Fe} = 3.31 \mu_B$. Notice that the Rh moments are
no longer enhanced as in the smaller FeRh$_{N-1}$ but significantly 
reduced: $\mu_{\rm Rh} = 0.62\mu_B$ for the apex atoms and 
$\mu_{\rm Rh} = 0.52\mu_B$ for the Rh atoms sharing a triangle with the Fe.
This is of course related to the fact that the ground-state $S_z$ 
is relatively low. The effect is even stronger in the case of 
the SP isomer of FeRh$_4$. Here we find two Rh moments 
$\mu_{\rm Rh} = 0.43\mu_B$ that couple parallel to the Fe moment, 
one very small Rh local moment $\mu_{\rm Rh} = 0.05\mu_B$, 
and an antiparallel moment $\mu_{\rm Rh} = -0.48\mu_B$. This explains
the reduced average total moment $\overline\mu_5 = 0.8 \mu_B$ and
the very small average Rh moment $\mu_{\rm Rh} = 0.15 \mu_B$ found in SP isomer of FeRh$_4$.
The present example illustrates the subtle competition between 
cluster structure and magnetism in $3d$-$4d$ nanoalloys.

Further increase in the Fe content does not change the 
topology of the optimal structure. Moreover, we start to see that
for nearly equal concentrations of Fe and Rh 
(i.e., Fe$_2$Rh$_3$ and Fe$_3$Rh$_2$)
the low-lying isomers are the result of changes on the chemical order, 
i.e., changes in the distribution of the Fe and Rh atoms within the cluster, 
rather than the result of changes in the cluster topology.
The most stable configuration 
corresponds to the case where the 3 Rh atoms (in Fe$_2$Rh$_3$)
or the 3 Fe atoms (in Fe$_3$Rh$_2$) are all NNs of each other
(see Table \ref{tab:5}). This is understandable from a 
single-particle perspective, since the band energy is lower 
when orbitals having nearly the same energy levels are hybridized.
In addition, the most stable configurations maximize first the number
of FeRh NN pairs, followed by the number of RhRh pairs.\cite{foot:bonds5} 
Finally, in the Fe-rich limit, for example in Fe$_4$Rh, the lowest-energy 
structure remains a TBP but the closest isomer corresponds to the SP, 
which has a different topology, rather than a different position of 
the Rh atom in the TBP.

The trends in the magnetic properties are dominated by the Fe content.
As for smaller clusters the average magnetic moment per atom 
$\overline\mu_N$ increases monotonously with increasing Fe concentration. 
This holds for all optimal structures and in most of the first 
excited isomers. In fact the latter show in general the same 
$\overline\mu_N$ as the optimal structure. The only exception is
FeRh$_4$, which is also the only case where an antiparallel alignment 
of Rh local moments is found. In all other investigated cases 
the magnetic order was found to be FM-like.
The local Fe moments show the usual enhancement with respect 
to pure Fe$_N$, due to an increase in the number of 
Fe $d$-holes. This effect is stronger for Rh-rich clusters,
since the larger the number of Rh atoms is, the stronger is
the FeRh charge transfer (see Table \ref{tab:5}).
In contrast, the substitution of Rh by Fe 
does not always enhances the Rh local moments, as we observed 
systematically for smaller sizes. Finally, it is interesting to
observe that the different chemical orders found in the 
low lying isomers of Fe$_2$Rh$_3$ and Fe$_3$Rh$_2$ 
correspond to different local 
magnetic moments. The environment dependence of $\mu_\alpha$ follows 
in general the well-known trend of higher spin polarization
at the lowest coordinated sites.

\subsection{FeRh hexamers}
\label{sec:6} 

In Table \ref{tab:6} the results for FeRh hexamers are summarized. 
For each composition all possible cluster topologies 
(63 different graphs\cite{phm,Wang}) and all
non-equivalent distributions of Fe and Rh atoms
were taken into account as initial guess for the
{\em ab initio} optimization of the cluster geometry. 
As in previous cases, all relevant values of the total magnetic moment
$2S_z$ are scanned. Despite the diversity of starting topologies
most low coordinated structures relax into compact ones
in the course of the unconstrained relaxations. In the end, 
the square bipyramid (SBP), in general somehow slightly distorted, 
yields the lowest energy regardless of composition. 

\begin{table}
\caption{(Color online)\label{tab:6} 
Structural, electronic and magnetic properties of FeRh hexamers as in 
Table \protect\ref{tab:3}.
        }
\renewcommand{\arraystretch}{0.05}
\begin{tabular}{|c|c|c|c|c|c|c|c|c|} 
\hline 
\hline
             &                                    &     &             &     &      &     &      &                                          \\
Cluster      &Struct. &$E_B$&$\Delta{E_m}$&$d_{\alpha\beta}$&$\overline\mu_{N}$&$\mu_{\rm Fe}$&$\mu_{\rm Rh}$&$\nu_{0}$                    \\ \hline
             &                                    &     &             &                  &      &      &     &                             \\ 
Fe$_6$       &\raisebox{-0.6cm}{\includegraphics[scale=0.2]{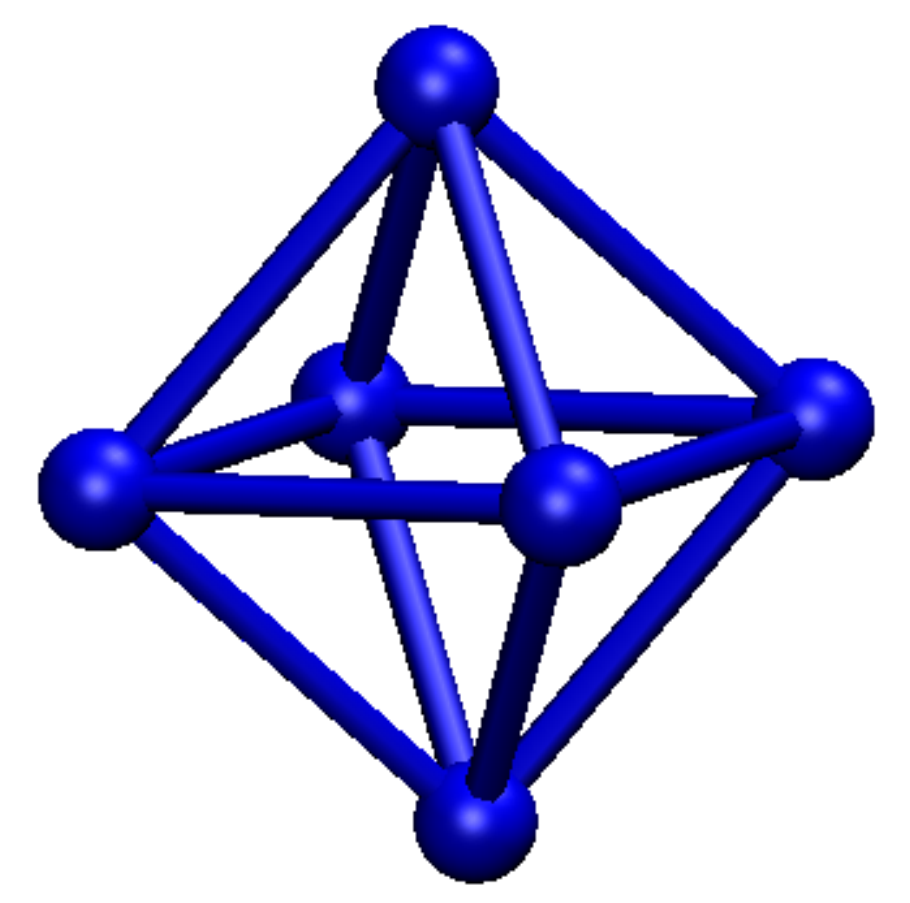}}&2.74    &1.03 &2.38  &3.33  &2.95 &       &206                  \\
             &                                    &     &             &     &      &     &      &                                          \\[0.2cm]
Fe$_5$Rh     &\raisebox{-0.6cm}{\includegraphics[scale=0.18]{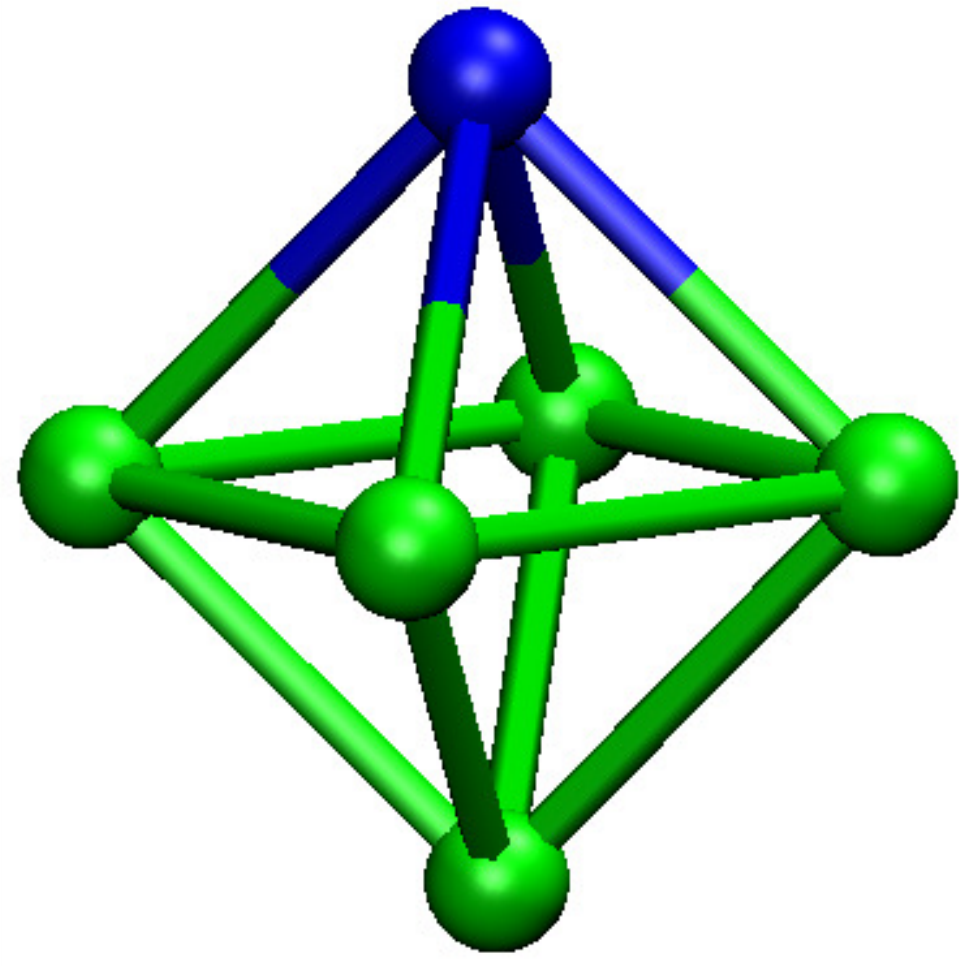}}&2.94  &1.07 &2.38  &3.16  &3.10 & 1.20  &210                  \\[-0.5cm]
             &                                    &     &                                &2.46  &      &     &      &                     \\[0.4cm]
Fe$_4$Rh$_2$ &\raisebox{-0.7cm}{\includegraphics[scale=0.18]{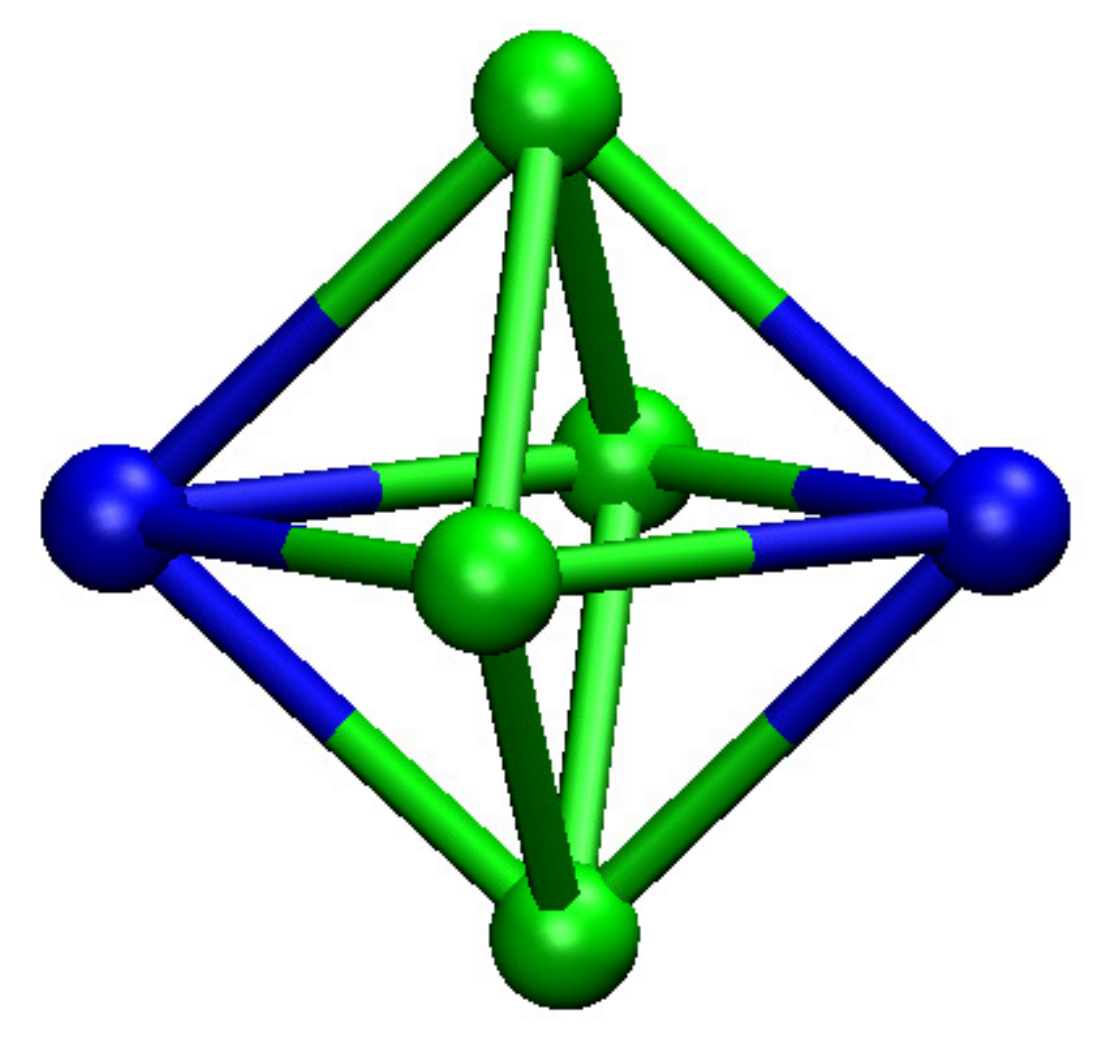}}&3.14  &0.85 &2.38  &3.00     &3.28 & 1.22  & 214                 \\[-0.6cm]
             &                  &     &                                                  &2.45  &      &     &      &                      \\[0.6cm]
Fe$_3$Rh$_3$ &\raisebox{-0.8cm}{\includegraphics[scale=0.19]{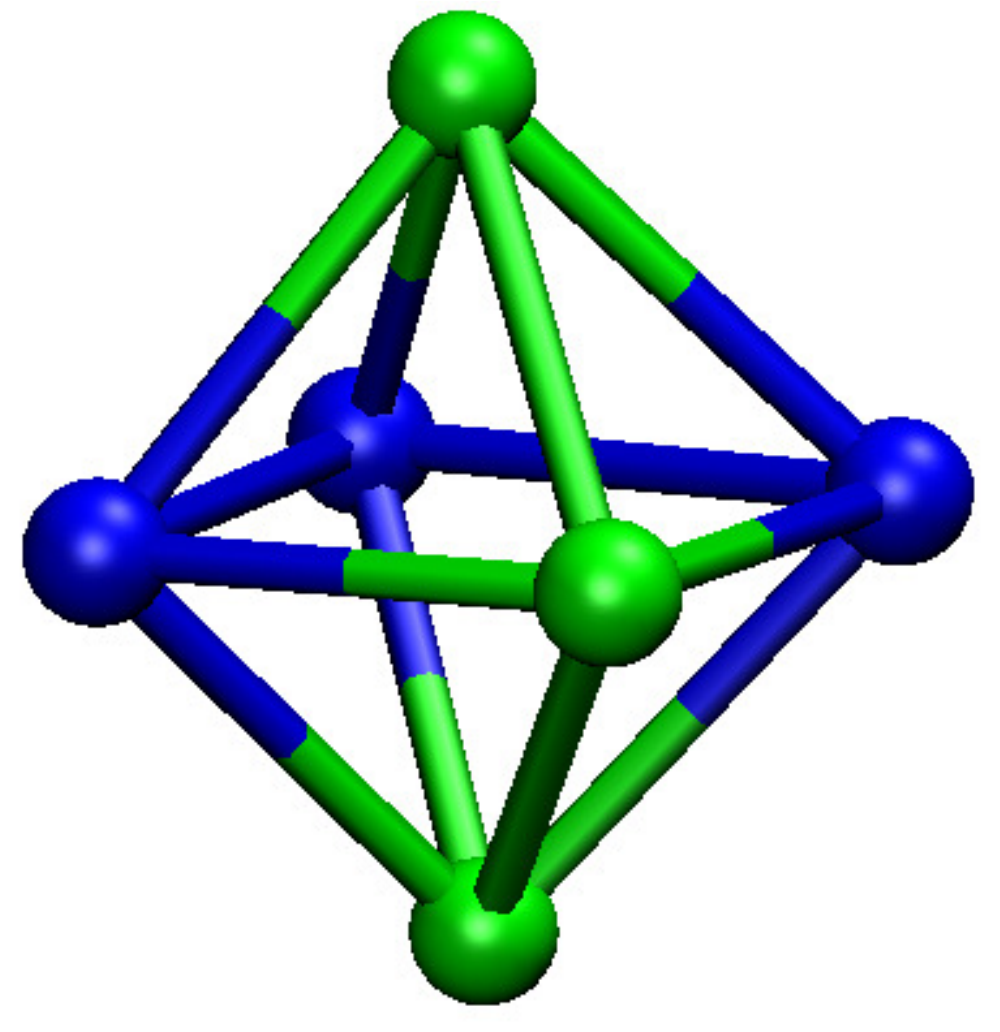}}&3.21  &0.56 &2.62  &2.50  &3.32 & 1.14  &239                  \\[-0.7cm]
             &                                    &     &                                &2.38  &      &     &      &                        \\[0.1cm]
             &                                    &     &                                &2.61  &      &     &      &                         \\[0.4cm]
Fe$_2$Rh$_4$ &\raisebox{-0.6cm}{\includegraphics[scale=0.22]{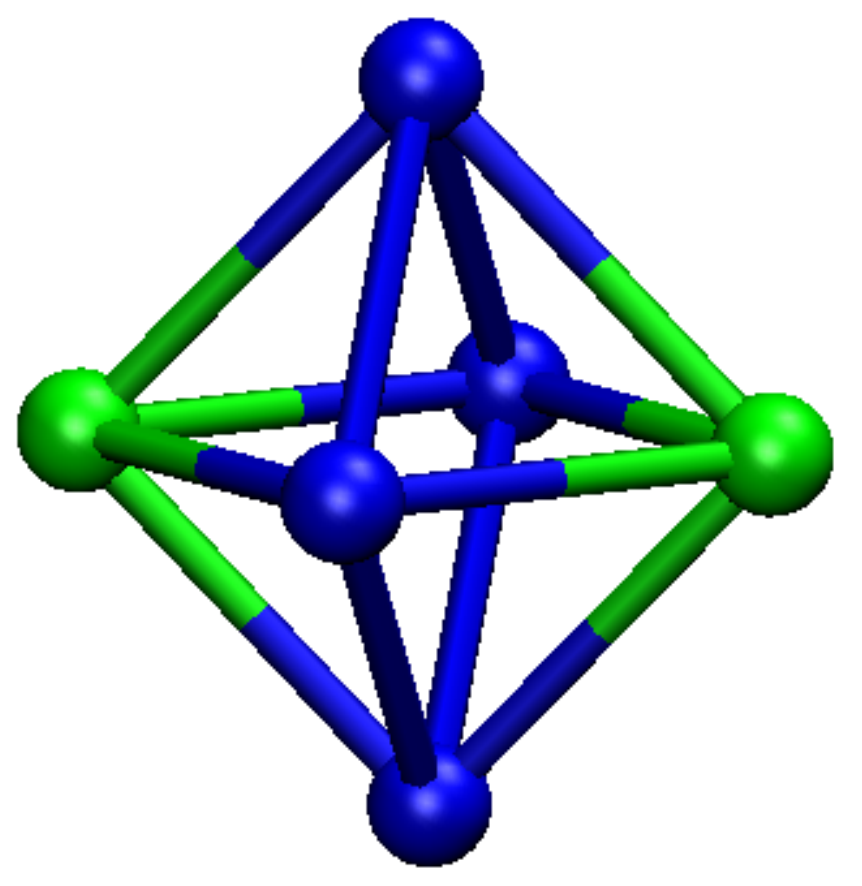}}&3.26  &0.38 &2.46  &2.33  &3.38 & 1.39  &207                  \\[-0.5cm]
             &                                    &     &                                &2.51  &      &     &      &                       \\[0.4cm]
FeRh$_5$     &\raisebox{-0.7cm}{\includegraphics[scale=0.2]{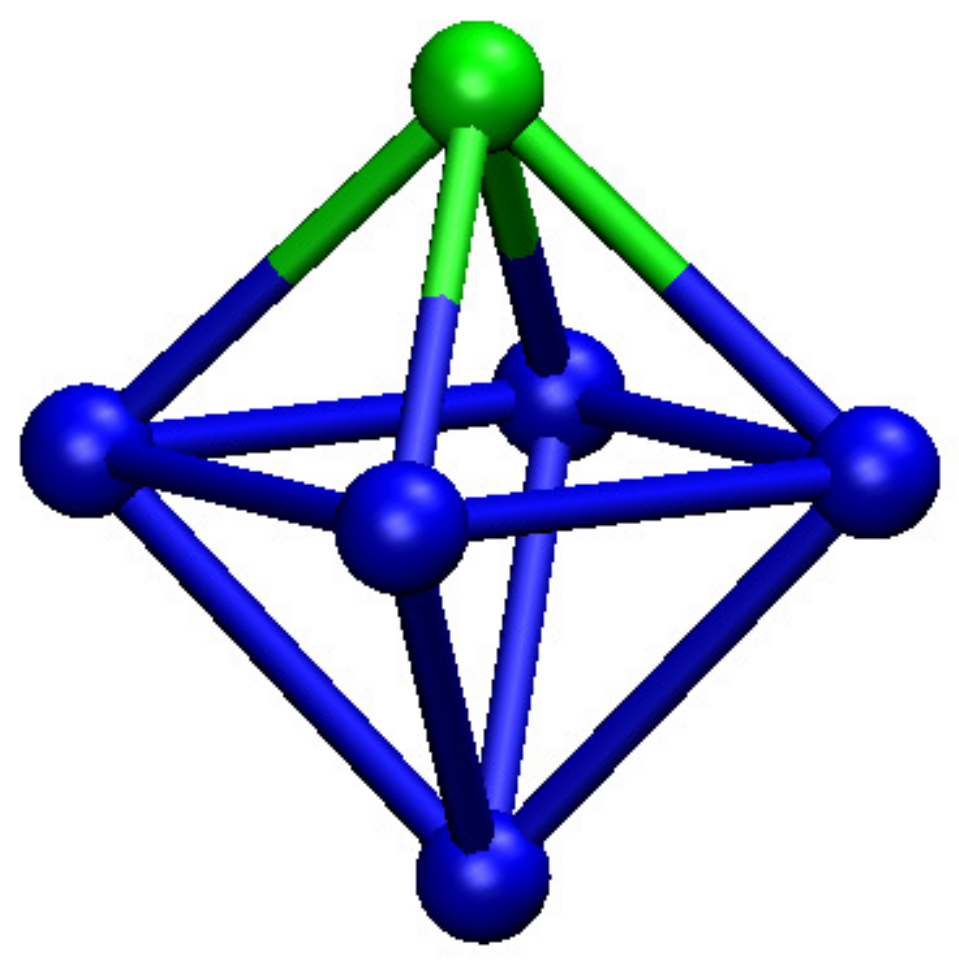}}&3.24&0.28 &2.44  &1.83  &3.37 & 1.33  &192                  \\[-0.6cm]
             &                                    &     &                                &2.54  &      &     &      &                         \\[0.6cm]
Rh$_6$       &\raisebox{-0.7cm}{\includegraphics[scale=0.14]{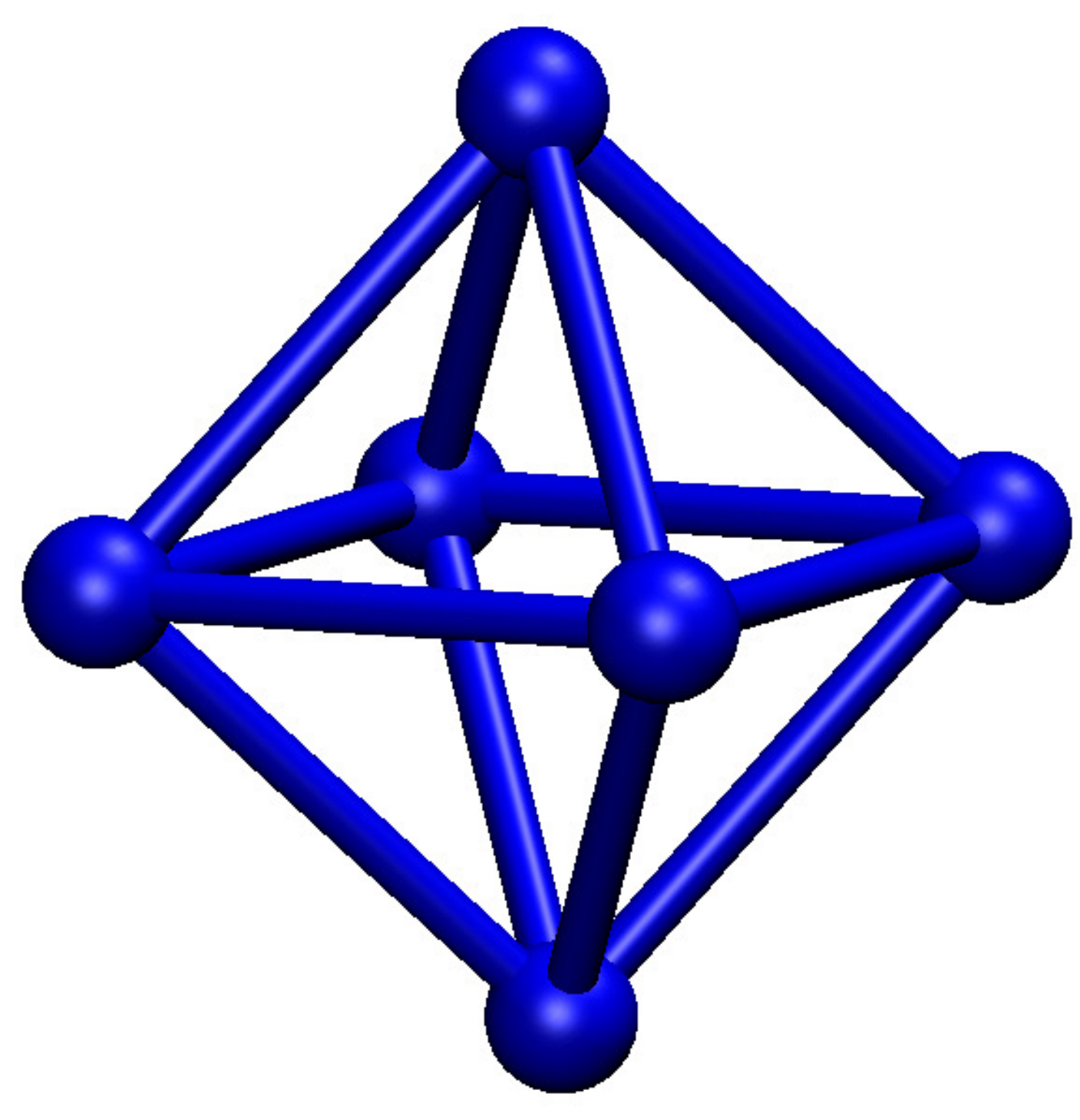}}&3.20     &0.19 &2.54  &1.00   &           & 0.91 &188                  \\
             &                                    &     &             &     &      &     &      &                                            \\
\hline \hline
\end{tabular}
\end{table}

The binding energy per atom $E_B$ shows a similar composition dependence as for pentamers. For Fe-rich clusters $E_B$ 
increases steadily with increasing Rh content, by about
0.2~eV each time a Rh replaces an Fe 
(see Table \ref{tab:6}). Qualitatively, this confirms that the 
bonding between Fe and Rh is stronger than between Fe atoms.
However, for nearly equal concentrations and in the Rh-rich clusters (Fe$_m$Rh$_{6-m}$ with $m \le 2$)
$E_B$ becomes almost independent of $m$. This seems to be the result 
of a compensation of bonding and magnetic contributions. In fact, on the one side, the 
magnetic energy $\Delta E_m$ continues to decrease with increasing
Rh content, by about $0.1$--$0.2$~eV per Rh substitution, even
for high Rh content. And on the other side, this is compensated by an 
increase of the bonding energy with increasing number of Rh atoms. 

In the case of Rh$_6$ an octahedron with an average moment $\overline\mu_6 = 1 \mu_B$ and average 
bond length $d = 2.54$~{\AA} yields the lowest energy. The first isomer, a trigonal biprism (TBP), lies only 
$28$ meV above the optimum, showing a somewhat shorter average bond 
length $d = 2.46$~{\AA} and a higher average moment 
$\overline\mu_6 = 1.67 \mu_B$. 
These results are consistent with previous DFT calculations.\cite{Rh_lwang} 
A single Fe substitution enhances the average moment to 
$\overline\mu_6 = 1.83 \mu_B$ but does not change the topology 
of the optimal FeRh$_5$.
The RhRh distances remain essentially unchanged and the FeRh distances are somewhat 
shorter. The important increase in 
the ground-state spin polarization ($5 \mu_B$ in all) is not only 
due to the larger Fe moment ($\mu_{\rm Fe} = 3.37 \mu_B$ in the PAW sphere) 
but also results from the enhancement of the local Rh moments 
($\mu_{\rm Rh} = 1.33 \mu_B$, see Table \ref{tab:6}).
The first isomer of FeRh$_5$ corresponds to a distorted trigonal prism with significantly contracted FeRh and RhRh 
bond lengths. 
In Fe$_2$Rh$_4$ the Fe atoms occupy the opposite apex positions 
of the octahedron. In this way each Fe is four-fold coordinated 
with all Rh atoms. Heterogeneous bonds are favored over FeFe ones. 
The local Fe moments in Fe$_2$Rh$_4$ are the largest among all hexamers: 
$\mu_{\rm Fe} = 3.38 \mu_B$, slightly beyond the value found in FeRh$_5$. 
This corresponds to a large number of $d$ holes. In addition, particularly 
important spin polarizations are induced at the 
neighboring Rh atoms ($\mu_{\rm Rh} = 1.39 \mu_B$).
The first isomer of Fe$_2$Rh$_4$ corresponds to a capped trigonal bipyramid (CTBP) having  a 
short FeFe NN bond. This structure lies 
only $0.11$~eV higher in energy and has the same total moment as 
the optimal geometry. Replacing a further Fe atom yields 
Fe$_3$Rh$_3$, whose optimal structure is an octahedron. Here we find 
two isosceles open Fe$_3$ and Rh$_3$ triangles that form a
$\pi/2$ angle with respect to each other (see Table \ref{tab:6}).
Out of the 12 NN pairs in the Fe$_3$Rh$_3$ octahedron, 8 are FeRh and only 
4 are homogeneous (2 FeFe and 2 RhRh). The local Fe magnetic moments are similar to 
the other clusters but the Rh moments are 
somewhat smaller in average ($\mu_{\rm Rh} = 1.14 \mu_B$).  
The first excited isomer of Fe$_3$Rh$_3$ is 
a CTBP that lies $25$~meV per atom above the ground state. 
The lowest-energy structure found for Fe$_4$Rh$_2$ is an octahedron, 
while a distorted CTBP is an isomer lying $0.14$~eV per atom above. 
In the former the Rh atoms are far apart occupying the apical 
positions, whereas in the latter they are NNs. The situation is 
thus similar to what we find in Fe$_3$Rh$_2$. For low Rh or Fe 
concentrations the atoms are distributed in order to favor
the FeRh bonds rather than homogeneous NN pairs between the atoms 
in the minority. 
The octahedron and a distorted CTBP remain the two most stable 
structures as one further reduces the Rh content 
(see Table \ref{tab:6} for Fe$_5$Rh and Fe$_6$). 

Concerning the magnetic properties one observes 
qualitatively similar trends as in the smaller clusters.
The average magnetic moment per atom $\overline\mu_N$ increases 
monotonously with Fe content. Accordingly, the 
energy gain $\Delta E_m$ associated to magnetism also
increases with the number of Fe atoms. There are in general very 
little differences in $\overline\mu_N$ between the optimal structure
and the first low-lying isomer. The largest part of the 
spin polarization (about 90\%) can be traced back to the 
local $d$ magnetic moments with the PAW sphere of the atoms.
As expected, the $s$ and $p$ spin polarizations are 
almost negligible in comparison to the $d$-orbital contributions.
A significant increase of the Fe moments is observed  upon Rh doping, 
which result from the larger number 
of available Fe $d$ holes and the low coordination number. Moreover, the Rh moments in Fe$_m$Rh$_n$ are 
stabilized by the proximity of the Fe atoms. In the alloy hexamers
the values of $\mu_{\rm Rh}$ are larger than in the pure
Rh$_6$. However, this is not a general trend, since the magnetic moments
in small Rh$_n$ are often quite important due to the extremely reduced 
coordination numbers.

\subsection{Exploring heptamers and octamers}
\label{sec:7} 

For Fe$_m$Rh$_n$ clusters having $m + n = N \ge 7$ we did not attempt 
to perform a systematic sampling of initial topologies for 
further unconstrained structural relaxation, as was done for the smaller
sizes. Instead of aiming at a true global optimization, only
a few compact and open starting structures are considered.
For $N = 7$, the topologies include 
bicapped trigonal bipyramid (BCTBP), capped octahedra (CO), and pentagonal bipyramid (PBP), 
while for $N = 8$, they are the
tricapped trigonal bipyramid (TCTBP), bicapped octahedra (BCO), capped pentagonal bipyramid (CPBP), 
and cube (C). This choice is motivated by previous results for pure clusters and by the trend to compact 
geometries observed for smaller sizes $ N \le 6$. Although far from exhaustive, 
the considered geometries allow to explore
various relevant growth patterns with a reasonable computational effort. 
Certainly, a more complete study would be necessary in order to draw definitive 
conclusions about the optimal topologies. For each composition, all possible distributions 
of the Fe and Rh atoms within the cluster, as well as all
relevant values of the total magnetization $S_z$ 
are taken into account (from the non-magnetic state to saturation).
Therefore, the trends on the interplay between chemical 
order and magnetic behavior remain rigorous within the framework of 
the sampled topologies.

\begin{table}
\centering
\caption{(Color online)\label{tab:7} 
Structural, electronic and magnetic properties of FeRh heptamers as 
obtained from a restricted representative sampling of cluster topologies (see text). 
        }
\renewcommand{\arraystretch}{0.15}
\begin{tabular}{|c|c|c|c|c|c|c|c|c|} 
\hline 
\hline
               &                                              &        &         &      &     &     &        &                      \\ 
Cluster        &Struct. &$E_B$&$\Delta{E_m}$&$d_{\alpha\beta}$&$\overline\mu_{N}$&$\mu_{\rm Fe}$&$\mu_{\rm Rh}$&$\nu_{0}$           \\ \hline
               &                                              &        &         &      &     &      &       &                      \\ 
Fe$_7$         &\raisebox{-0.6cm}{\includegraphics[scale=0.17]{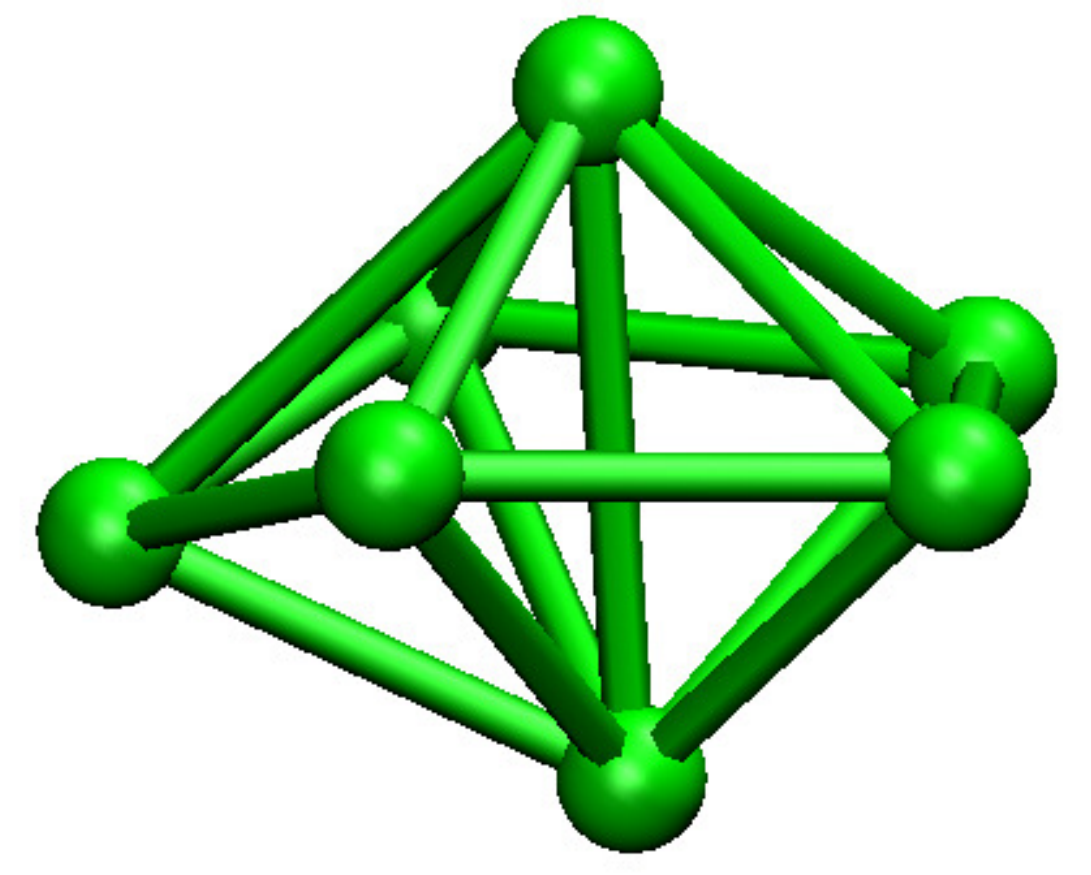}}&2.95    & 0.84 & 2.47& 3.14   & 2.88 &      & 209    \\[0.2cm]
Fe$_6$Rh       &\raisebox{-0.8cm}{\includegraphics[scale=0.14]{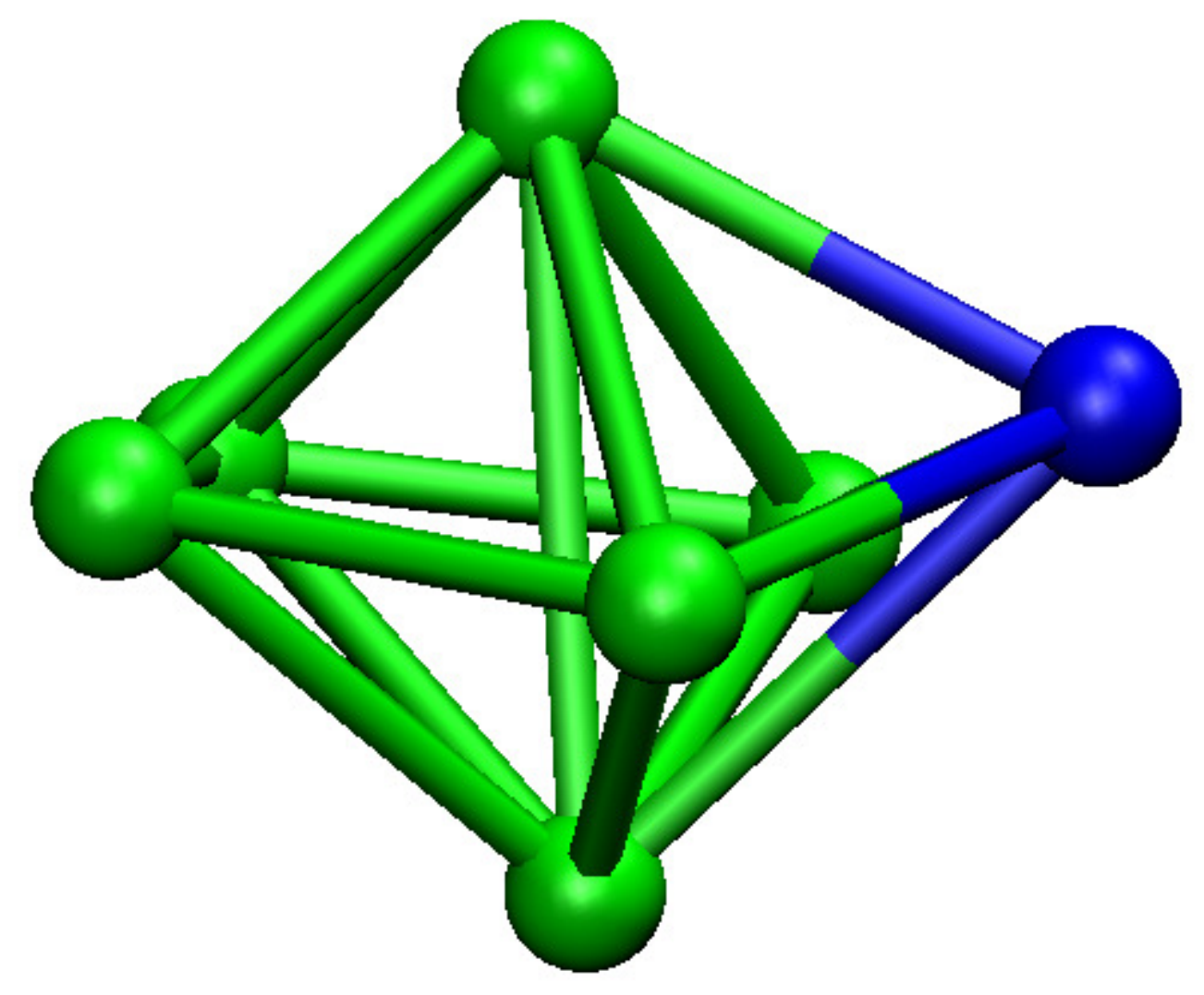}}&3.11  & 0.78 & 2.51& 3.00    & 2.98 & 1.17 & 226    \\[-0.6cm]
               &                                              &        &                      & 2.43 &     &      &       &       \\[0.5cm]   
Fe$_5$Rh$_2$   &\raisebox{-0.5cm}{\includegraphics[scale=0.13]{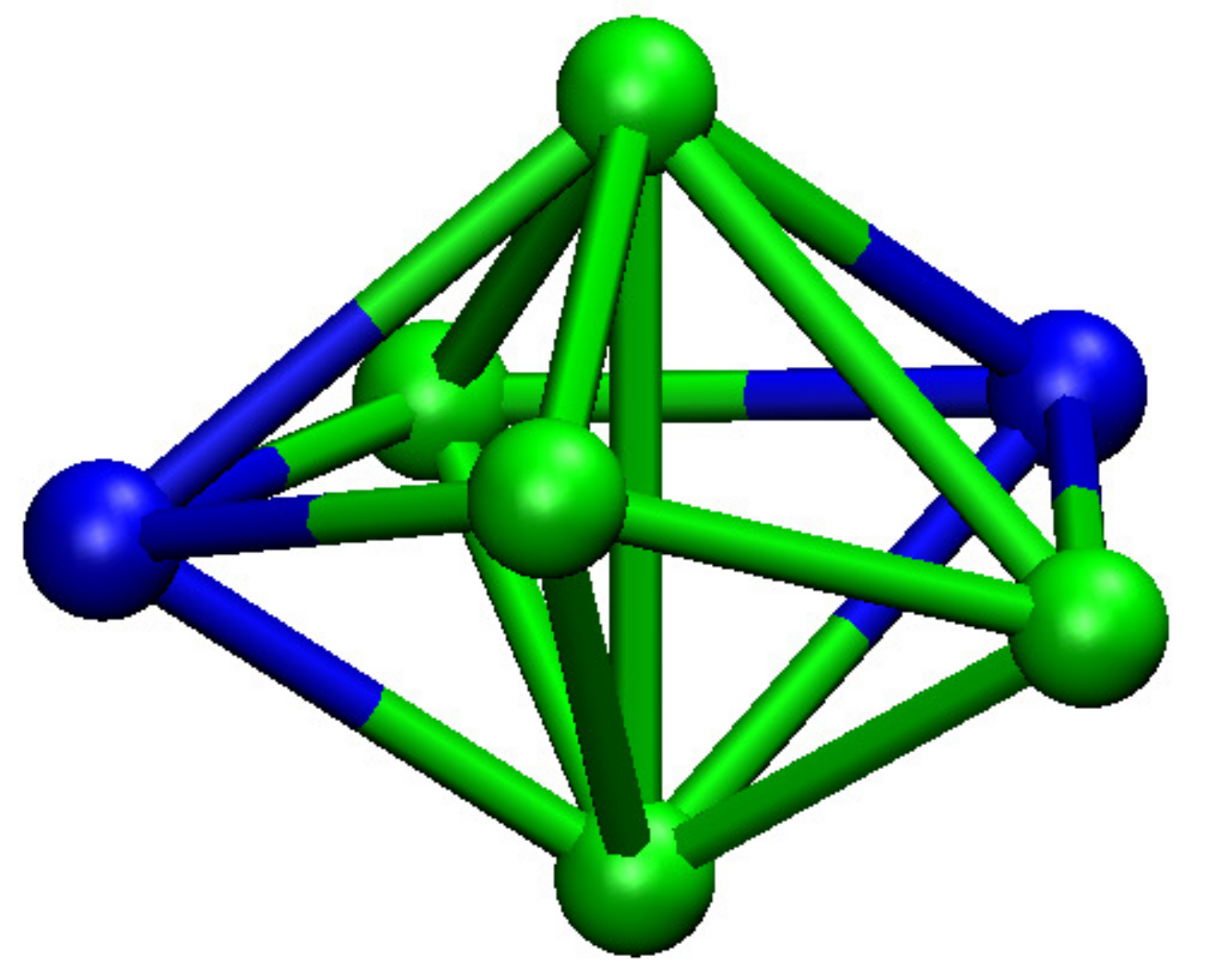}}& 3.25 & 0.73 & 2.45& 2.86 & 3.12 & 1.19 & 236    \\[-0.3cm]
               &                                              &        &                      & 2.46 &     &      &       &                      \\[0.2cm]
Fe$_4$Rh$_3$   &\raisebox{-0.8cm}{\includegraphics[scale=0.13]{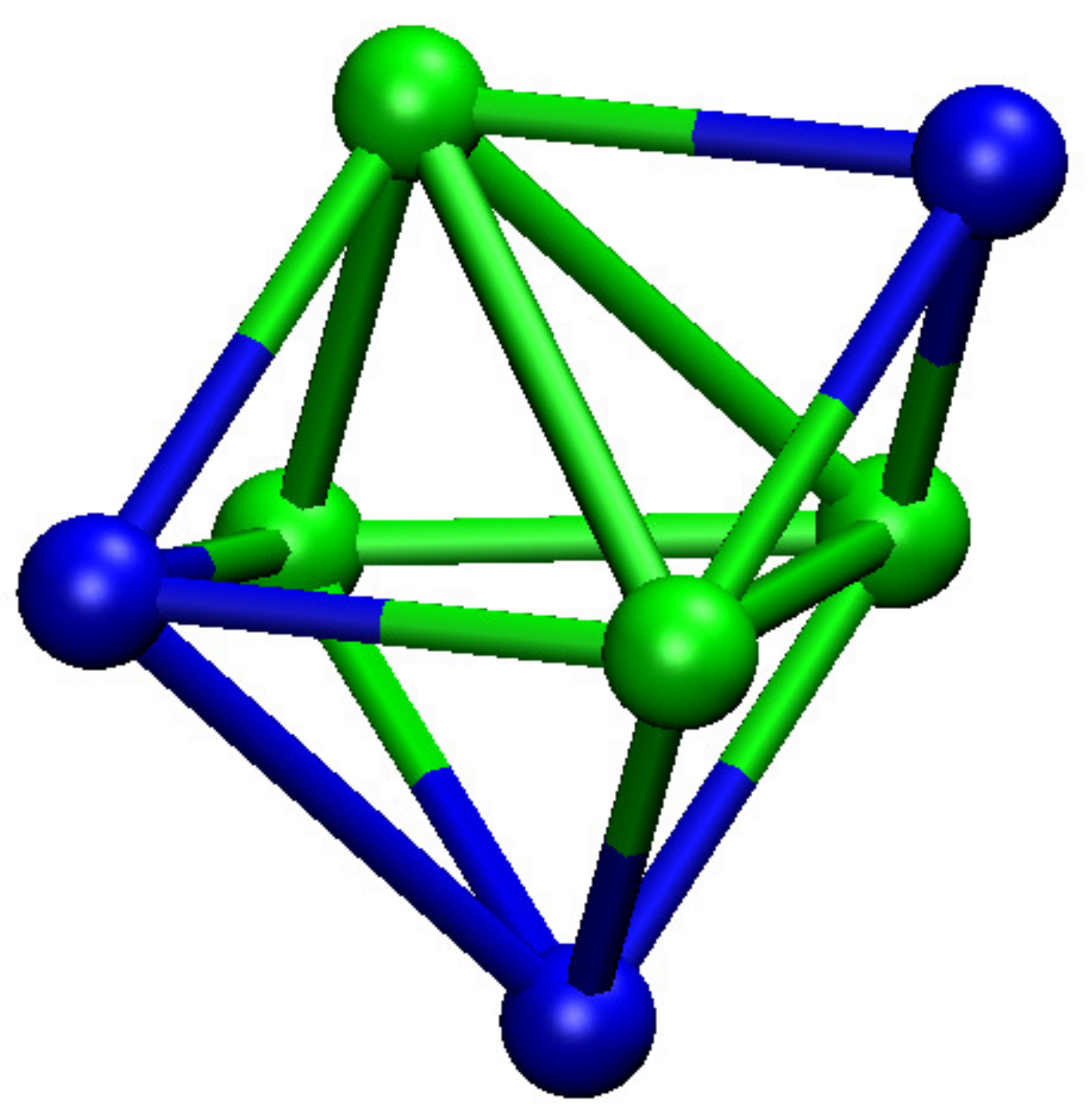}}& 3.36 & 0.62& 2.42& 2.71 & 3.27 & 1.26 & 222    \\[-0.65cm]
               &                                              &        &                      & 2.55 &     &      &       &                      \\[0.1cm]
               &                                              &        &                      & 2.71 &     &      &       &                      \\[0.1cm]
Fe$_3$Rh$_4$   &\raisebox{-0.7cm}{\includegraphics[scale=0.13]{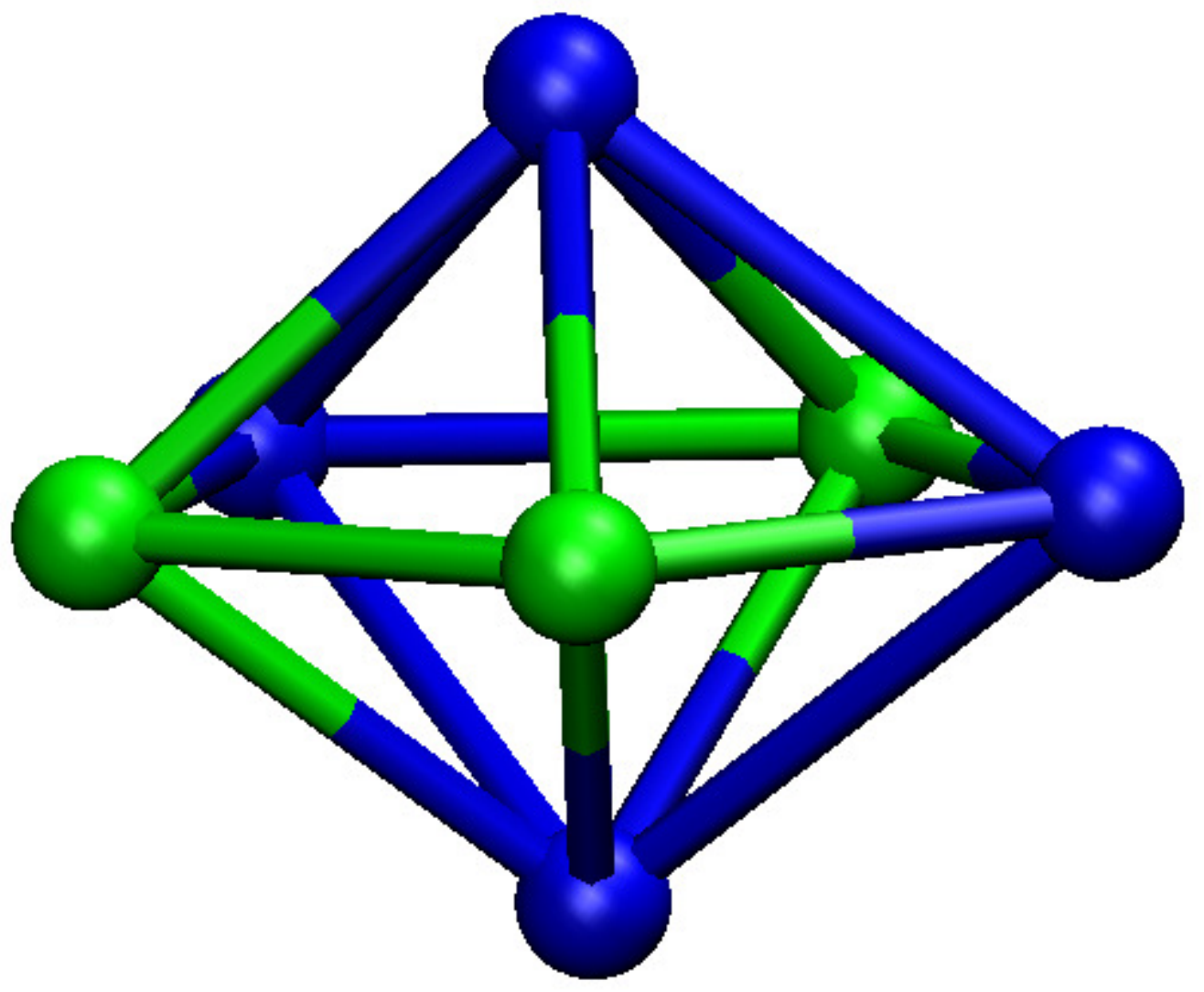}}& 3.38  & 0.57 & 2.48& 2.28 & 3.25 & 1.16 & 218    \\[-0.6cm]   
               &                                              &        &         & 2.37 &     &      &       &                      \\[0.1cm]
               &                                              &        &         & 2.67 &     &      &       &                      \\[0.1cm]
Fe$_2$Rh$_5$   &\raisebox{-0.7cm}{\includegraphics[scale=0.14]{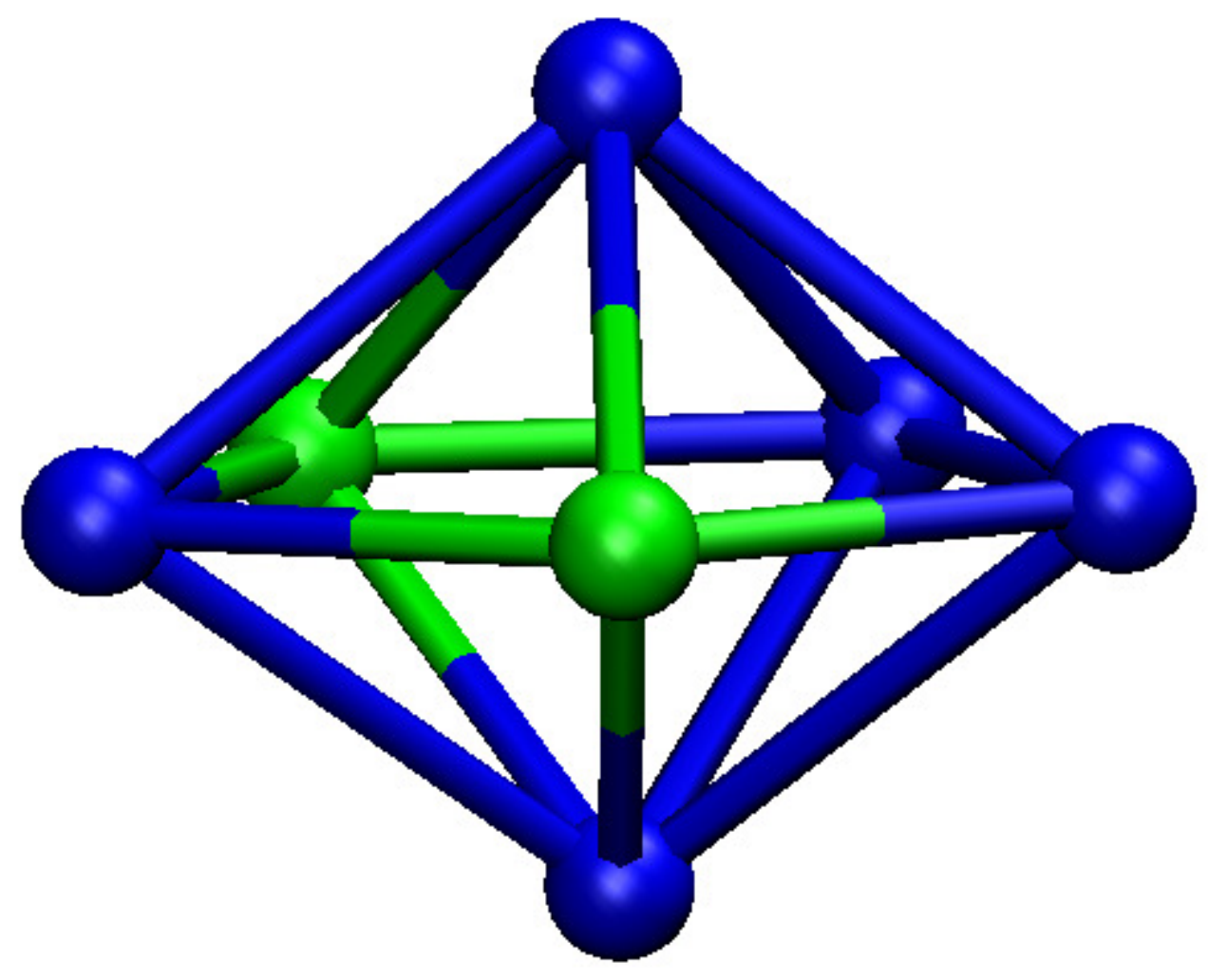}}& 3.41  & 0.45 & 2.40& 2.14 & 3.34 & 1.33 & 205    \\[-0.5cm]
               &                                              &        &         & 2.62 &     &      &       &                      \\[0.3cm]
FeRh$_6$       &\raisebox{-0.7cm}{\includegraphics[scale=0.15]{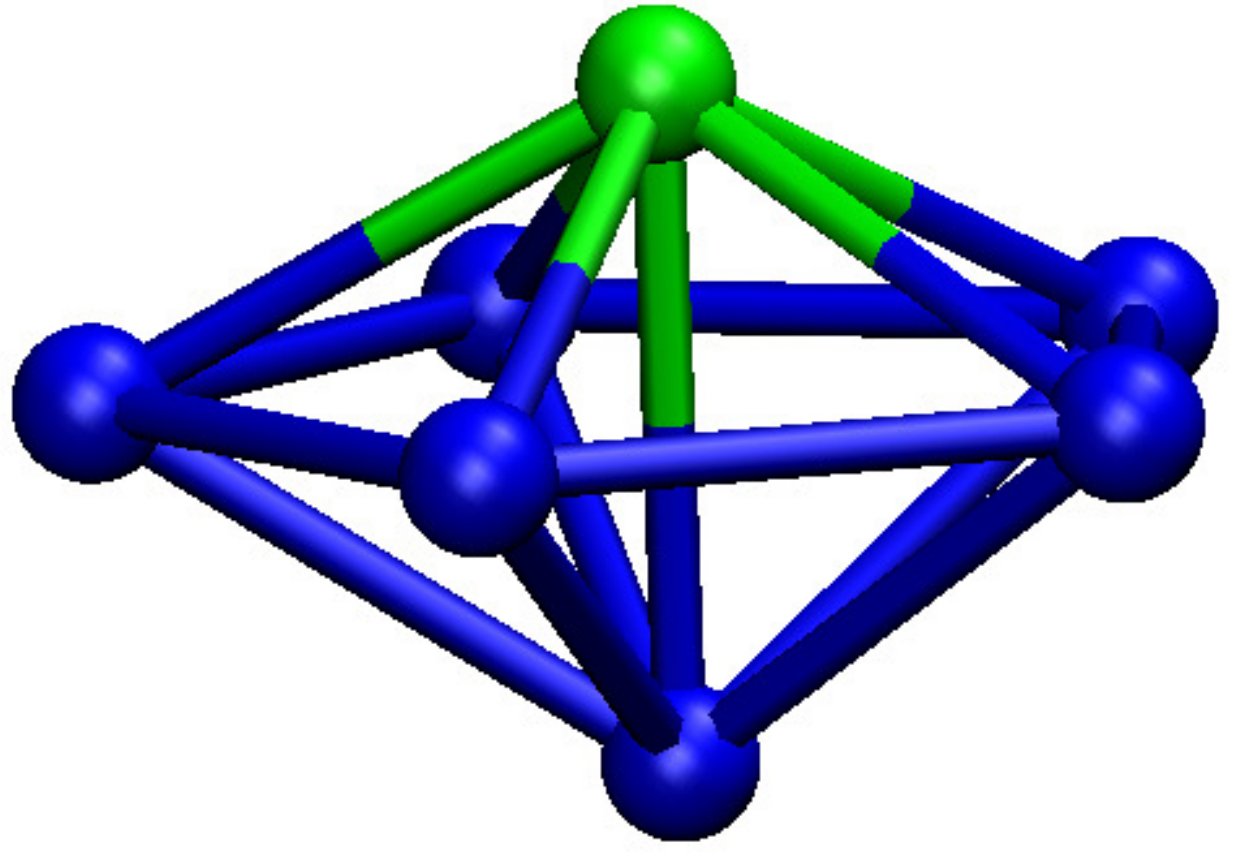}}& 3.37  & 0.28 & 2.52& 1.71 & 3.20 & 1.29 & 220    \\[-0.5cm]
               &                                              &        &         & 2.57 &     &      &       &                      \\[0.3cm]
Rh$_7$         &\raisebox{-0.7cm}{\includegraphics[scale=0.15]{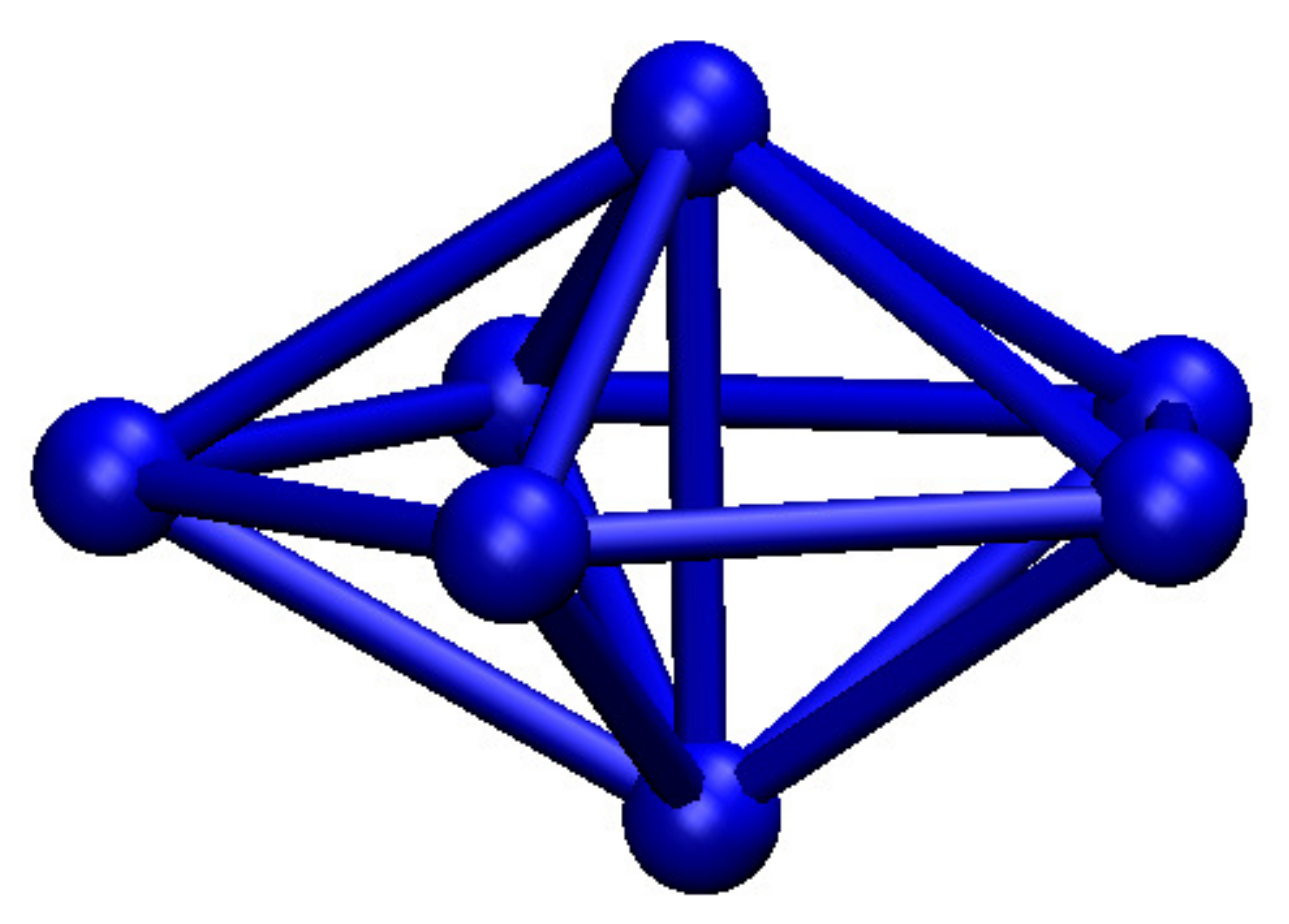}}& 3.33   & 0.22 & 2.61& 1.86 &  & 1.62        & 203    \\
               &                                              &        &         &      &     &      &       &                      \\
\hline \hline
\end{tabular}
\end{table} 

The results for $N=7$ are summarized in Table \ref{tab:7}. 
As in smaller Fe$_m$Rh$_n$ the binding energy per atom increases first with 
increasing Rh content and becomes essentially independent of composition
in the Rh-rich limit ($5 \le n \le 7$). 
For pure Rh$_7$ the PBP is the most stable structure among the  
considered starting geometries. This result is 
consistent with some earlier DFT studies.\cite{Rh4_ref} However it contrasts with the calculations
by Wang {\em et al.},\cite{Rh_lwang} who used the GGA functional of Ref.~\protect\onlinecite{pw91}~(PW91) and obtained a capped octahedra, 
and with the calculations of Bae {\em et al.},\cite{ycbau} who found a prism plus an atom on a square face. 
According to our results, these structures are, respectively 20 and 4 meV per atom higher in energy
than the PBP. In the case of a prism plus an atom on the square face, the energy difference with the ground state seems too small
to be able to draw definitive conclusions.  

FeRh heptamers with high Rh concentrations also favor a PBP topology.  
In FeRh$_6$ the Fe atom occupies an apex site, while in 
Fe$_2$Rh$_5$ and Fe$_3$Rh$_4$ the Fe atoms belong to the pentagonal ring.
Notice that the distances between the two apex atoms in Rh$_7$
and between the Fe and Rh apex atoms in FeRh$_6$ are relatively short. 
In Fe$_2$Rh$_5$ and Fe$_3$Rh$_4$ the Fe atoms are as far as possible 
from each other and the distance between the Rh apex atoms is larger.
This is consistent with the previously discussed trend to favor 
the stronger FeRh bonds. 
For example, the energy involved in changing the position of the
Fe atom in FeRh$_6$ from the apex (7 FeRh bonds) to the pentagonal 
ring (5 FeRh bonds) is $0.016$~eV per atom. 

As the Fe content increases, the topology of Fe$_m$Rh$_n$ changes. In fact
Fe$_4$Rh$_3$ corresponds to a CO, while for $m \ge 5$ the configuration
yielding the lowest energy can be regarded as a strongly distorted
PBP (see Table \ref{tab:7}). Already in Fe$_4$Rh$_3$, but also in 
Fe$_5$Rh$_2$, one observes a tendency of the Fe atoms to group in 
subclusters, bringing the Rh atoms to outer positions, so that the number
of FeRh bonds is largest. 
Concerning the shape of the Fe rich heptamers, one observes important deformations  
of the pentagonal bipyramid (D$_{5h}$ symmetry) which are similar to the distortions found in pure Fe$_7$.\cite{Fe_rollman, Fe_Ballone} While the precise origin 
of the symmetry lowering is difficult to establish in the alloys, it is reasonable to expect that it is similar to the case of pure Fe$_7$.
According to Ref.~\protect\onlinecite{Fe_rollman}, the deformations found in Fe$_7$ are due to the presence of degenerate electronic states
in the undistorted PBP structure. In order to verify this hypothesis we have analyzed the Kohn-Sham spectrum 
of the symmetric structure (D$_{5h}$ symmetry) and found that it is highly degenerate
at $\varepsilon_F$. In contrast the spectrum of the distorted structure has a band gap about 0.4 eV at $\varepsilon_F$.
This suggests that the distortions in Fe rich FeRh clusters can be interpreted as a Jahn-Teller effect.

\begin{table}
\caption{(Color online)\label{tab:8}
Structural, electronic and magnetic properties of FeRh octamers as 
obtained from a restricted representative sampling of cluster topologies (see text).
        }
\renewcommand{\arraystretch}{0.06}
\begin{tabular}{|c|c|c|c|c|c|c|c|c|} 
\hline 
\hline
               &                                              &        &         &       &      &     &        &                      \\ 
Cluster        &Struct. &$E_B$&$\Delta{E_m}$&$d_{\alpha\beta}$&$\overline\mu_{N}$&$\mu_{\rm Fe}$&$\mu_{\rm Rh}$&$\nu_{0}$             \\ \hline
               &                                              &        &         &       &      &     &        &                      \\ 
Fe$_8$         &\raisebox{-0.5cm}{\includegraphics[scale=0.11]{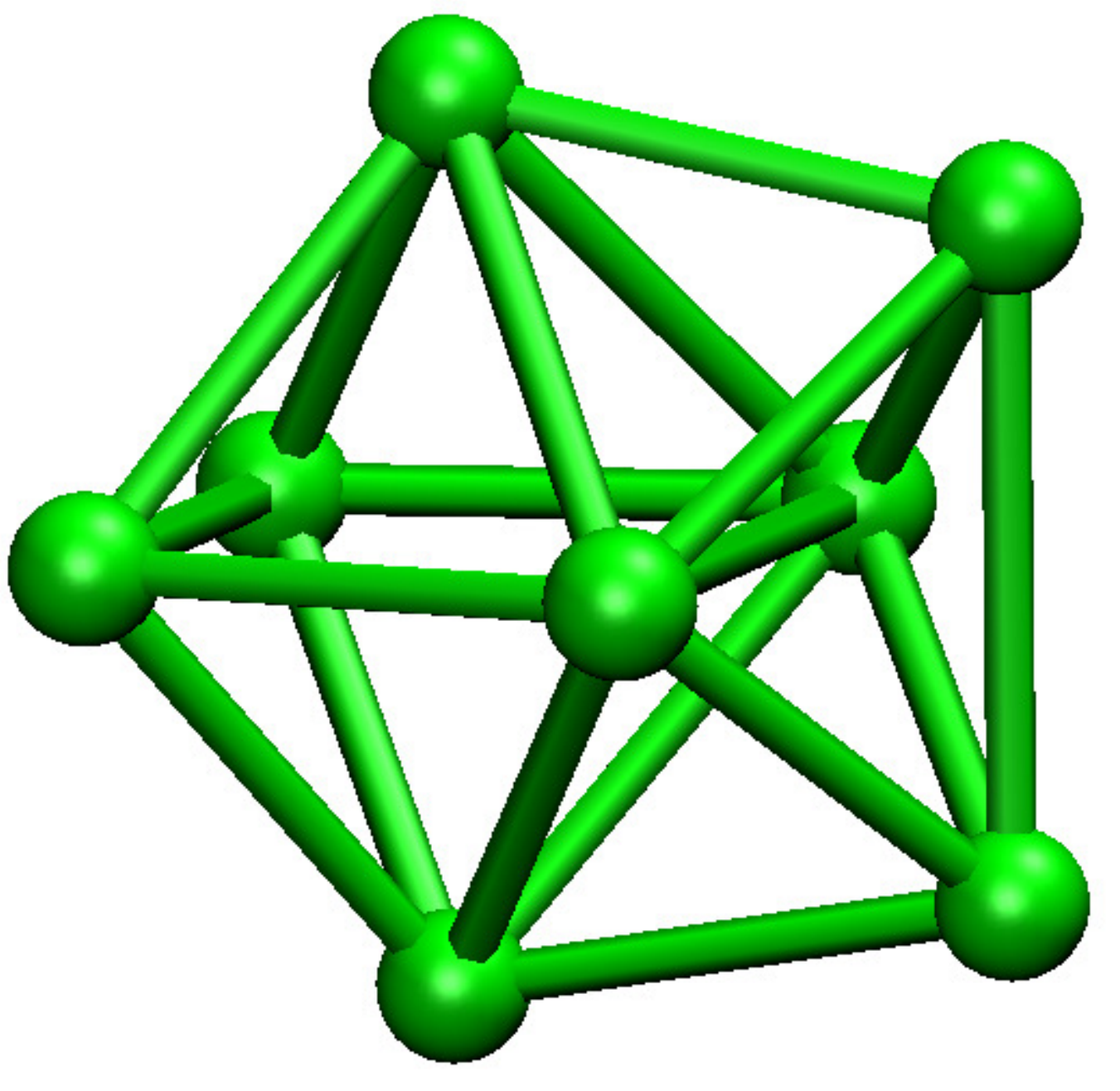}}&3.03& 0.76  & 2.42 & 3.00& 2.77   &               & 209  \\
               &                                              &        &         &       &      &     &        &                      \\
Fe$_7$Rh       &\raisebox{-0.5cm}{\includegraphics[scale=0.12]{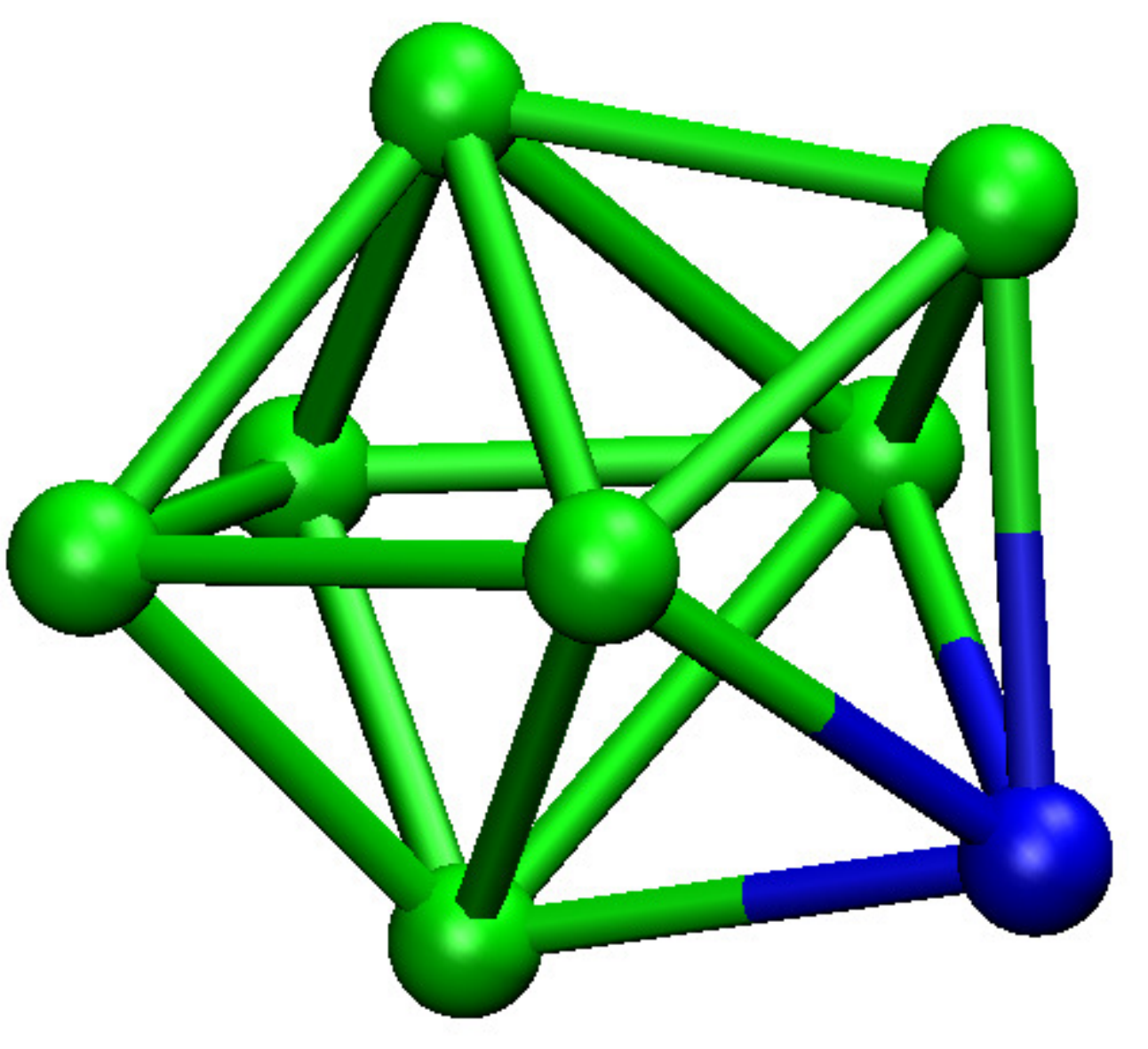}}  &3.19    & 0.73 & 2.43& 2.87   & 2.88   & 1.04 & 208  \\[-0.3cm]
               &                                              &        &         & 2.47  &      &     &        &                      \\[0.3cm]
Fe$_6$Rh$_2$   &\raisebox{-0.5cm}{\includegraphics[scale=0.12]{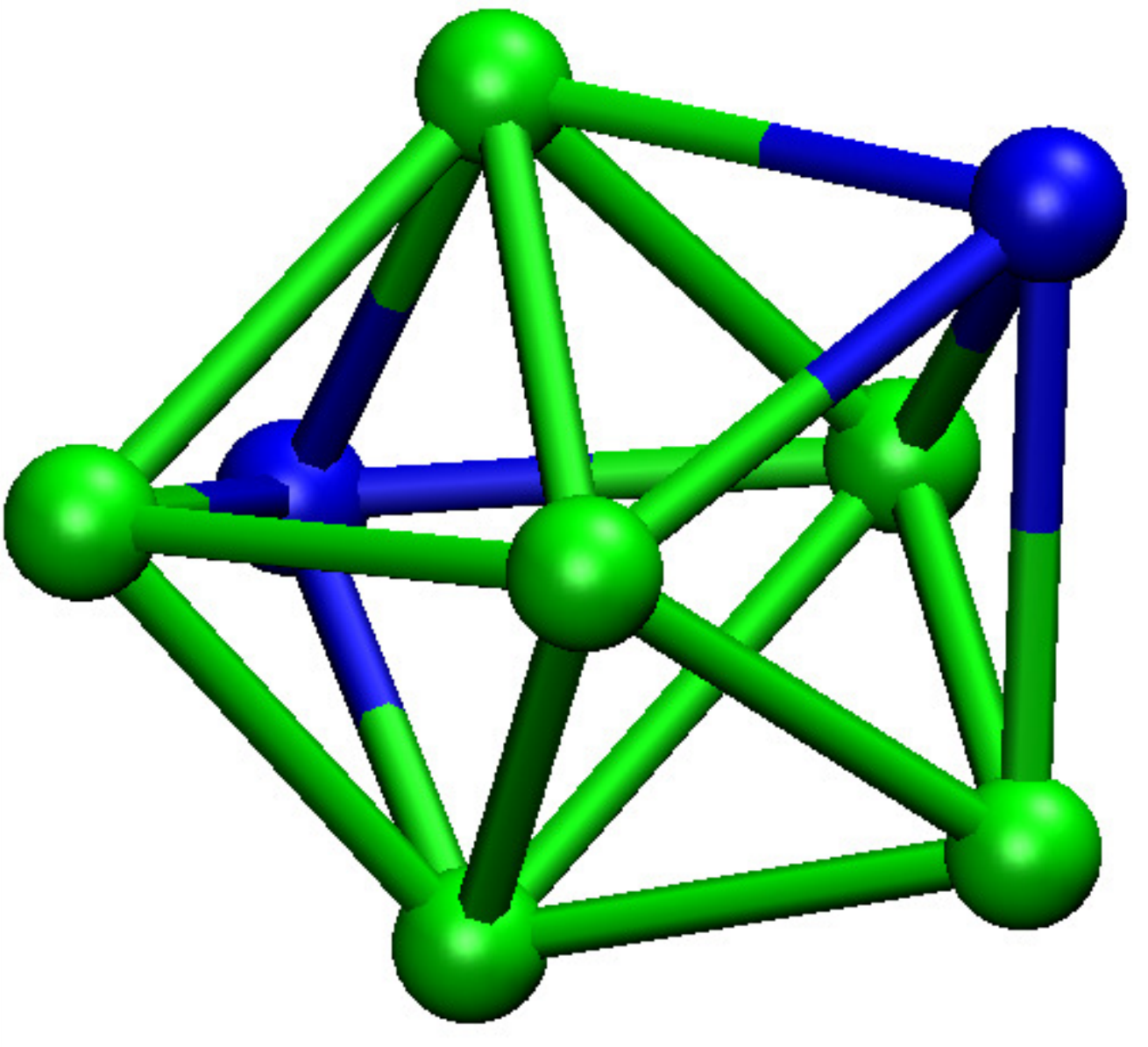}}  &3.32   & 0.66 & 2.43& 2.75   & 3.18   & 1.04 & 190  \\[-0.3cm]
               &                                              &        &         & 2.47  &      &     &        &                      \\[0.2cm]
Fe$_5$Rh$_3$   &\raisebox{-0.5cm}{\includegraphics[scale=0.14]{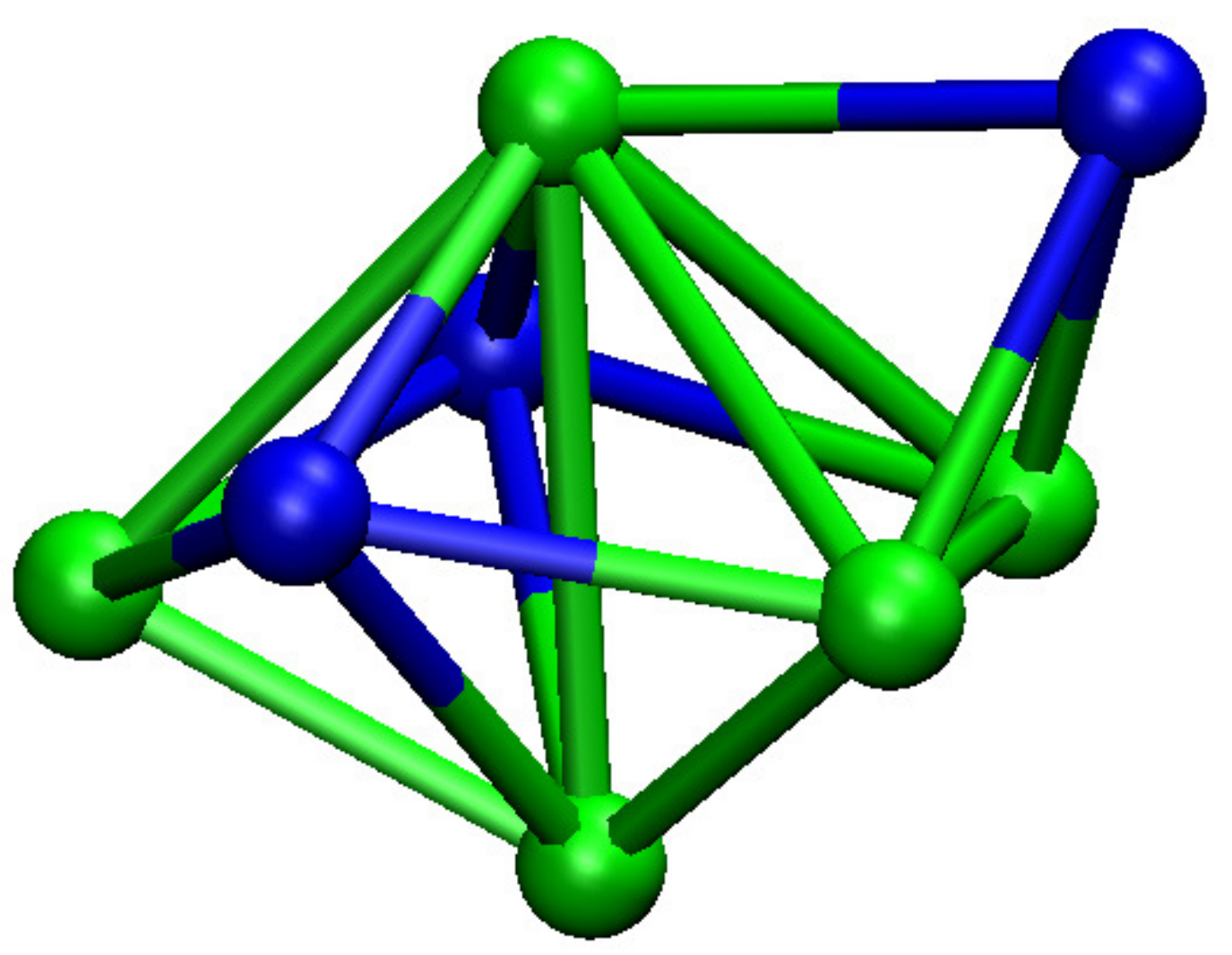}}& 3.42   & 0.69 & 2.43& 2.62   &3.15    & 1.10 & 189  \\[-0.4cm]
               &                                              &        &         & 2.57  &      &     &        &                      \\[0.3cm]
Fe$_4$Rh$_4$   &\raisebox{-0.5cm}{\includegraphics[scale=0.14]{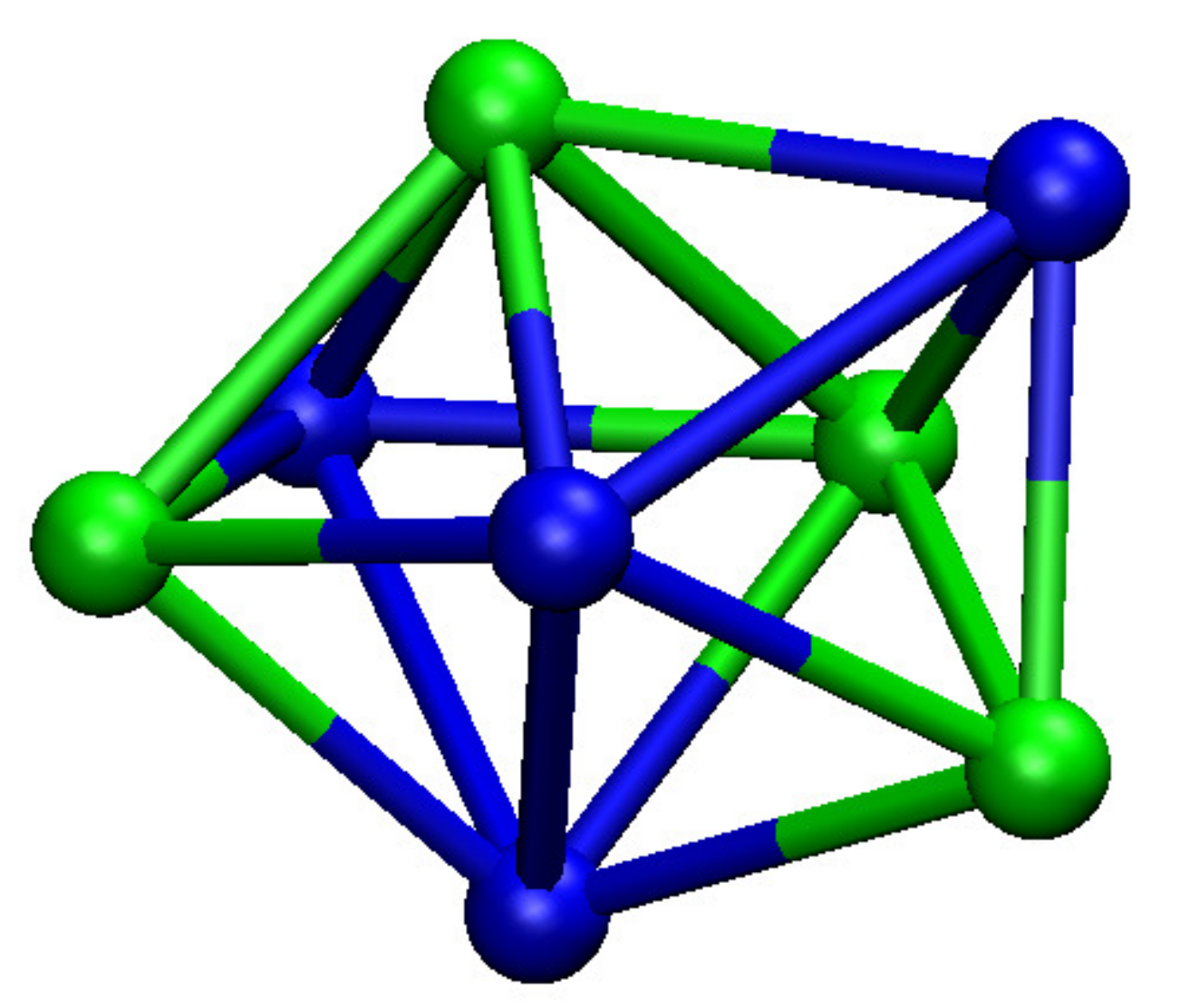}}& 3.47   & 0.53 & 2.45& 2.50   & 3.20   & 1.25 & 197  \\[-0.4cm]
               &                                              &        &         & 2.42  &      &     &        &                      \\[0.1cm]
               &                                              &        &         & 2.70  &      &     &        &                      \\
Fe$_3$Rh$_5$   &\raisebox{-0.6cm}{\includegraphics[scale=0.14]{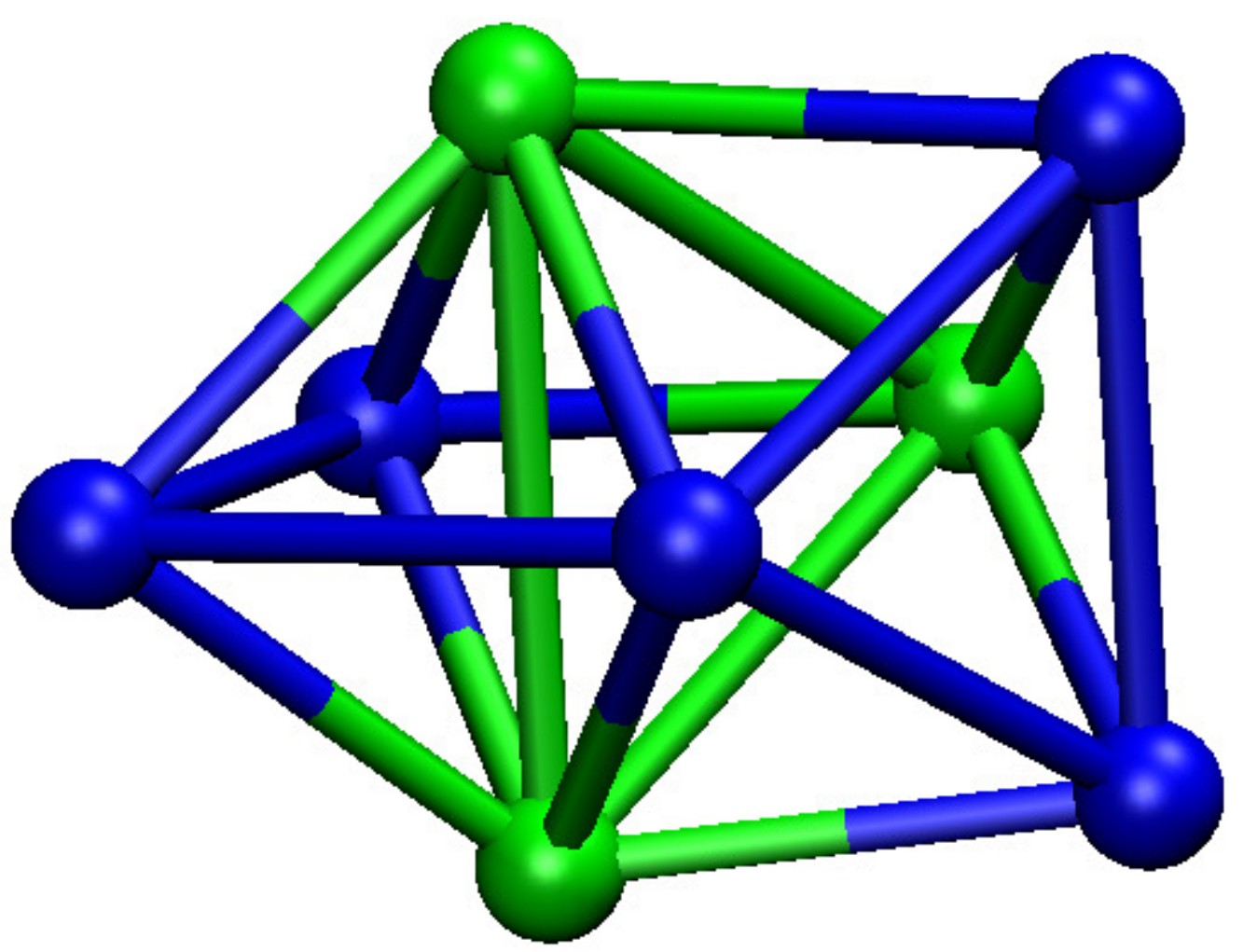}}  &3.54  & 0.42 & 2.77& 2.37   &3.36    & 1.34 & 214  \\[-0.5cm]
               &                                              &        &         & 2.39  &      &     &        &                      \\[0.1cm]
               &                                              &        &         & 2.65  &      &     &        &                      \\[0.2cm]
Fe$_2$Rh$_6$   &\raisebox{-0.5cm}{\includegraphics[scale=0.14]{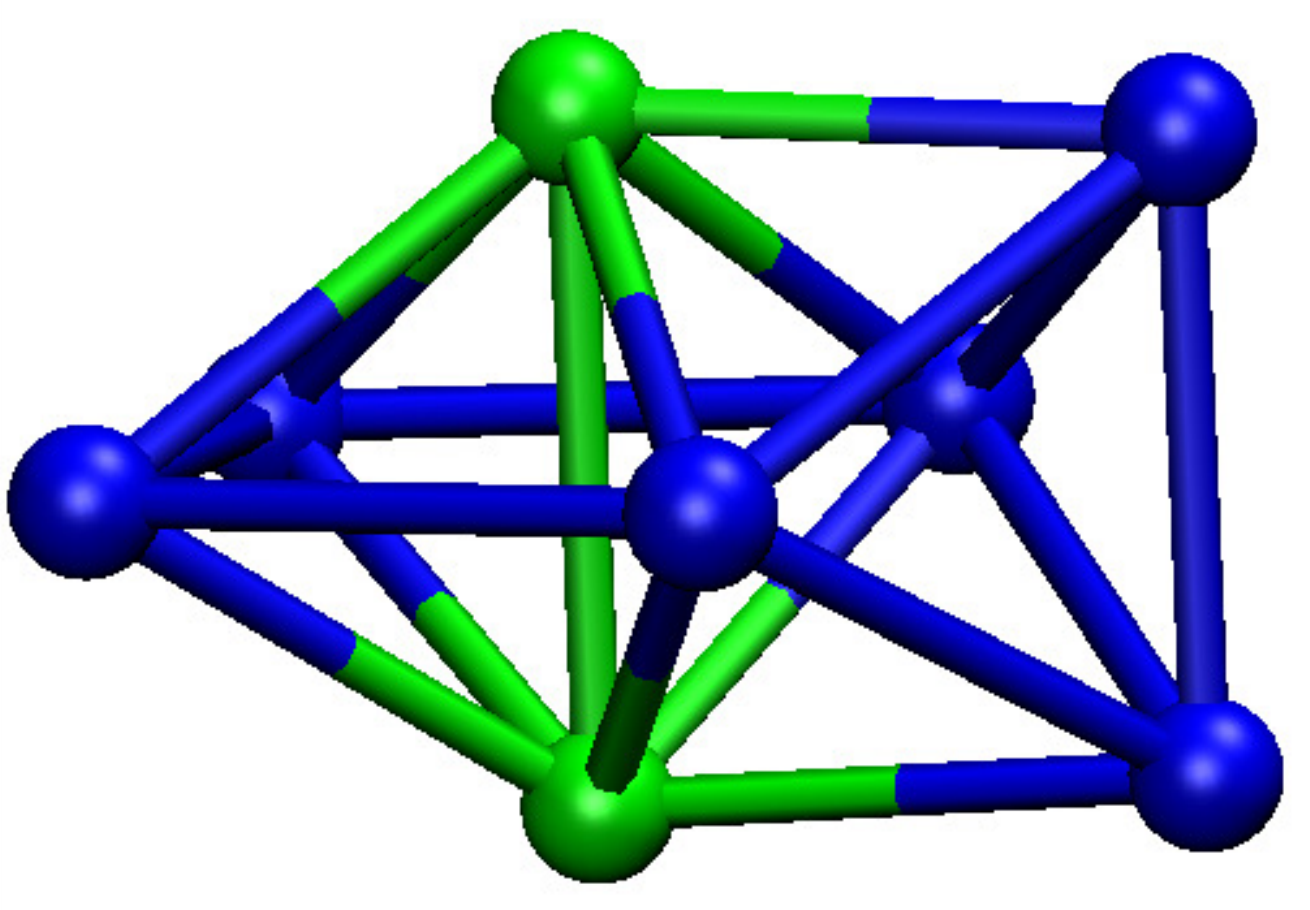}}& 3.53   & 0.32 & 2.70& 2.00   & 3.25   & 1.30 & 208  \\[-0.4cm]
               &                                              &        &         & 2.59  &      &     &        &                      \\[0.1cm]
               &                                              &        &         & 2.57  &      &     &        &                      \\
FeRh$_7$       &\raisebox{-0.5cm}{\includegraphics[scale=0.14]{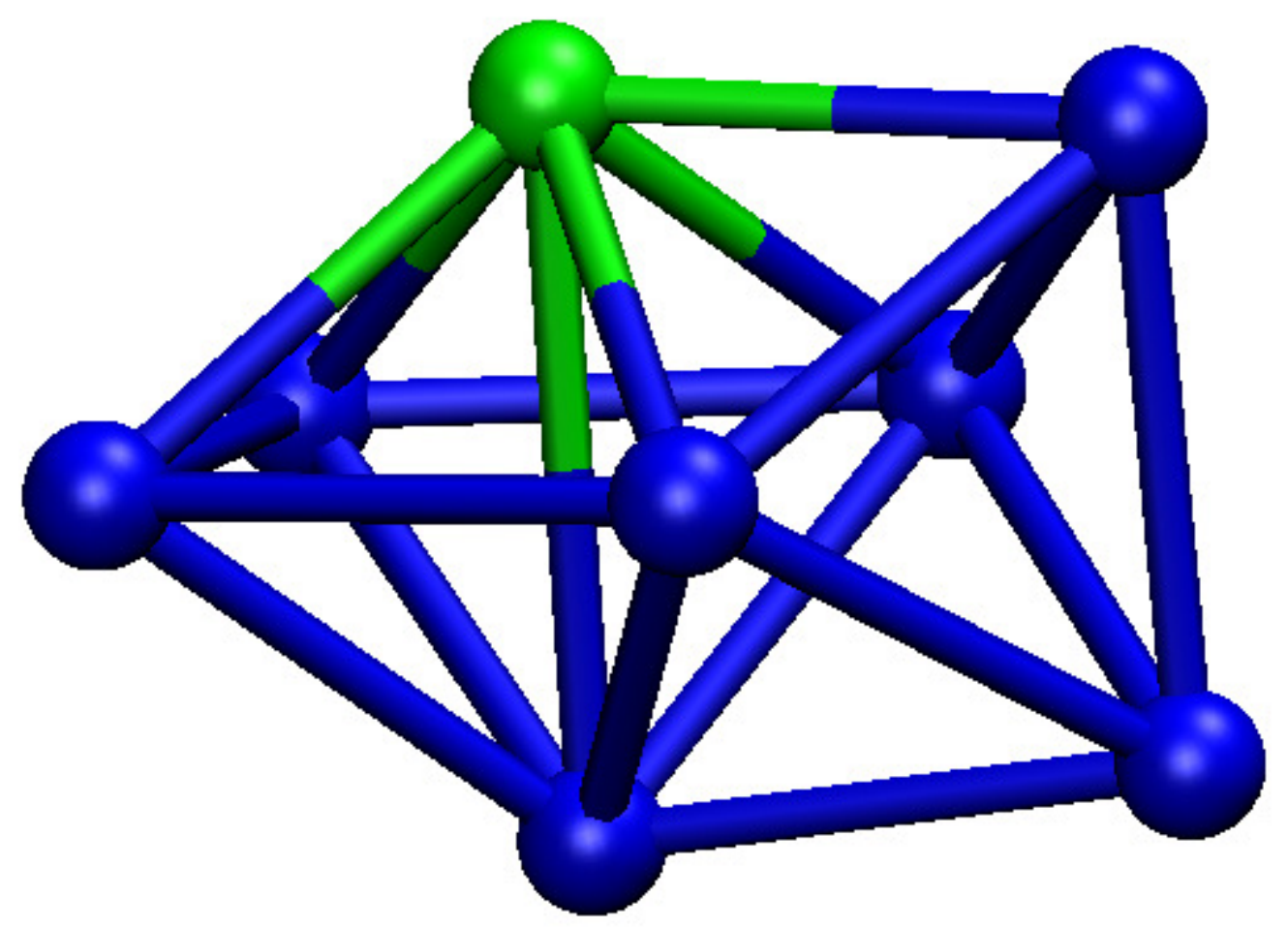}}& 3.49   & 0.23 & 2.45& 1.62   & 3.27   & 1.18 & 202  \\[-0.35cm]
               &                                              &        &         & 2.57  &      &     &        &                      \\[0.2cm]
Rh$_8$         &\raisebox{-0.4cm}{\includegraphics[scale=0.10]{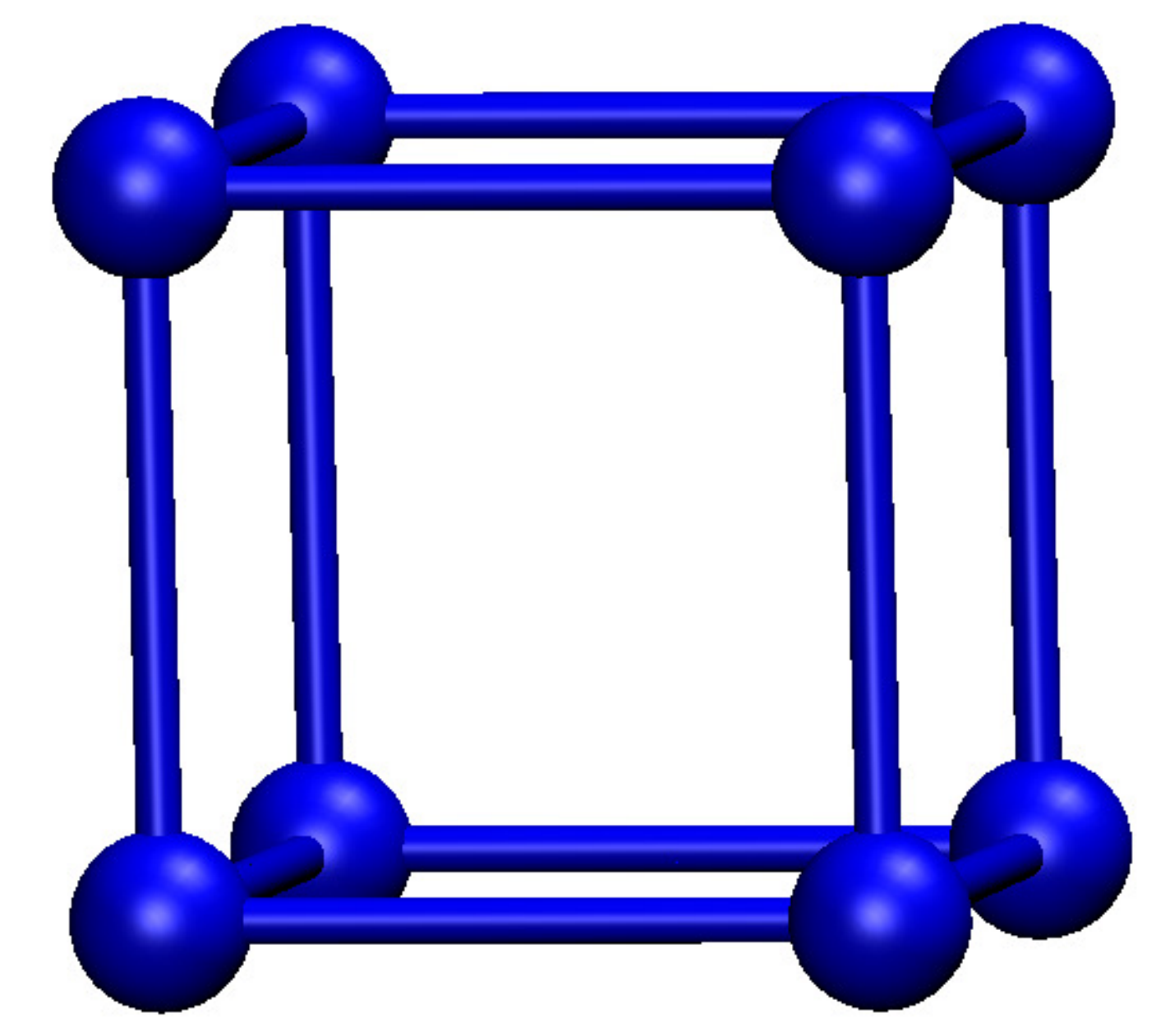}}& 3.59& 0.09 & 2.40& 1.50   &        & 1.33 & 160  \\
               &                                              &        &         &       &      &     &        &                      \\
\hline \hline
\end{tabular}
\end{table} 
%

%
%
In Table \ref{tab:8} results for FeRh octamers are reported.
The general trends concerning the composition dependence of the 
binding energy, chemical order, as well as the average and 
local magnetic moments are very similar to smaller clusters.
The most stable structure that we obtain for ${\rm Fe_8}$
is a BCO having $E_B = 3.03$~eV, an average magnetic moment $\overline\mu_8 = 3 \mu_B$, and a relatively 
short average bond-length $d = 2.42$~{\AA} (see 
Table~\ref{tab:8}). A similar structure is also found in previous spin-polarized LDA calculations,
where $E_B = 4.12$~eV and $\overline\mu_8 = 3 \mu_B$ were obtained.\cite{Fe_Dieguez} 
We have repeated these calculation for the BCO structure with our computational parameters and atomic reference energies and found  
$E_B = 3.51$~eV.
The discrepancies between LDA and GGA results reflect the importance of 
exchange and correlation to the binding energy. 
In the other extreme, for pure Rh$_8$,
the structure that we find with the considered starting
topologies is a regular cube having $E_B = 3.59$~eV,  
an average magnetic moment $\overline\mu_8 = 1.5 \mu_B$,
and all bond lengths equal to $2.40$~{\AA}.
These results are in good agreement with
previous calculations by Bae {\em et al.}.\cite{vijay05}
It is interesting to observe that the substitution of a single Rh atom by Fe in FeRh$_7$ results in a compact topology, which is more stable than
the relatively open (relaxed) cube-like structures derived from pure Rh$_8$. The same trend holds for higher Fe content (i.e., Fe$_m$Rh$_{8-m}$ with m $\geq 1$). 
The dominant structure for non-vanishing Fe content 
is a BCO with slight distortions. Only for Fe$_5$Rh$_3$
we find a different topology, namely, a distorted CPBP.
The typical isomerization energies between the BCO and
the TCTBP are $\Delta E_{\rm iso} = 10$--$30$~meV per atom. The average magnetic
moments in the lowest lying isomers are either the same or very similar.

The magnetic properties of heptamers and octamers follow qualitatively the
behavior observed in smaller clusters. In most cases the 
average magnetic moment per atom $\overline\mu_N$ and the magnetic 
energy $\Delta E_m$ increase with Fe concentration. 
The only exception is the pure Rh heptamer, for which $\overline\mu_7$ 
is somewhat larger than in FeRh$_6$. This is not due to AF-like coupling 
between Fe impurity moment and the remaining Rh atoms but rather to a reduction of the Rh local moments in FeRh$_6$ 
($\mu_{i}^{\rm Rh} \simeq 1.61$--$1.63 \mu_B$ in Rh$_7$, 
while $\mu_{i}^{\rm Rh} \simeq 1.25$--$1.30 \mu_B$ in FeRh$_6$). Remarkably, the Rh local moments 
in Rh$_7$ are the largest among all the heptamers. They amount to 87\%
of the total moment, which stresses the importance of the local $d$-electron
contributions. Also in Rh$_8$ one finds quite large local moments, 
which are actually larger than the Rh moments in most Fe doped clusters. This shows 
that for these sizes the Fe atoms do not necessarily increase 
the Rh moments by simple proximity effects (see Tables \ref{tab:7} 
and \ref{tab:8}). Nevertheless, a different behavior is expected for
larger $N$, where pure Rh clusters are no longer magnetic on their own. 
The local Fe moments are 
strongly enhanced with respect to pure Fe$_N$ 
($\mu_{\rm Fe} \simeq 2.8\mu_B$ in Fe$_7$ or Fe$_8$) 
reaching values up to $3.36 \mu_B$, 
particularly when the Fe atoms are in a Rh rich environment. As in the smaller clusters, this is a 
consequence of a charge transfer from the Fe to the Rh atoms, which increases the number 
of polarizable Fe $d$-holes. Notice that some kind of interaction between the Fe atoms 
seems to favor this effect, since the largest $\mu_{\rm Fe}$ are found 
for clusters having 2 or 3 Fe atoms rather than for the single Fe impurity.  
Large Fe moments are also found in bulk FeRh alloys.\cite{shira,gruner08} 

%
To conclude this section it is interesting to compare the cluster results 
with available experiments and calculations for macroscopic 
alloys.\cite{shira,moru,gruner08} 
Band structure calculations for the periodic Fe$_{0.5}$Rh$_{0.5}$ alloy
having a CsCl structure yield an antiferromagnetic (AF) ground state, 
which is more stable than the ferromagnetic solution.\cite{moru}
This is qualitatively in agreement with experiments showing AF 
order when the Rh concentration is above or equal to 50\%.\cite{shira}
In contrast our results for small clusters show
a FM-like order for all Rh concentrations, even for the pure 
Rh clusters. This is a consequence of the reduction of local coordination
number and the associated effective $d$-band narrowing, which renders 
the Stoner criterion far easier to satisfy, and which tends to stabilize 
the high-spin states with respect to the low-spin AF states. 
In fact, even in the bulk calculations
on FeRh, the energies of the AF and FM states are not very different, and
a coexistence of both solutions is found over a wide range of
volumes.\cite{moru} Moreover, experiment shows an AF to FM transition 
with increasing temperature, which is accompanied by an enhanced thermal 
expansion.\cite{shira}
Recent {\em ab initio} calculations have revealed 
the importance of competing FM and AF exchange interactions 
in stoichiometric $\alpha$-FeRh.\cite{gruner08} 
Moreover, neutron diffraction experiments\cite{shira} on Fe$_{1-x}$Rh$_x$
for $0.35 < x <0.5$ and calculations\cite{gruner08} for $x=0.5$ show that the Fe moments 
$\mu_{\rm Fe}$ are significantly enhanced with respect 
to $\mu_{\rm Fe}$ in pure $\alpha$-Fe, particularly in the FM 
state where it reaches values of about $3.2 \mu_B$.\cite{shira,gruner08}
These bulk results are remarkably similar to the trends found 
in Fe$_m$Rh$_n$ clusters over a wide range of compositions.
As in the clusters, the induced Rh moments $\mu_{\rm Rh}$ play 
an important role in the stability of the FM phase. Bulk experiments\cite{shira} on Fe$_{1-x}$Rh$_x$
yield $\mu_{\rm Rh}\simeq 1\mu_B$ for $0.35 < x <0.5$
which is comparable to, though somewhat smaller than the present 
cluster results.

\section{Size and composition dependence}
\label{sec:comp}

The main purpose of this section is to focus on the dependence
of the electronic and magnetic properties of Fe$_m$Rh$_n$ clusters 
as a function of size and composition. In the following we present 
and discuss results for the binding energy, average and local spin 
moments, and electronic densities of states for $N = m + n \le 8$.

\subsection{Binding energy and magnetic moments} 
\label{sec:EB}

\begin{figure}
\begin{center}
  \includegraphics[width=8cm,height=5.2cm,angle=0,clip=true]{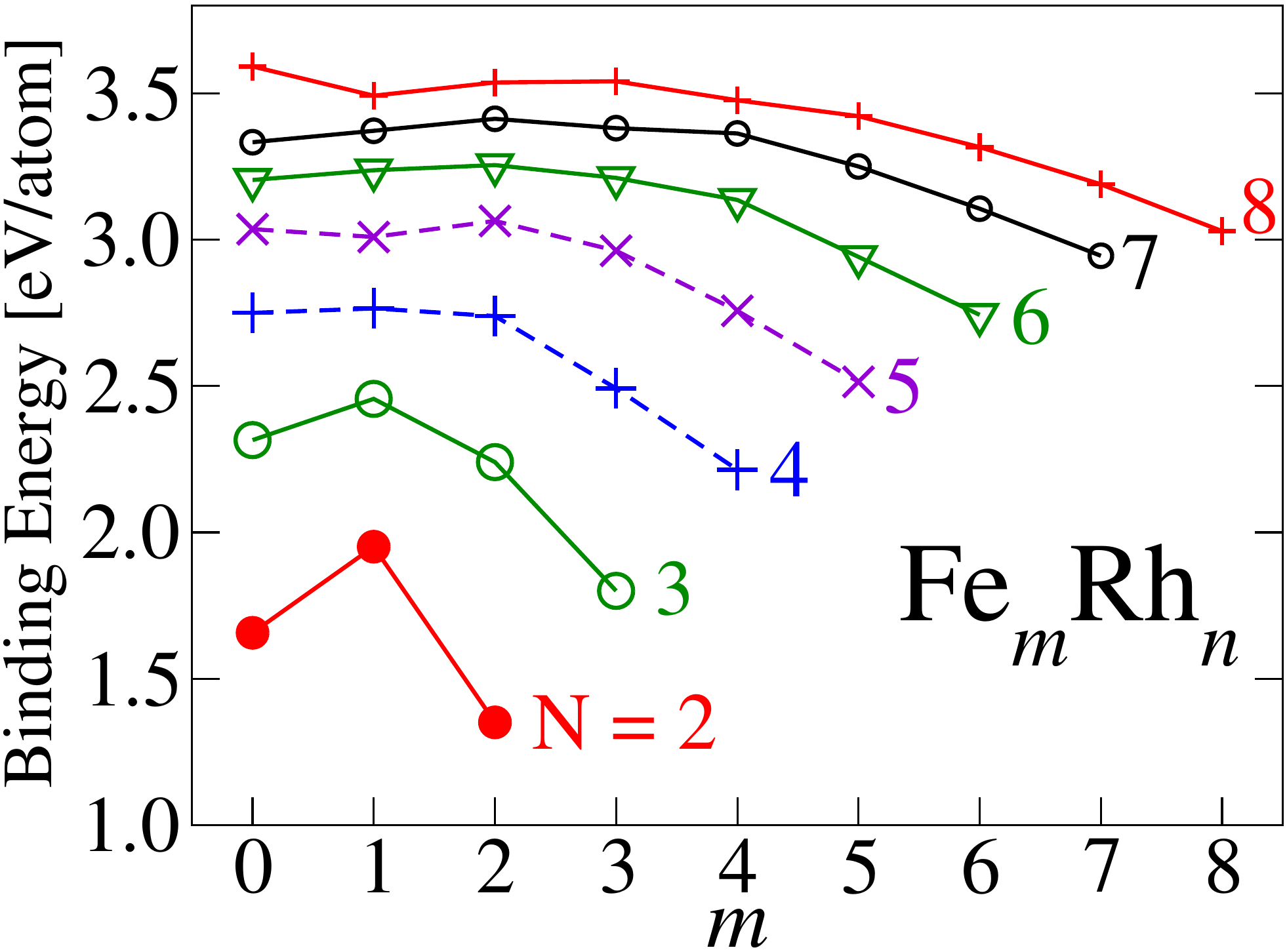}
\end{center}
\caption{(Color online) 
Binding energy per atom $E_B$ of Fe$_m$Rh$_n$ 
clusters as a function of the number of Fe atoms. The lines connecting 
the points for each $N = m + n$ are a guide 
to the eye.
        }
\label{fig:EB}
\end{figure}

In Fig.~\ref{fig:EB} the binding energy per atom $E_B$ is given as
a function of the number of Fe atoms $m$. Besides the expected 
monotonic increase of $E_B$ with increasing $N$, an interesting 
concentration dependence is observed. For very small sizes
($N\le 4$) $E_B$ is maximal for $m=1$ or $2$, despite the fact
that $E_B$ is always larger for pure Rh than pure Fe clusters.
This indicates that in these cases the bonding resulting
from FeRh pairs is stronger than RhRh bonds. Only for $m \ge  N - 1$,
when the number of weaker FeFe bonds dominates, one observes that 
$E_B$ decreases with increasing $m$. For larger sizes ($N\ge 5$)
the strength of RhRh and FeRh bonds becomes very similar, so that 
the maximum in $E_B$ is replaced by a range of
Fe concentrations $x = m /N \lesssim 0.5$ where $E_B$ depends
very weakly on $m$.

\begin{figure}
\includegraphics[width=8cm,height=5.2cm,angle=0,clip=true]
                {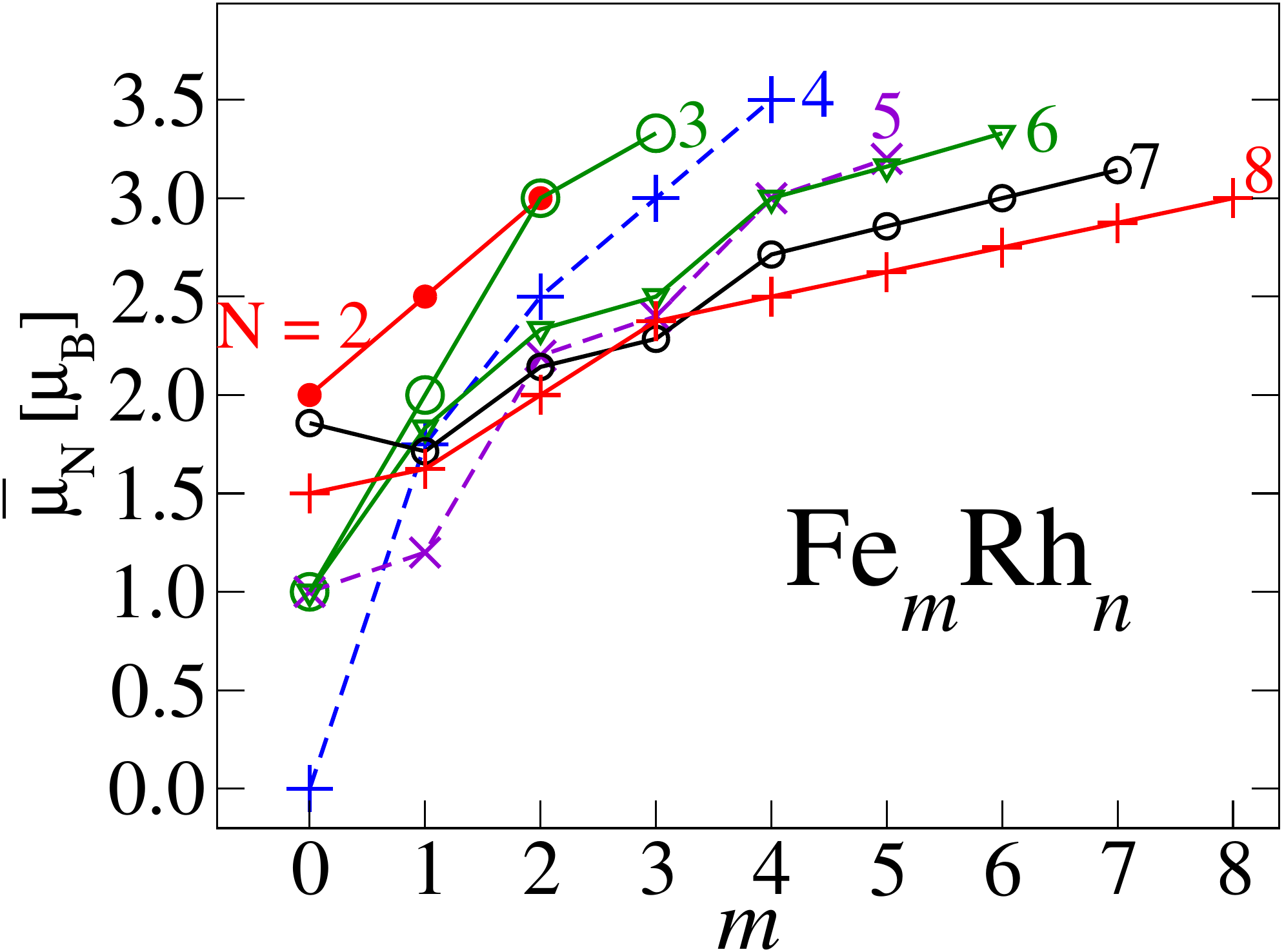}
\caption{(Color online) 
Total magnetic moment per atom $\overline\mu_{N}$ of Fe$_m$Rh$_n$ 
clusters as a function of number of Fe atoms. The symbols corresponding to each size are the same as in Fig.~\ref{fig:EB}.
The lines connecting the points for each $N = m + n$ are a guide to the eye. 
        }
\label{fig:moma}
\end{figure}

In Fig.~\ref{fig:moma} the average magnetic moments $\overline\mu_N$
of Fe$_m$Rh$_n$ are shown as a function of $m$ for $N\le 8$.
As already discussed in previous sections, $\overline\mu_N$ increases 
monotonously, with the number of Fe atoms. 
This is an expected consequence of the larger Fe local moments and
the underlying FM-like magnetic order. 
The average slope of the curves tends to increase with decreasing $N$, since 
the change in concentration per Fe substitution is more important the 
smaller the size is. The typical increase in $\overline\mu_N$
per Fe substitution is about ($1/N$)$\mu_B$ per Fe substitution.
Notice, moreover, the enhancement of the magnetic moments of the pure 
clusters in particular for Fe$_N$ ($m=N$), which go well beyond 
$3 \mu_B$, the value corresponding to a saturated $d$-band in the $d^7s^1$ 
configuration. In contrast, the moments of pure Rh$_N$ are far from saturated except for $N = 2$ and $7$ 
(see Fig.~\ref{fig:moma} for $m=0$). 
In this context it is important to recall that a thorough global optimization, for example, by considering a large number
of initial topologies, could affect the quantitative values of the magnetic moments for $N = 7$ and $8$.

\begin{figure}
\includegraphics[width=8cm,height=5.2cm,angle=0,clip=true]
                {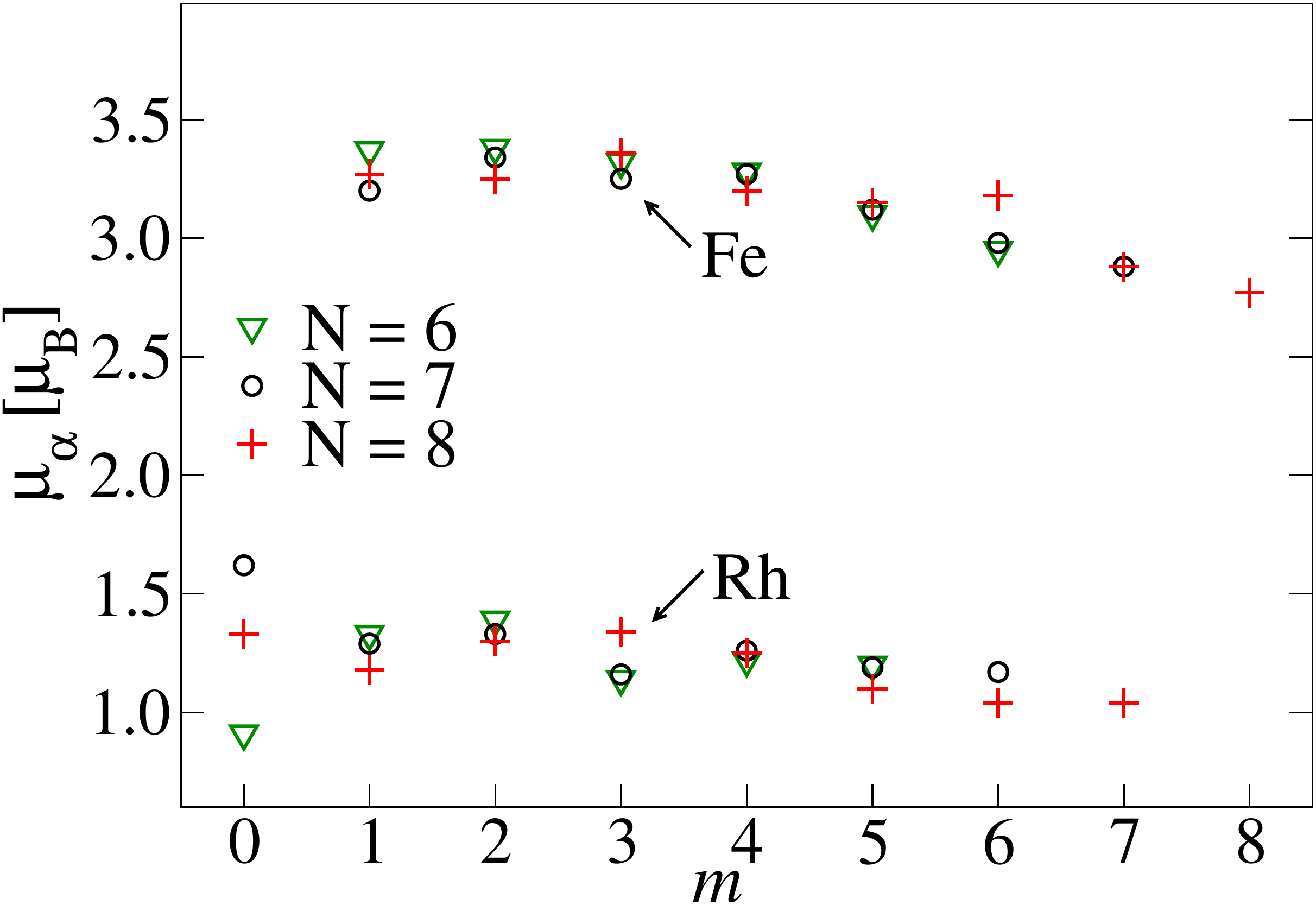}
\caption{(Color online) 
Local magnetic moment $\mu_{\alpha}$ at the Fe and Rh atoms as a function 
of the number Fe atoms $m$.}
\label{fig:moml}
\end{figure}

The local magnetic moments in the PAW sphere of the Fe and Rh atoms 
provide further insight on the interplay between $3d$ and $4d$ 
magnetism in Fe$_m$Rh$_n$. In Fig.~\ref{fig:moml} $\mu_{\rm Fe}$ and $\mu_{\rm Rh}$ are shown as a function 
of $m$ for $N = 6$--$8$. The Fe moments are 
essentially given by the saturated $d$-orbital contribution.
For pure Fe clusters the actual values of $\mu_{\rm Fe}$ within the PAW sphere 
are somewhat lower than $3\mu_B$ due to a partial spill-off of
the spin-polarized density. Notice that the Fe moments increase as
we replace Fe by Rh atoms showing some weak oscillations as a function of $m$.
The increase is rather weak for a single
Rh impurity in Fe$_{N-1}$Rh but becomes stronger reaching a more or less
constant value as soon as the cluster contains 2 or more 
Rh ($m \le N-2$, see Fig.~\ref{fig:moml}). This effect can be traced 
back to a $d$ electron charge transfer from Fe to Rh which, together with
the extremely low coordination number, yields a full polarization of the 
larger number of Fe $d$ holes. On the other side the Rh moments 
are not saturated and therefore are more sensitive to size, structure
and composition. The values of $\mu_{\rm Rh}$ are in the range of 
$1$--$1.5\mu_B$ showing some oscillations as a function of $m$. 
No systematic enhancement of $\mu_{\rm Rh}$ with increasing Fe content
is observed. This behavior could be related to charge transfers effects
leading to changes in the number of Rh $d$ electrons as a function of $m$.

Finally, it is interesting to analyze the role played by
magnetism in defining the cluster structure by comparing
magnetic and non-magnetic calculations. For the smallest
FeRh clusters ($N=3$ and $4$) the magnetic energy $\Delta E_m = E(S_z \!=\!0) - E(S_Z)$ 
gained upon magnetization is higher in the first excited isomer than in
the most stable structure. This implies that the contribution of magnetism
to the structural stability is not crucial, since the non-magnetic calculations
yield the same ordering, at least concerning the
two best structures. This suggests that for the smallest sizes
the kinetic or bonding energy dominates the structural stability,
which also explains that the two most stable isomers
have different topologies.
The situation changes for large clusters. For $N\ge 5$ one finds a number of
FeRh clusters for which the optimal structure is actually stabilized by magnetism.
For example, in Fe$_4$Rh, Fe$_3$Rh$_2$, and FeRh$_4$ the energy ordering
of the two most stable isomers would be reversed if magnetism were neglected.
It should be noted that in these cases the structures differ only in the 
chemical order, not in the topology which is a TBP. In the FeRh hexamers 
the energy differences between the low-lying isomers are more important and only
in one case, Fe$_4$Rh$_2$, magnetism appears to be crucial for stabilizing the
actual optimal structure. A similar strong interplay between structure, 
chemical order and magnetism is expected for larger FeRh clusters.

\subsection{Electronic structure} 
\label{sec:elst}

\begin{figure*}
\includegraphics[width=15cm,height=13.6cm,angle=0,clip=true]
                {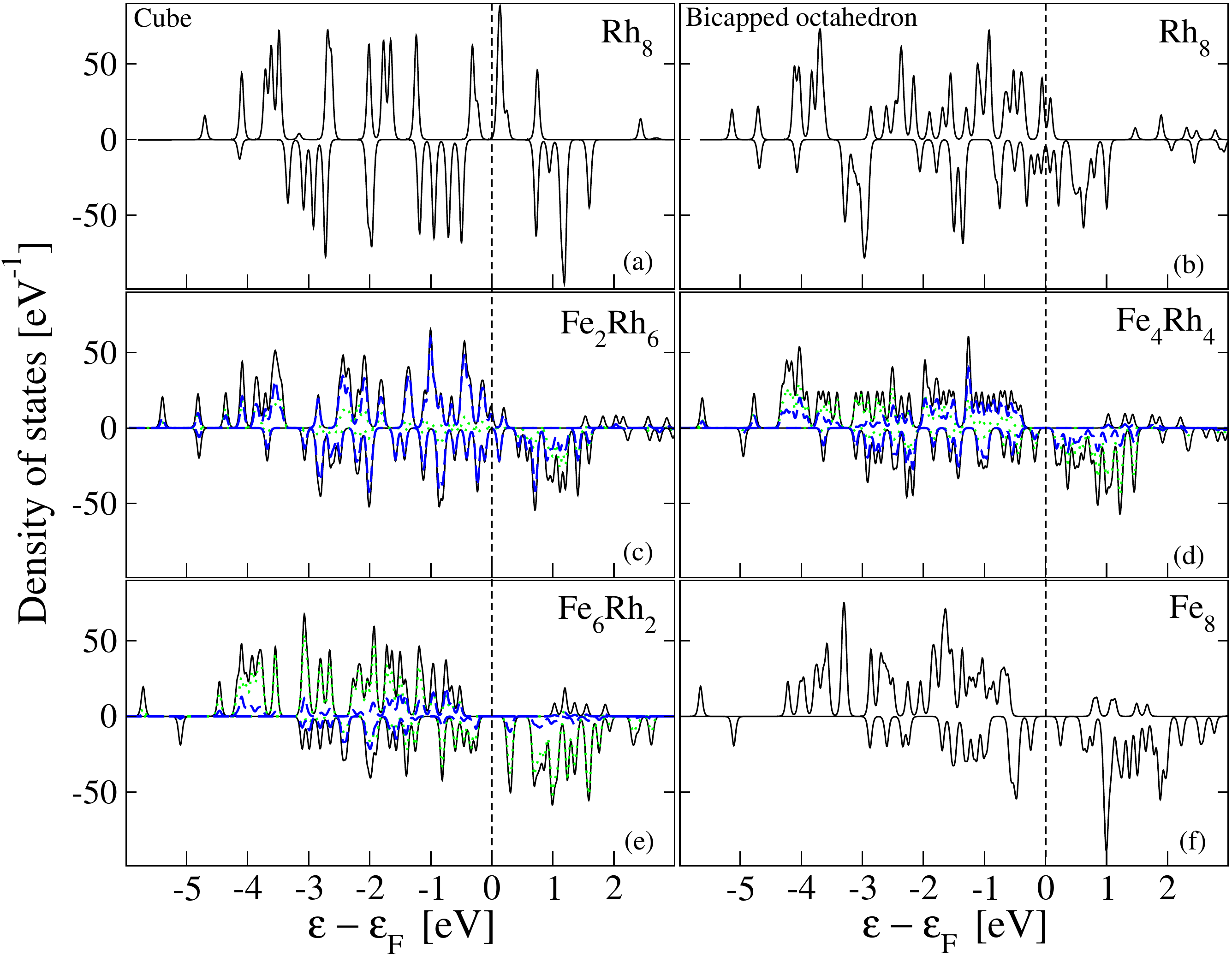}
\caption{(Color online) 
Electronic density of states (DOS) of FeRh octamers.
Results are given for the total (solid), the Fe-projected (dotted), 
and the Rh-projected (dashed) $d$-electron DOS. Positive (negative) 
values correspond to majority (minority) spin. 
A Lorentzian width $\lambda = 0.02$~eV has been used to
broaden the discrete energy levels. The corresponding structures 
are illustrated in Table~\protect\ref{tab:8}.
        }
\label{fig:dos}
\end{figure*}

In the previous sections the structure and spin moments of FeRh 
alloy clusters have been discussed as a function of 3$d$/4$d$ concentration. 
Although these properties are intimately related to the size and composition 
dependence of electronic structure, it is in general 
very difficult to achieve a physical transparent correlation between 
global and microscopic behaviors. Nevertheless, it is very interesting 
to analyze, at least for some representative examples, how the 
electronic structure depends on the composition of magnetic 
nanoalloys. To this aim we report in Fig.~\ref{fig:dos}
the spin-polarized $d$-electron density of states (DOS) of 
representative FeRh octamers having the relaxed
structures illustrated in Table~\ref{tab:8}. Results
for pure Fe$_8$ and Rh$_8$ are also shown for the sake of comparison.
In all the clusters, the dominant peaks in the relevant energy range
near $\varepsilon_F$ correspond either to the Fe-$3d$ or to
the Rh-$4d$ states. The valence spectrum is 
largely dominated by these $d$-electron contributions. In fact the
total DOS and the $d$-projected DOS are difficult to tell apart. 

First of all, let us consider the DOS of the pure clusters. Our results for Rh$_8$ with a cube structure are similar to those 
of previous studies.\cite{vijay05} They show the dominant $d$-electron contribution near $\varepsilon_F$, with the characteristic  
ferromagnetic exchange splitting between the minority and majority spin states.
In Fig.~\ref{fig:dos} we also included the DOS for Rh$_8$ with a BCO structure, since it allows us
to illustrate the differences in the electronic structure of compact and 
open geometries. Moreover, the DOS of pure Rh$_8$ with BCO structure is very useful in order to demonstrate
the dependence of DOS on Fe content, since the structures of 
Fe$_m$Rh$_{8-m}$ with $m \geq 1$ are similar to the BCO. 
Both Fe$_8$ and Rh$_8$ show relatively narrow $d$-bands
which dominate the single-particle energy spectrum in the range 
$-5 {\rm eV} \le \varepsilon - \varepsilon_F \le 3 {\rm eV}$. 
The spin polarization of the DOS clearly reflects the ferromagnetic order
in the cluster. Putting aside the exchange splitting, the peak 
structure in the up and down DOS $\rho_\sigma(\varepsilon)$
are comparable. There are even  
qualitative similarities between the two elements. However, looking in 
more detail, one observes that the effective $d$-band width in
Fe$_8$ (about 4~eV) is somewhat smaller than in Rh$_8$ (about 5~eV).
Moreover, in Rh$_8$ the DOS at $\varepsilon_F$ is non-vanishing for both spin 
directions and the finite-size gaps are very small 
(see Fig.~\ref{fig:dos}). In contrast, the majority $d$-DOS 
is fully occupied in Fe$_8$, with the highest majority state
lying about 0.5~eV below $\varepsilon_F$. In addition there
is an appreciable gap (about 0.1~eV) in the corresponding 
minority spectrum. These qualitative differences are of course consistent
with the fact that Fe$_8$ is a strong ferromagnet with saturated moments, 
while Rh$_8$ should be regarded as a weak unsaturated ferromagnet.

The trends as a function of concentration reflect the crossover between
the previous contrasting behaviors.
For low Fe concentration (e.g., Fe$_2$Rh$_6$)
we still find states with both spin directions close to 
$\varepsilon_F$. The magnetic moments are not saturated, 
although the Fermi energy tends to approach the 
top of the majority band. Moreover, the majority-spin states 
close to $\varepsilon_F$ have dominantly Rh character. Small Fe doping
does not reduce the $d$-band width significantly. Notice the rather
important change in the shape of the DOS in Fe$_2$Rh$_6$ as compared
to the DOS in Rh$_8$. This is a consequence of the change in topology
from cubic to bicapped octahedron (BCO).  
 
For equal concentrations (Fe$_4$Rh$_4$) the first signs of 
$d$-band narrowing and enhanced exchange splitting 
start to become apparent. The spin-up states (majority band) 
which in Fe$_2$Rh$_6$ contribute to the DOS at $\varepsilon_F$ 
now move to lower energies (0.3 eV below $\varepsilon_F$) 
so that the majority band 
is saturated. Only spin-down (minority) states are found around $\varepsilon_F$,
although there is a significant gap in $\rho_\downarrow(\varepsilon)$  
(see Fig.~\ref{fig:dos}). In the majority band Rh dominates over Fe 
at the higher energies (closer to $\varepsilon_F$), while Fe dominates 
in the bottom of the band. In the minority band the participation of Rh (Fe)
is stronger (weaker) below $\varepsilon_F$ and weaker (stronger) 
above $\varepsilon_F$. This is consistent with the fact that the Rh 
local moments are smaller than the Fe moments. 

Finally, in the Fe rich limit (e.g., Fe$_6$Rh$_2$), 
the majority-band width becomes as narrow as in Fe$_8$,
while the minority band is still comparable to Rh$_8$.
The exchange splitting is large, the majority band  
saturated and only minority states are found close
to $\varepsilon_F$. As in Fe$_8$, $\rho_\downarrow(\varepsilon)$  
shows a clear gap at $\varepsilon_F$ (see Fig.~\ref{fig:dos}).
However, the Rh contribution to the minority states below
$\varepsilon_F$ remains above average despite the relative
small Rh content. The Fe contribution largely
dominates the unoccupied minority-spin DOS, in agreement with
the larger local Fe moments. 

\section{Discussion}
\label{sec:concl}

The structural, electronic and magnetic properties of small  
Fe$_m$Rh$_n$ clusters having $N = m+n \le 8$ atoms have been investigated 
systematically in the framework of a generalized gradient approximation 
to density-functional theory.
For very small sizes ($N\le 4$ atoms) 
the binding energy $E_B$ shows a non-monotonous dependence on concentration,
which implies that the FeRh bonds are stronger than the homogeneous ones. 
However, for larger sizes the FeRh and RhRh bond strengths 
become comparable, so that $E_B$ depends weakly on concentration for
high Rh content.

The magnetic order of the clusters having the most stable structures is found to be FM-like. 
Moreover, the average magnetic moment per atom $\overline\mu_N$ increases 
monotonously, which is almost linear over a wide range of concentration with Fe content.
Consequently, the energy gain $\Delta E_m$ associated to magnetism also
increases with the number of Fe atoms. The largest part of the 
spin polarization (about 90\%) can be traced back to the 
local $d$ magnetic moments within the PAW sphere of the atoms.
The $s$ and $p$ spin polarizations are almost negligible in general.
A remarkable enhancement of the local Fe moments is
observed as result of Rh doping. This is a 
consequence of the increase in the number of Fe $d$ holes, 
due to charge transfer from Fe to Rh, combined with 
the extremely reduced local coordination.
The Rh local moments are important already in the
pure clusters ($N\le 8$). Therefore, they are not significantly enhanced
by Fe doping. However, the overall stability of magnetism,
as measured by the total energy gained by the onset of spin 
polarization, is found to increase with increasing Fe content.

A further interesting aspect, particularly for future 
studies is the role of spin-orbit (SO) interactions on the magnetism of nanoalloys.
We have performed some representative calculations by taking into account spin-orbit coupling (SOC) in order to explore their effect on the ground-state structure, 
chemical order and spin moments. For example in Fe$_6$, Fe$_3$Rh$_3$ and Rh$_6$ we find that the changes in the ground-state energy resulting from SO interaction are typically of the order of 0.2 eV for the whole cluster.
This is often comparable to or larger than the energy differences between the low-lying isomers. However, 
the SO energies are very similar for different structures, so that the ground-state structures remain essentially the same as in the scalar relativistic (SR) calculations. The 
changes in the bond lengths and in the spin moments resulting from SOC are also very small (e.g., $|\overline\mu^{\rm SOC}-\overline\mu^{\rm SR}| \simeq$ 0.01$\mu_B$ and $|d_{ij}^{\rm SOC}-d_{ij}^{\rm SR}| \simeq$ 0.001{\AA} in Fe$_3$Rh$_3$).
As a result, the conclusions drawn from our SR calculations
on the relative stability and local spin moments seem to be unaffected by the spin-orbit contributions.

FeRh clusters are expected to develop a variety of further interesting 
behaviors, which still remain to be explored. For instance, 
larger cluster should show a more complex dependence of the magnetic 
order as a function of concentration. In particular for large Rh content
one should observe a transition from FM-like to 
AF-like order with increasing cluster size, in agreement
with the AF phase found in solids for more than 50\% Rh concentration. 
Moreover, the metamagnetic transition observed in bulk FeRh alloys 
also puts forward
the possibility of similar interesting phenomena in nanoalloys as 
a function of temperature. Finally, the contributions of orbital 
magnetism and magnetic anisotropy deserve to be explored in 
detail as a function of composition and chemical order, even for 
the smallest sizes, particularly because of their implications 
for potential applications.\cite{apl}

\begin{acknowledgments}

It is pleasure to thank Dr.~J.~L.~Ricardo-Ch\'avez and Dr. L. D\'iaz-S\'anchez for helpful discussions
and useful comments. Computer resources from ITS (Kassel) and CSC (Frankfurt) are gratefully acknowleged.

\end{acknowledgments}

\begin{thebibliography}{99}
\bibitem{foot-acc}
A number of tests have been performed in order to assess the 
numerical accuracy of the calculations. Increasing the cut-off energy E$_{max}$ = 268 eV and supercell size $a$ = 12~{\AA} to E$_{max}$ = 500 eV and $a$ = 22~{\AA}
in Rh$_4$ increases the computation time
by a factor $4$--$7$. This yields a total energy differences of 1.75 meV
and 0.25 meV, respectively. In the above calculations the changes in average 
bond length (bond angle) amounts to 10$^{-3}$ {\AA} (10$^{-4}$ degrees). These difference are not significant 
for our physical conclusions. In fact, typical isomerization energies in these clusters 
are an order of magnitude larger, of the order of $10$--$30$ meV. We also found that the total energy 
is nearly independent of the choice of the smearing parameter $\sigma$, provided it is not too large ($\sigma$ $\leq$ 0.05~eV). Values from $\sigma$ = 0.01 to 0.1 eV have been checked. Therefore, we judge
that our set of standard parameters (E$_{max}$ = 268 eV,
supercell size $a$ from 10 to 22 {\AA}, and $\sigma$ = 0.02 eV)
offers a sufficiently good accuracy at a reasonable computational costs.
\bibitem{tobe}
Results for lower lying isomers and excited magnetic configurations
will be reported elsewhere.
\bibitem{foot_R0}
The precise choice of the NN distances is not very important, since this
concerns merely the configuration for starting the unconstrained 
structural relaxation.
\end{thebibliography}
\end{document}